\documentclass[sn-mathphys-num]{sn-jnl}% Math and Physical Sciences Numbered Reference Style

\usepackage{graphicx}%
\usepackage{multirow}%
\usepackage{amsmath,amssymb,amsfonts}%
\usepackage{amsthm}%
\usepackage{mathrsfs}%
\usepackage[title]{appendix}%
\usepackage{xcolor}%
\usepackage{textcomp}%
\usepackage{manyfoot}%
\usepackage{booktabs}%
\usepackage{algorithm}%
\usepackage{algorithmicx}%
\usepackage{algpseudocode}%
\usepackage{listings}%
\usepackage{makecell}%
\usepackage{graphbox}%
\usepackage{threeparttable}%
\usepackage{setspace}%
\usepackage{geometry}%
\usepackage[inkscapelatex=false]{svg}%
\usepackage{subfig}%
\usepackage{adjustbox}%
\usepackage{array}

\newtheorem{definition}{Definition}%

\begin{document}
	
\title[Article Title]{Heterogeneous Graph Representation of Stiffened Panels with Non-Uniform Boundary Conditions and Loads}

\author*[1]{\fnm{Yuecheng} \sur{Cai}}\email{ycai05@mail.ubc.ca}
\author[1,2]{\fnm{Jasmin} \sur{Jelovica}}\email{jasmin.jelovica@ubc.ca}

\affil[1]{\orgdiv{Department of Mechanical Engineering}, \orgname{The University of British Columbia}, \orgaddress{\city{Vancouver}, \state{BC}, \postcode{V6T 1Z4}, \country{Canada}}}
\affil[2]{\orgdiv{Department of Civil Engineering}, \orgname{The University of British Columbia}, \orgaddress{\city{Vancouver}, \state{BC}, \postcode{V6T 1Z4}, \country{Canada}}}

\abstract{Surrogate models are essential in structural analysis and optimization. We propose a heterogeneous graph representation of stiffened panels that accounts for geometrical variability, non-uniform boundary conditions, and diverse loading scenarios, using heterogeneous graph neural networks (HGNNs). The structure is partitioned into multiple structural units, such as stiffeners and the plates between them, with each unit represented by three distinct node types: geometry, boundary, and loading nodes. Edge heterogeneity is introduced by incorporating local orientations and spatial relationships of the connecting nodes. Several heterogeneous graph representations, each with varying degrees of heterogeneity, are proposed and analyzed. These representations are implemented into a heterogeneous graph transformer (HGT) to predict von Mises stress and displacement fields across stiffened panels, based on loading and degrees of freedom at their boundaries. To assess the efficacy of our approach, we conducted numerical tests on panels subjected to patch loads and box beams composed of stiffened panels under various loading conditions. The heterogeneous graph representation was compared with a homogeneous counterpart, demonstrating superior performance. Additionally, an ablation analysis was performed to evaluate the impact of graph heterogeneity on HGT performance. The results show strong predictive accuracy for both displacement and von Mises stress, effectively capturing structural behavior patterns and maximum values.}


\maketitle

\clearpage % Start a new page for highlights
\section*{Article Highlights}
\begin{itemize}
	\item A novel heterogeneous graph approach is proposed to represent stiffened panels
	\item Node and edge heterogeneities are explored to achieve optimal modeling performance
	\item Non-uniform boundary conditions and loads are accurately captured for diverse structural geometries
\end{itemize}

\section{Introduction}\label{sec1}

Stiffened panels are structural components used in thin-walled structures in aerospace, marine and civil engineering. These panels consist of a plate with attached stiffeners, offering excellent load-bearing capacity while allowing relatively simple manufacturing for structures such as bridges, ship hulls, offshore platforms, aircraft wings and fuselages. Finite element method (FEM) is often used for the analysis of real-life structures made of stiffened panels \cite{clough1990original}. Coupling FEM with optimization algorithms can lead to significant advancements in design of large-scale engineering structures \cite{samanipour2020adaptive,jelovica2022improved,chu2021design}. However, this approach can be computationally very intensive, especially when dealing with complex geometries or multiple design configurations, since the number of function evaluations for optimization of real life structures with a metaheuristic typically from $10^4$ to $10^6$ \cite{klanac2009optimization, cai2023neural}. This cost becomes prohibitive when nonlinear FEM is required to assess large-sale structures, such as ship hulls. To reduce the computational cost of structural analysis, various well-established structural theories are available, such as generalized beam theory (GBT) \cite{davies1994second}, asymptotic beam sectional analysis (VABS) \cite{cesnik1997vabs,yu2002validation}, carrera unified formulation (CUF) \cite{carrera2014finite,carrera2011beam}, and equivalent single layer (ESL) \cite{avi2015equivalent,reddy2003mechanics,putranto2021ultimate,putranto2022ultimate}. While powerful, these theories often necessitate simplifying assumptions about geometry, material behavior, or loading conditions.

Surrogate models are employed in engineering to reduce the computational cost of analyzing complex systems, particularly when the original model is too time-consuming or resource-intensive to run repeatedly, as in design optimization. Traditional surrogate models include multivariate adaptive regression splines (MARS), kriging (KRG), radial basis functions (RBF), and response surface method (RSM) \cite{chen2006review}. However, applying the aforementioned models to complicated engineering problems is limited due to their inherent assumptions. Moreover, they may be limited in terms of handling the geometrical complexity of a structure. Among these, neural networks (NNs) have been gaining more interest lately, showing exceptional capabilities to emulate nonlinear physical behavior, particularly in engineering \cite{mai2022robust,shojaeefard2013modelling,kabir2021failure}. 

Multilayer perceptron (MLP) is one of the most commonly used surrogate models, largely due to its ability to approximate any continuous function as demonstrated by the universal approximation theorem \cite{hornik1989multilayer}. By modeling the engineering problems parametrically, researchers can build up the model by identifying problem variables as MLP inputs. Early studies of MLP-based surrogate models for structural analysis can be found in Refs. \cite{papadrakakis1998structural,bisagni2002post,sun2021prediction,limbachiya2021application,ferreira2022lateral}, where researchers have used one or more hidden layers for predicting load carrying capacity, buckling load, shear and lateral-torsional resistance, etc. This approach also proved to be efficient in many recent studies, e.g., Refs. \cite{pham2022free,shamass2022web,zarringol2023artificial,zhu2023artificial,xu2023novel}.
For structures that can be represented as 2D or 3D matrices, researchers have explored more advanced NNs such as convolutional neural networks (CNNs) and generative adversarial networks (GANs) to capture more complex structural features, such as various mechanical characteristics of composite materials \cite{ramkumar2021unconventional}, and nonlocal response for fexoelectric structures \cite{wang2023cnn}.

Unlike CNNs, which operate on Euclidean grids, graph neural networks (GNNs) are designed to handle the complex, non-Euclidean structures prevalent in many real-world applications, such as computer vision \cite{xu2017scene}, chemistry \cite{gilmer2017neural}, biology \cite{fout2017protein}, and recommender systems \cite{ying2018graph}, etc. Recent advancements of GNNs in mechanical engineering have also achieved great success. In fluid mechanics, GNNs are used as reduced order models (ROMs) to replace the expensive computational fluid dynamics (CFD) analysis \cite{lino2022multi,shao2023pignn,pfaff2020learning,gao2022finite}. Benefiting from GNN's scalability, efficiency and propagation property, researchers adopted different types of GNNs to model truss and truss-like problems, where the joints are modelled as nodes, and each bar is represented as an edge \cite{zheng2023tso,chou2024structgnn,whalen2022toward,cao2024vertex}. Similar techniques have also been adopted to metamaterials \cite{xue2023learning} and lattice structures \cite{jiang2024gnns,jain2024latticegraphnet}.

Recent work has pushed GNNs beyond homogeneous‐graph surrogates and pure interpolation into both optimization and physics‐informed frameworks. On the design side, Zhang et al. developed an automated GNN‐based tool for 2D structural‐layout generation \cite{zhang2024end}, and Zhao et al. proposed an intelligent beam‐layout synthesis method for frame structures, demonstrating robust plan creation for complex building systems \cite{zhao2023intelligent}. Li et al. combined GNNs with exploratory genetic algorithms to automate clash‐free rebar detailing, achieving up to a 90\% reduction in computation time \cite{li2023automated}. Furthermore, Zhang et al. (2024) introduced a fully differentiable framework that embeds structural layouts as graphs and computes performance gradients end-to-end, enabling real-time, gradient-based design updates \cite{zhang2024differentiable}. On the analysis side, some researchers embed problem-related physics into the neural network loss functions to reduce data requirements while improving fidelity. This approach is generally conceptualized as Physics-Informed Neural Networks (PINNs) \cite{raissi2019physics}. Song et al. (2023) introduced StructGNN-E, a physics-informed GNN that performs elastic structural analysis without labeled data, enforcing equilibrium via a modified graph‐isomorphism network \cite{song2023elastic}. Parisi et al. (2024) further investigated mechanics-injected GNNs, embedding stress‐energy residuals directly into the network’s architecture to improve prediction fidelity for large civil‐engineering systems \cite{parisi2024use}. However, applications of GNNs to thin-walled structural components (e.g., stiffened steel panels) remain scarce. Recently, \cite{cai2024efficient} modeled stiffened panels using coarse homogeneous graphs where each rectangular-shaped plate was modeled as a node. However, they oversimplified boundary conditions and loads by collapsing each edge to a single scalar value, limiting the method’s ability to capture non-uniform effects.

To account for more realistic (non-uniform) loading and edge boundary conditions, we propose a heterogeneous graph representation of stiffened panels in this study. Heterogeneous representations enable more accurate information encoding by distinguishing between different types of nodes and edges, thereby increasing the expressiveness of the graph. The superiority of heterogeneous graph representations has been demonstrated in several studies \cite{hu2020heterogeneous, wang2019heterogeneous, fu2020magnn}, showing improved accuracy over homogeneous GNNs on multiple datasets. Leveraging these properties, our approach represents stiffened panels heterogeneously by defining the model into three distinct node types: geometry, boundary, and loading nodes. This approach aims to preserve important structural information that might be lost in homogeneous representations. To accurately capture the interactions between these node types, we also incorporate edge heterogeneity, where distinct edge types connect the corresponding nodes based on their local orientations and spatial relationships. We investigate various types of heterogeneous graph representations variants to investigate the influence of different levels of heterogeneity on model performance. By introducing heterogeneity into the graph representation, the proposed approach can effectively predict the stress and displacement fields of structures subjected to non-uniformly distributed displacements and rotations at structural boundaries and loads. Case studies involve stiffened panels subjected to patch loads and three box beams subjected to different loading conditions. Using these test cases, we compare the new model with the homogeneous graph representation proposed in Ref. \cite{cai2024efficient}, analyze the proposed representations at different levels of heterogeneity, and demonstrate the performance using heterogeneous graph transformer (HGT) for predicting stress and displacements in stiffened panels.

\section{Preliminaries and related works}\label{sec2}

\subsection{Graph definitions}\label{sec2_1}
Unlike CNNs, GNNs can be used to model systems with non-Euclidean distances. One of the simplest forms of a graph is the homogeneous graph, which can be formally defined as:
\begin{definition}[Homogeneous Graph]
A homogeneous graph can be defined as $\mathcal{G} = \{\mathcal{V}, \mathcal{E}, \mathcal{A}\}$, where $\mathcal{V}$ represents the node set, $\mathcal{E}$ denotes the edge set, and $\mathcal{A}$ is the adjacency matrix. An edge $e_{ij} = (v_i, v_j) \in \mathcal{E}$ indicates a connection from node $v_i$ to node $v_j$. The adjacency matrix $\mathcal{A} \in \mathbb{R}^{|\mathcal{V}| \times |\mathcal{V}|}$ is a convenient way to represent the graph structure, where $\mathcal{A}_{ij} = 1$ if $e_{ij} \in \mathcal{E}$, and $\mathcal{A}_{ij} = 0$ otherwise. For an undirected graph, $\mathcal{A}_{ij} = \mathcal{A}_{ji}$.
\end{definition}
To model more complex systems, heterogeneous graphs, also referred to as heterogeneous information networks, can be used. These graphs consist of various types of entities (i.e. nodes) and multiple types of relationships (i.e. edges), which can be defined as follows:
\begin{definition}[Heterogeneous Graph]
A heterogeneous graph can be defined as $\mathcal{G} = \{\mathcal{V}, \mathcal{E},\\ \mathcal{A},\mathcal{T_V},\mathcal{T_E} \}$, where $\mathcal{V}$ represents the node set, $\mathcal{E}$ denotes the edge set, and $\mathcal{A}$ is the adjacency matrix. Each node $v\in\mathcal{V}$ and edge $e\in\mathcal{E}$ in the heterogeneous graph are associated with a mapping function $\phi(v):\mathcal{V}\rightarrow \mathcal{T_V}$ and $\psi(e):\mathcal{E}\rightarrow \mathcal{T_E}$ that maps each node $v$ and edge $e$ to a node type $t_v$ and edge type $t_e$, respectively. $\mathcal{T_V}$ and $\mathcal{T_E}$ denote the node type set and edge type set, respectively. In a heterogeneous graph, $|\mathcal{T_V}|+|\mathcal{T_E}|>2$.
\end{definition}
The heterogeneous graph allows a more versatile representation of multiple types of nodes and relationships, making it suitable for modeling more complex structures/systems where nodes and edges have different roles or attributes.

\subsection{Graph neural networks}\label{sec2_2}
In recent years, there have been a lot of advancements in the development of graph neural networks (GNNs) and heterogeneous graph neural networks (HGNNs) \cite{khemani2024review,morris2024future}. The core mechanism underlying GNNs and HGNNs is the message-passing process, where the representation of a node is aggregated from its one-hop neighborhoods. This process typically involves two key steps: message aggregation and node update \cite{gilmer2017neural}. 

1. Message Aggregation: Assume $m_v^{(l)}$ is the aggregated message of node $v$ at layer $l$. These messages encapsulate information about the one-hop local neighborhood, and can be mathematically expressed as:
\begin{equation}
	\mathbf{m}_v^{(l)} = \text{Aggregate} \left( \alpha_{u,e,v}\cdot \mathbf{h}_u^{(l-1)} : \forall u \in \mathcal{N}(v),\forall v \in\mathcal{V},\forall e \in \mathcal{E}(u,v) \right)
\end{equation}
\begin{equation}
	\alpha_{u,e,v} = \phi_{\Theta}
	\left(\mathbf{h}_v^{(l-1)}, \mathbf{h}_u^{(l-1)},\mathbf{e}_{u,v}\right) 
\end{equation}
where $h_u^{(l-1)}$ represents the hidden representation of neighbor $u$ at layer $(l-1)$, and $\mathcal{N}(v)$ denotes the set of neighbors of node $v$. The attention value $\alpha_{u,v,e}$ is computed based on the source node $u$, the target node $v$, and the associated edge $e$ using a differentiable function $\phi_{\Theta}$. Graph neural networks often incorporate attention mechanisms, such as the node-level attention used in the graph attention network (GAT) \cite{velivckovic2017graph} and its improved variant, GATv2 \cite{brody2021attentive}. In the context of heterogeneous graphs, various networks have been developed to dynamically calculate neighborhood weights, such as the heterogeneous graph attention network (HAN) \cite{wang2019heterogeneous} and the heterogeneous graph transformer (HGT) \cite{hu2020heterogeneous}, among others. The neighborhood information are often multiplied by the attention value and aggregated by $\text{Aggregate}(\cdot)$, which is defined by a permutation-invariant function, such as, sum, mean, or max \cite{xu2018powerful}.

2. Node Update: After aggregating messages, the node updates its state using the aggregated information. This is typically done using a neural network function, such as a multilayer perceptron (MLP), to ensure the node representation captures higher-order interactions. The updated state $h_v^{l}$ at iteration $l$ is given by:
\begin{equation}
	\mathbf{h_v}^{(l)} = \text{Update} \left( \mathbf{h_v}^{(l-1)}, \mathbf{m_v}^{(l)} \right)
\end{equation}
where $\text{Update}(\cdot)$ is often parameterized by a neural network. Many models demonstrated the effectiveness of this framework, such as graph sampling and aggregation (GraphSAGE) \cite{hamilton2017inductive}:
\begin{align}
	\mathbf{m}_v^{(l)}
	&= \frac{1}{|\mathcal{N}(v)|}\sum_{u\in\mathcal{N}(v)} \mathbf{h}_u^{(l-1)}, \\
	\mathbf{h}_v^{(l)}
	&= \sigma\!\Bigl(\mathbf{W}^{(l)}\,[\,\mathbf{h}_v^{(l-1)} \,\|\, \mathbf{m}_v^{(l)}]\Bigr),
\end{align}
where \(\|\) denotes concatenation, \(\mathbf{W}^{(l)}\) is a learned weight matrix, and \(\sigma\) an activation function. This model has been applied to predict von Mises stress fields in stiffened panel with various geometries \cite{cai2024efficient}. 
Other GNN variants such as graph convolutional network (GCN) \cite{kipf2016semi}, HAN \cite{wang2019heterogeneous}, and HGT \cite{hu2020heterogeneous} can also be modelled under this framework. The versatility and ability to model complex relational data of this framework make it a fundamental building block in the design of modern GNN and HGNN architectures.

\subsection{Homogeneous graph representation}\label{sec2_3}
To represent physical structures as graphs, researchers have developed mesh-based representations by replacing structural mesh information with a graph representation. In this approach, each FE element is represented by a graph node, and graph edges define the connectivity between FE elements. This method captured the topology of complex structures while reducing computational complexity \cite{gao2024finite,pfaff2020learning,block2024fast,li2023machine}. However, one inherent limitation of such an approach is that the number of nodes in the graph is the same as the number of elements in the corresponding FE model, which could be computationally expensive if dealing with a large graph. To further save computational time, researchers employ the graph sparsification \cite{spielman2008graph,spielman2011spectral} and graph coarsening technique \cite{loukas2018spectrally,loukas2019graph,cai2021graph}, which can reduce the complexity of a graph by reducing the number of edges and nodes, respectively. These methods aim to use a sparser graph that can preserve the essential structural properties of the original graph.

To incorporate more engineering biases, Ref. \cite{cai2024efficient} chose to manually conduct the coarsening process for graphs on stiffened panels. While not algorithmically driven, this work effectively aggregates multiple finite elements (which could be seen as smaller nodes in a more detailed graph) from a rectangular-shaped plate into a super-node. Each node’s output vector, containing stress values at different locations in the corresponding plate, ensures that the detailed stress information is preserved. However, Ref. \cite{cai2024efficient} has stacked all structural geometry information, edge boundary conditions, and external loadings into a single vector, neglecting the joint dependencies among these factors. Additionally, the boundary conditions were oversimplified as either fixed or simply supported, which fails to represent real-world conditions. In this work, we address these limitations by introducing a novel heterogeneous graph representation technique, detailed in Section \ref{sec3}.

\section{Methodology}\label{sec3}

\subsection{Heterogeneous graph representation for stiffened panels}\label{sec3_1}

\begin{figure}
    \centering
    \includegraphics[width=1\linewidth]{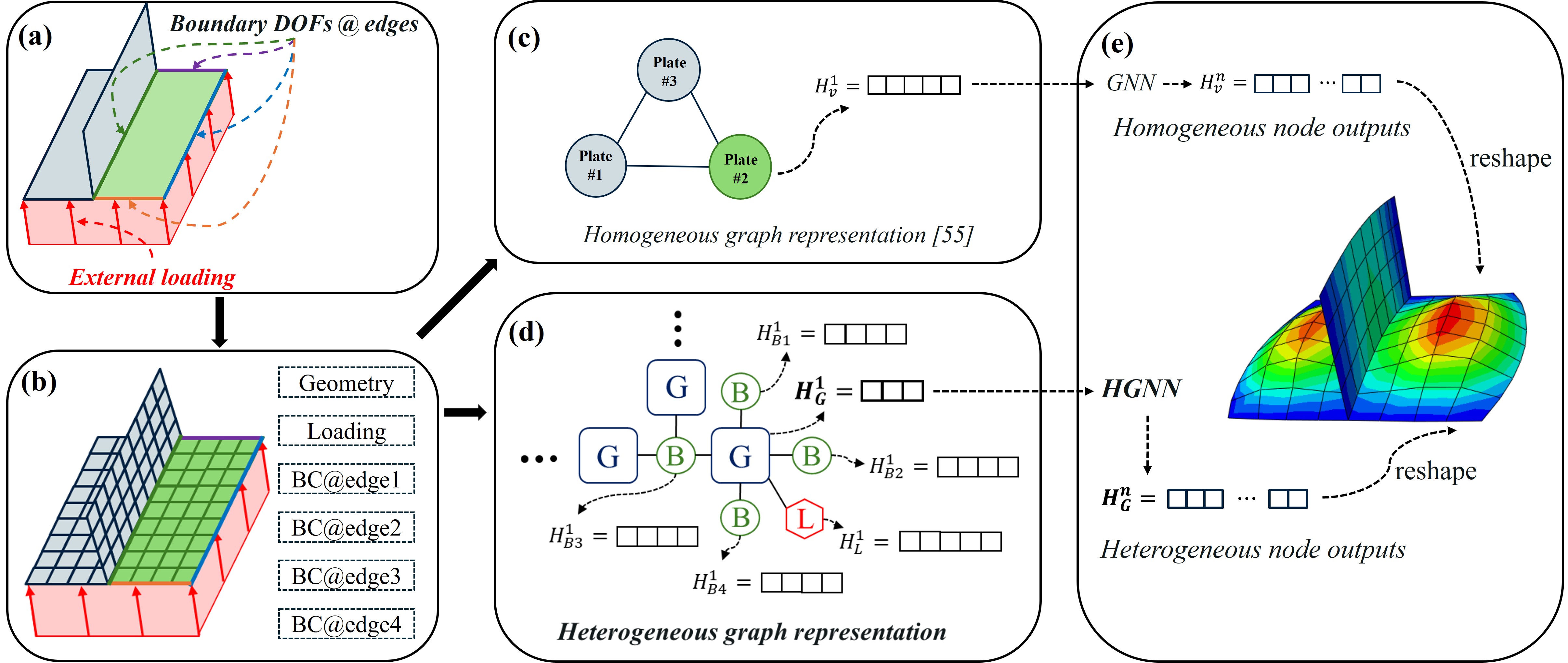}
    \caption{Comparison of the proposed heterogeneous graph representation with the homogeneous graph representation. (a) Physical model of an example panel structure. (b) Corresponding finite element model. (c) Homogeneous graph representation. (d) Proposed heterogeneous graph representation of the example panel structure. (e) Predicted fields by either GNN or HGNN. In the proposed heterogeneous graph representation, geometry, external loading, boundary conditions for each structural unit (e.g., stiffener or plate between stiffeners) are defined as separate node types. In contrast, the homogeneous graph representation concatenates all this information into a single node.}
    \label{fig: Heterogeneous_conceptual_graph}
\end{figure}

\begin{figure}
    \centering
    \includegraphics[width=1\linewidth]{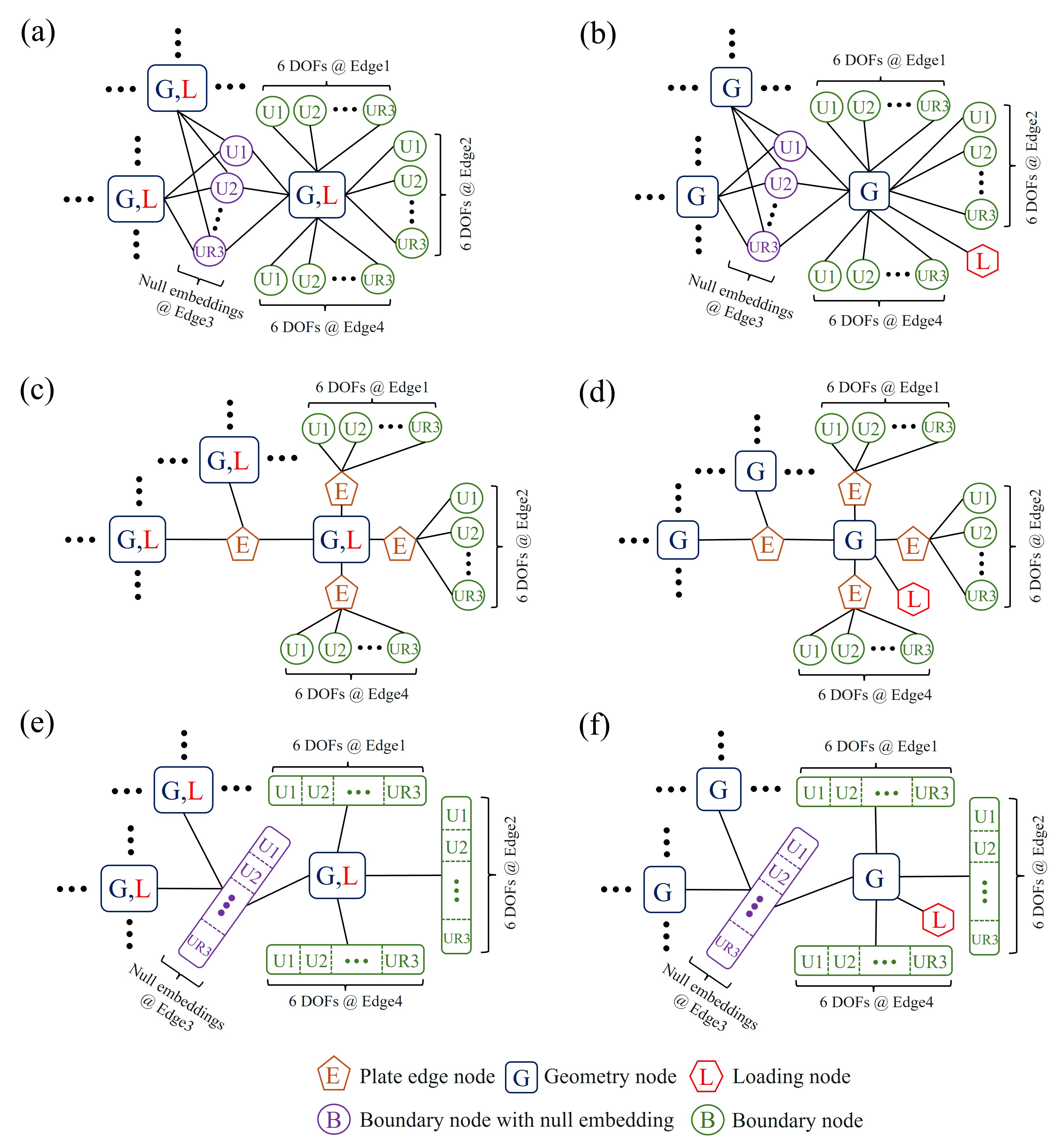}
    \caption{Heterogeneous graph representation with different levels of heterogeneity (a) Separate DOFs (b) Separate DOFs + Isolated loading node (c) Separate DOFs + Edge node (d) Separate DOFs + Edge node + Isolated loading node (e) Combined DOFs (f) Combined DOFs + Isolated loading node.}
    \label{fig: Heterogeneous_graph_alternatives}
\end{figure}

\begin{figure}
    \centering
    \includegraphics[width=0.8\linewidth]{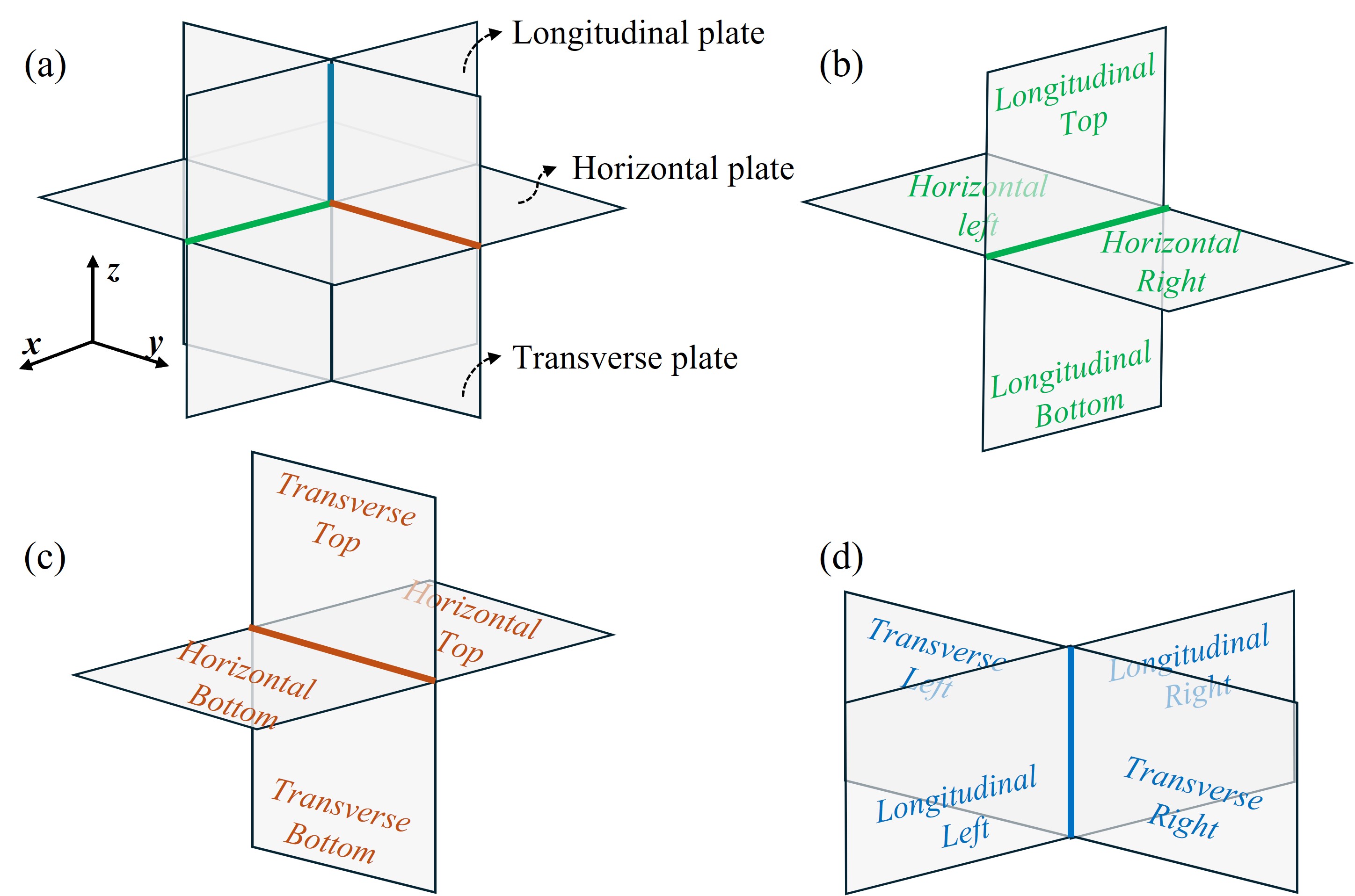}
    \caption{Edge definition illustration. (a) 3D overview of the edge possible locations. (b) Local locations of edges on the x-axis. (c) Local locations of edges on the y-axis. (d) Local locations of edges on the z-axis.}
    \label{fig: Edge_definition}
\end{figure}

This work extends the approach in Ref. \cite{cai2024efficient} from uniform boundary conditions and loadings to non-uniform distributions for both. Representing non-uniformly distributed features typically requires explicit formula definitions. In this study, we discretized the non-uniform boundary conditions (imposed edge displacements and rotations) and loadings into vectors, with values sampled evenly along the edge and through the plate, respectively. Incorporating all this information into a single node becomes less efficient and may affect the quality of neural network predictions. A comparison of the different approaches is provided in Section \ref{sec5_1}.

\subsubsection{Node heterogeneity}\label{sec3_1_1}
To more accurately describe the interactions between external loads, edge boundary conditions, and structures, we introduce heterogeneity to the graph representations, as shown in Fig. \ref{fig: Heterogeneous_conceptual_graph}. Specifically, we represent plate geometries, plate edge boundaries, and external loads of each rectangular-shaped plate as distinct node types.  The geometry node contains the geometric information of the plate; the boundary node represents the displacements and rotations imposed to the corresponding edge, and the loading node encodes the external load applied to the plate in vector form. The connectivity between nodes naturally reflects the structural layout: loading nodes are directly connected to geometry nodes, and each geometry node is linked to four boundary nodes, with each edge treated individually due to its local orientation. This explicit differentiation allows for clearer and more specialized feature integration and learning, allowing for a direct flow of information changes between different structural components. 

Note that since each edge contains displacement and rotation information as six degrees of freedom (DOFs), careful structuring of the graph is necessary to efficiently represent these edge boundary conditions. The heterogeneous graph representation in Fig. \ref{fig: Heterogeneous_conceptual_graph} simplifies each plate edge to a single boundary node, which does not fully reflect the heterogeneous graph employed in this study. Fig. \ref{fig: Heterogeneous_graph_alternatives} provides a detailed illustration of the heterogeneous graph representation, presenting six configurations with varying levels of node heterogeneity. The DOFs of an edge can be represented either by separate nodes for each DOF or by a single node that concatenates all DOFs. Moreover, because the displacements of interior edges are unknown, we substitute them with a learnable null embedding plus a binary mask. Additionally, to remove the need for null embedding, the separated boundary nodes can be connected to the geometry node through an additional edge node (Fig. \ref{fig: Heterogeneous_graph_alternatives} c and d). Finally, the external loading can either be integrated into the geometry node or retained as a separate node. We conducted an ablation study on these different representations, with the results presented in Section \ref{sec5_2}.

\subsubsection{Edge heterogeneity}\label{sec3_1_2}
To accurately represent the interactions between different node types, we utilized distinct edges to connect corresponding nodes. This variation in edge definition is referred to as edge heterogeneity in this study. Each unique node type is connected by a specific edge type, with the goal of maximizing the interactions within the structure. However, the connections between geometry nodes and boundary nodes vary depending on the local orientation of the edge relative to the plate.

This variation arises because non-uniform boundary condition definitions incorporate spatial information along the edge. When an edge is shared by multiple plates, its local orientation differs for each plate. Fig. \ref{fig: Edge_definition} illustrates this situation. Assuming all plates are positioned in 3D space without any inclination, their edges can occupy one of three possible global orientations, indicated by green, brown, and blue lines. Each edge is connected to a maximum of four plates, each with a distinct local spatial relationship to the edge. Consequently, the current configuration contains 12 different types of geometry-boundary edges. In contrast, the geometry-loading edge remains consistent across all geometry and loading node connections, as these interactions do not involve local spatial relationships or orientations.

Edge heterogeneity can also impact the number of parameters required to train the network, as most HGNNs assign a distinct learnable matrix to each meta-relation triplet $(u, e, v)$. Representing separate DOFs on edges as distinct nodes enables a more detailed learning flow but significantly increases the number of matrices assigned to these triplets. Introducing an additional edge node between geometry and boundary nodes can reduce the number of learnable matrices by transferring spatially oriented edges from geometry-boundary connections (Fig. \ref{fig: Heterogeneous_graph_alternatives}a and b) to edge-boundary connections (Fig. \ref{fig: Heterogeneous_graph_alternatives}c and d). In this configuration, only one spatially oriented edge is defined between the geometry and edge nodes, reducing the number of spatially oriented edges from six to one. The boundary nodes are then connected to the edge node through spatially invariant edges.

\subsection{HGT architecture}

\begin{figure}
    \centering
    \includegraphics[width=1\linewidth]{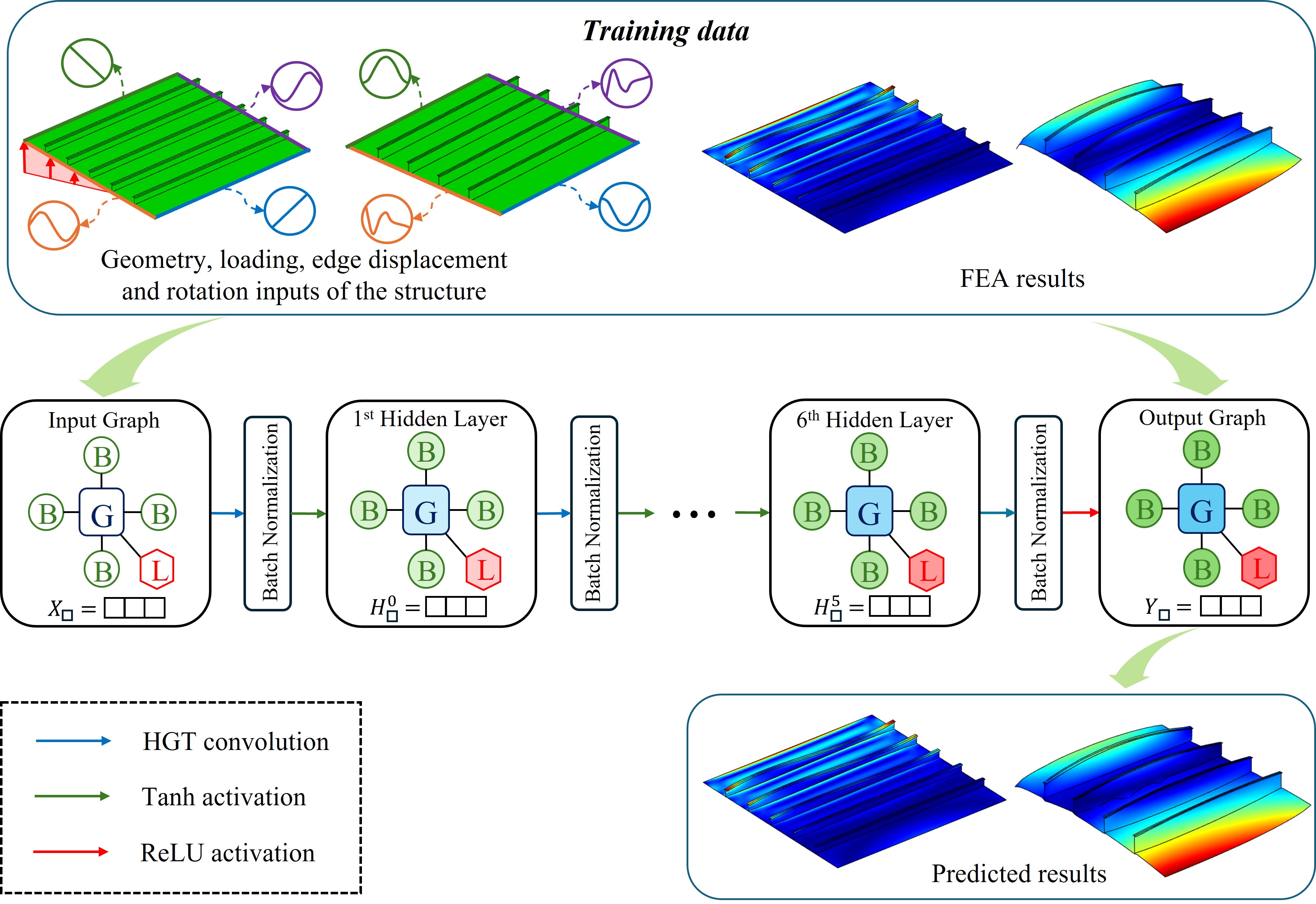}
    \caption{Heterogeneous graph transformer architecture, with inputs and outputs of the network.}
    \label{fig: HGT_architecture}
\end{figure}

In this study, we utilized the Heterogeneous Graph Transformer (HGT) \cite{hu2020heterogeneous} to learn features at each layer of the heterogeneous graph, as HGT emerged as the best-performing model based on our preliminary evaluation of various HGNNs. The HGT convolutional layer is defined as:

\begin{equation}
h_v^{(l)} = \sigma(\tilde{h}_v^{(l)})W_{\mathcal{T_V}(v)}^A + h_v^{(l-1)}
\end{equation}

where \(h_v^{(l)}\) represents the message (embedding) for node \(v\) at layer \(l\), which combines the message from the previous layer \(l-1\) with the updated message \(\tilde{h}_v^{(l)}\) multiplied by a linear projection \(W_{\mathcal{T_V}(v)}^A\). Following the framework introduced in Section \ref{sec2_2}, the updated message for node \(v\) at layer \(l\) is calculated as:

\begin{equation}
\tilde{h}_v^{(l)} = \bigoplus_{\forall u \in N(v)} \left( \alpha_{u,e,v} \cdot \textbf{m}_{u,e,v}^{(l)} \right)
\end{equation}

where \(\textbf{m}_{u,e,v}^{(l)}\) is the multi-head message of the node pair \((u,v)\), calculated by:

\begin{equation}
\textbf{m}_{u,e,v}^{(l)} = \parallel_{i \in [1,h]}  h_u^{(l-1)} W^{MSG}_{\mathcal{T_V}(u)} W^{MSG}_{\mathcal{T_E}(e)}
\end{equation}

Here, \(h\) denotes the number of heads, and \(W^{MSG}_{\mathcal{T_V}(u)}\) and \(W^{MSG}_{\mathcal{T_E}(e)}\) are the source node-specific and edge-specific matrices, respectively. The attention value \(\alpha_{u,e,v}\) in HGT is inspired by the self-attention mechanism used in transformers \cite{vaswani2017attention}. Specifically, for each meta-relation triplet \( (u, e, v) \), the attention coefficients \( \alpha_{u,e,v} \) are calculated using multi-head attention with \( h \) heads as:

\begin{equation}
\alpha_{u,e,v} = \text{Softmax}\left( \big\Vert_{i=1}^{h} \text{ATT-Head}_{i}^{(l)}(u, e, v) \right),
\end{equation}

where \( \big\Vert \) denotes concatenation over the \( h \) attention heads. Each attention head is computed as:

\begin{equation}
\text{ATT-Head}_{i}^{(l)}(u, e, v) = \left( K_u^{(l)} W^{\text{ATT}}_{\mathcal{T}_E(e)} Q_v^{(l)} \right) \cdot \frac{\mu_{\langle \mathcal{T}_V(u), \mathcal{T}_E(e), \mathcal{T}_V(v) \rangle}}{\sqrt{d}},
\label{eq: att-head}
\end{equation}

with the key and query vectors defined as:

\begin{equation}
K_u^{(l)} = h_u^{(l)} W^{K}_{\mathcal{T}_V(u)},
\end{equation}

\begin{equation}
Q_v^{(l)} = h_v^{(l)} W^{Q}_{\mathcal{T}_V(v)}.
\end{equation}

The query \( Q_v^{(l)} \) and key \( K_u^{(l)} \) are first computed by projecting the hidden states of nodes \( v \) and \( u \) using the type-specific matrices \( W^{Q}_{\mathcal{T}_V(v)} \) and \( W^{K}_{\mathcal{T}_V(u)} \), respectively. Unlike the vanilla self-attention mechanism, the attention score in HGT incorporates an edge-type-specific matrix \( W^{\text{ATT}}_{\mathcal{T}_E(e)} \) to better model distributional differences, as shown in Eq. \ref{eq: att-head}. Additionally, the attention score is scaled by a prior tensor dependent on the meta-relation triplet \( (u, e, v) \) to further refine the attention.

The architecture of the HGT network used in this study is illustrated in Fig. \ref{fig: HGT_architecture}. Structures with varying geometries, loadings, and boundary conditions (non-uniform displacements and rotations along an edge) are converted into heterogeneous graphs. Let $\tau\in\mathcal T_V$ be a generic node‐type. Then $X_\tau$ is the input embedding at layer 1 for nodes of type $\tau$. These embeddings are then fed into the HGT model, which predicts the von Mises stress and displacement fields of the structures. To stabilize the training process, batch normalization layers were added after each HGT layer. The hyperparameters were fine-tuned using a quasi-random search algorithm, with the final values summarized in Table \ref{tab: NN hyperparameter}. A discussion on the influence of hyperparameters on HGT performance is provided in Appendix \ref{secA1}.
We adopted the Root Mean Squared Error (RMSE) as the loss function:

\begin{equation}
\text{RMSE} = \sqrt{\frac{1}{n}\sum_{i=1}^{n}(\textbf{y}_i - \hat{\textbf{y}}_i)^2}\label{Eq: RMSE}
\end{equation}

where \(n\) represents the number of data points, and \(\textbf{y}_i\) and \(\hat{\textbf{y}}_i\) denote the actual and predicted values for the \(i\)th data point (stress or displacement at a point in a panel), respectively. RMSE provides an estimate of the average magnitude of error, placing greater emphasis on larger discrepancies between predicted and actual values due to the squaring of the individual errors. This characteristic is crucial for surrogate models in structural analysis, since highest stress and displacement values are of primary importance.

\begin{table}[h]
\centering
\caption{Hyperparameter settings for the employed HGT network.}\label{tab: NN hyperparameter}%
\small
\begin{tabular}{@{}lc@{}}
\toprule
Category & Value  \\
\midrule
Number of hidden layers    & 6   \\
Number of hidden neurons for each layer   & 64  \\
Activation function    & tanh   \\
Optimizer   & adam \\
Learning rate  & 0.001  \\
Batch size  & 200 \\
L2 regularization  & 1e-5\\
\toprule
\end{tabular}
\end{table}

\section{Case studies and data preparation}\label{sec4}
\begin{figure}[htp]
\centering
\includegraphics[width=0.6\textwidth]{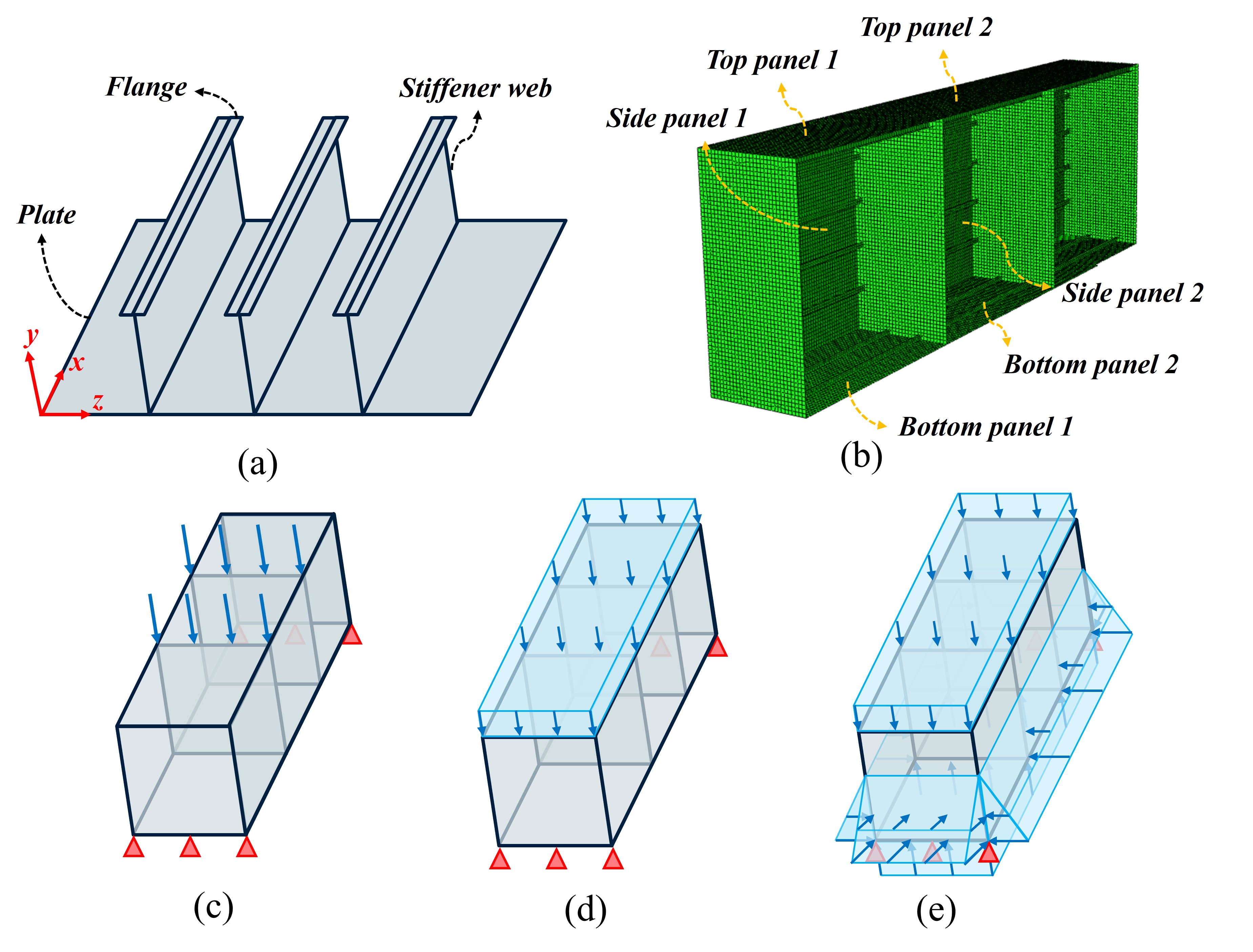}
\caption{(a) Schematic of a stiffened panel. (b) Sectional-view of the finite element box beam model. (c) Loading conditions for box beams in test case 2. (d) Loading conditions for box beams in test case 3. (e) Loading conditions for box beams in test case 4.}\label{fig: Stiffened_panel_and_box_beam}
\end{figure}

This study aims to accurately predict stress distribution in stiffened panels of various geometries under arbitrary boundary conditions and external loads. Four test cases are considered, one for isolated stiffened panels from Ref. \cite{cai2024efficient}, and three additional test cases involving a box beam made of stiffened panels. Main repeating unit of a stiffened panel is shown in Fig. \ref{fig: Stiffened_panel_and_box_beam}a. Fig. \ref{fig: Stiffened_panel_and_box_beam}b presents a sectional view of the finite element model of the box beam used in this study, highlighting the six unique panels based on structural and loading symmetry. The following is a brief summary of the four test cases:

\begin{itemize}
  \item Case 1: Isolated stiffened panel \cite{cai2024efficient}
  \item Case 2: Box beam subjected to four-point bending;
  \item Case 3: Box beam with uniform pressure applied to the top panel;
  \item Case 4: Box beam with uniform pressure on the top panel and water pressure on the bottom and side panels;
\end{itemize}

For cases 2, 3, and 4, we selected stiffened panels from a series of $9m\times3m\times3m$ box beams (Fig. \ref{fig: Stiffened_panel_and_box_beam}b), where the top, bottom, and side panels are all stiffened panels. Consequently, each dataset includes stiffened panels from various locations on the box beam.  All box beams contain four bulkheads with a distance of $3m$, with a random thickness between 40 and 60 mm. The geometries for panels on the top, bottom, and side of the box beam are set randomly and independently in the data sets. Taking into account all panels of the box beam, Table \ref{tabl: var limit} summarizes the range of geometry variables for each test case. We allow the height/width and thickness of all structural units to vary continuously within the specified domain, to enable a more granular representation of the design space.

\begin{table}[h]
\centering
\caption{Lower and upper limits of geometric variables for the stiffened panels in different cases.}\label{tabl: var limit}%
\small
\begin{tabular}{@{}lccc@{}}
\toprule
Category & Case 2 \& 3  & Case 4 & Unit\\
\midrule
Plate thickness    & 10 - 20   & 5 - 10  & mm  \\
Stiffener thickness    & 5 - 20   & 4 - 8  & mm  \\
Stiffener height    & 100 - 400  & 50 - 150  & mm  \\
Flange thickness  & 5 - 20 & 4 - 8 & mm \\
Flange width   & 50 - 150 & 50 - 100 & mm \\
Number of stiffeners  & 2 - 7 & 2 - 7 & $-$ \\
\toprule
\end{tabular}
\end{table}

To further differentiate those test cases, we applied different loading to the box beam structures, as shown in Fig. \ref{fig: Stiffened_panel_and_box_beam}. In test case 2, the line load is applied at the connections between bulkheads and top plates (4-point bending), ranging from 500 to 1000 kN/m. In test case 3, the pressure is applied uniformly across the top plate, ranging from \(1.11 \times 10^5\) to \(3.33 \times 10^5\) Pa. Test case 4 represents a scenario where the box beam is pressure-loaded and immersed in water. We considered both the structural weight and water buoyancy to better simulate real-world conditions. In this case, the structural density is set to \(7.85 \times 10^{-9}\) tonnes/mm\(^3\). The pressure loading applied to the top plate ranges from \(1.11 \times 10^5\) to \(3.33 \times 10^5\) Pa. The water pressure on the bottom and side panels is calculated based on the top pressure and the structural weight. Additionally, we adjusted the range of geometry variables in this test case to ensure that some structures could exhibit plastic material behavior. Under these specifications, the internal bending moments and shear forces vary markedly among the different test cases. For example, the four-point bending case results in linearly increasing moment from the supports to the bulkheads where the load is imposed while shear force is constant. Central region has a constant bending moment without shear force. However, since the beam is a hollow thin-walled structure, the distribution of shear forces at a cross-section varies from the neutral axis towards the top/bottom panels and finally to the centre line: shear forces reduce to zero at the centre line; axial stress is lower at the midspan of a top/bottom panel in comparison to corner points. Stress distribution becomes more complicated in cases 3 and 4. In case 3, uniform pressure generates a parabolic moment distribution along the length of the beam. In case 4, non-uniform pressure leads to asymmetric forces and moments. This variation in load distribution leads to very different nonlinear displacements and rotations along the edges of the stiffened panels for each box beam.

\begin{figure}[htp]
\centering
\includegraphics[width=0.5\textwidth]{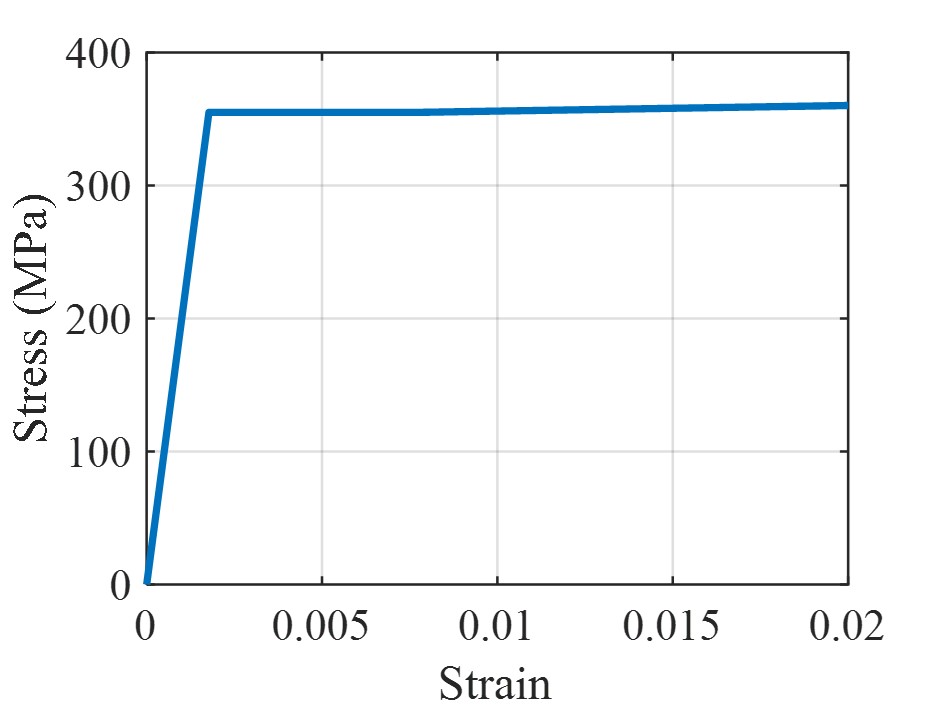}
\caption{Employed elastic-plastic steel stress-strain curve.}\label{fig: Material}
\end{figure}

The material employed in this study is elastic-plastic steel with Young's modulus of 200 GPa and a Poisson's ratio of 0.3. The initial yield stress is $\sigma_0 = 355 MPa$, and the plasticity of the employed nonlinear material follows the equation below:
\begin{equation}
	\sigma_f(\overline{\varepsilon}) =
	\begin{cases}
		\sigma_0 & \text{if $\overline{\varepsilon} \leq \varepsilon_L$}\\
		K(\overline{\varepsilon}_0 + \overline{\varepsilon})^n & \text{if $\overline{\varepsilon} > \varepsilon_L$}
	\end{cases}\label{Eq: Material}       
\end{equation}
where 
\begin{equation}
	\overline{\varepsilon}_0 = (\sigma_0 /K)^{1/n} - \varepsilon_L ,\label{Eq: Material2}       
\end{equation}
$\overline{\varepsilon}$ is the plastic strain, $\varepsilon_L$ is the plateau strain taken as 0.006 in this paper, work hardening parameters involve $K$ and $n$, which are 530 MPa and 0.26, respectively. The stress-strain curve of the employed elastic-plastic steel is represented in Fig.\ref{fig: Material} \cite{putranto2022ultimate}.

The data set is obtained using the ABAQUS FEM software, utilizing parametric modeling. `SR4' shell element is used for all structural parts. To ensure model accuracy, we utilize a FE mesh density of 80 elements along the stiffener direction, 10 elements between stiffeners, 6 elements for stiffener height, and 4 elements for flanges which was confirmed sufficient in a preliminary study and is very similar to Ref. \cite{cai2024efficient}. Due to geometric variations among structural units, we implement scatter interpolation to standardize the dimensions of NN inputs for loading and boundary nodes, as well as outputs for geometry nodes. Based on initial study here and in Ref. \cite{cai2024efficient}, we determine loading and boundary node features to be sampled on a uniform one-dimensional grid of $1\times 20$ points, while geometry node outputs are sampled on a two-dimensional grid of $10\times 20$ points (10 across the shorter dimension, 20 along the longer) for each structural unit (plate strip, web and flange). Denser sampling could improve accuracy, but with additional computational expenses. Material nonlinearity is discretized in a piecewise manner, employing Eq. \ref{Eq: Material} and Eq. \ref{Eq: Material2}, with updates made for every 0.001 plastic strain increment. The training procedure of all NNs in this study is executed using Pytorch Geometric and carried out on a computing device with a GTX 3090 GPU.

\section{Results and discussion}\label{sec5}

In this section, we employed HGT as the HGNN model, and GraphSAGE as the baseline GNN model for the performance evaluation and comparisons. Specifically, we first conduct an ablation analysis on the graph heterogeneity, specifically for graph representations shown in Fig.\ref{fig: Heterogeneous_graph_alternatives}, to demonstrate the effectiveness of each heterogeneous graph variant. Afterwards, we compare the homogeneous graph representation from Ref. \cite{cai2024efficient} with the heterogeneous graph representation proposed here, to demonstrate the versatility and generalization performance of the proposed approach. In the end, for each of the test cases described in Section \ref{sec4}, we evaluate the performance of the HGT with the best performing heterogeneous graph representation.

\subsection{Ablation analysis on graph heterogeneity}\label{sec5_1}

The ablation analysis in this study focuses on the impact of graph heterogeneity, using test case 4, which is the most complicated case here. As depicted in Fig. \ref{fig: Heterogeneous_graph_alternatives}, we explore different ways to represent the stiffened panel structure across varying levels of heterogeneity. Specifically, the six DOFs on each edge can be defined either separately or jointly, and the separately defined boundary nodes can also be connected to the geometry nodes via an additional edge node. For each of the three boundary node configurations, we further isolated the loading node to assess its impact on neural network performance. Note that different levels of graph heterogeneity may influence the number of parameters for training, which has been discussed in Section \ref{sec3_1_2}. Each model was trained five times using the same random seed. The mean and standard deviation are reported in Table \ref{tab: ablation}.

\begin{table}[h]
	\centering
	\caption{Ablation analysis on graph heterogeneity for test case 4.}\label{tab: ablation}%
	\footnotesize
	\begin{tabular}{@{}lccc@{}}
		\toprule
		Category & \#Param & RMSE (MPa)  & Difference \\
		&&(Avg. $\pm$ st.dev.)&(\% from best)\\
		\midrule
		(a) Separate DOFs & 3.98M    & 9.47$\pm$0.60  & 81.4\%  \\
		(b) Separate DOFs + Isolated loading node    & 4.19M   & 7.38$\pm$0.50 & 41.4\% \\
		(c) Separate DOFs + Edge node    & 2.08M  & 16.43$\pm$2.57  & 214.7\%  \\
		(d) Separate DOFs + Edge node + Isolated loading node  & 2.28M & \textbf{5.23$\pm$0.36} & - \\
		(e) Combined DOFs   & \textbf{0.81M} & 8.87$\pm$0.71 & 69.9\% \\
		(f) Combined DOFs + Isolated loading node  & 1.02M & 7.41$\pm$0.49 & 41.9\% \\
		\toprule
	\end{tabular}
\end{table}

The results indicate that introducing an additional node for external loading improves HGT performance across all three boundary configurations, despite a relatively small increase in the number of parameters. This improvement is attributed to the spatial information contained in the non-uniformly distributed loading, which can be more accurately captured when treated separately from the structural geometry information. Additionally, the choice of DOF definition has a considerable impact on the number of parameters in the HGT network. Compared to separating each DOF of the boundary, combining them can reduce the number of parameters by up to 79.6\%. However, this reduction may slightly compromise HGT performance for the test case considered here (case 4), as separating the DOFs into individual nodes might more accurately represent their effects on the structures. The best-performing graph was achieved by the heterogeneous graph type (d) illustrated in Fig. \ref{fig: Heterogeneous_graph_alternatives}, where external loading and each DOF were considered separately and connected via an additional edge node.

\subsection{Comparison between homogeneous and heterogeneous graph representation on stiffened panels}\label{sec5_2}

\begin{figure}[htp]
	\centering
	\includegraphics[width=1\textwidth]{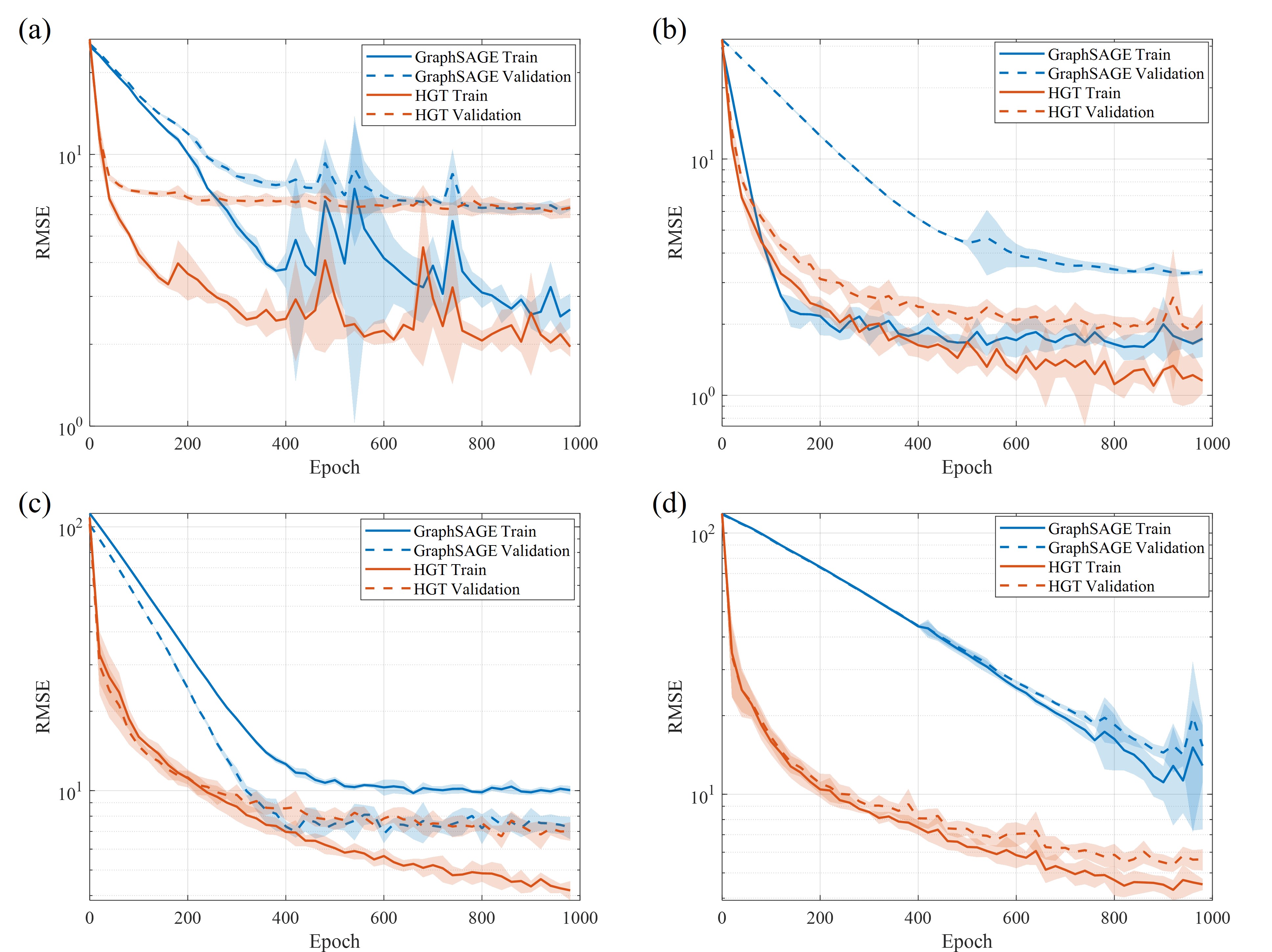}
	\caption{Training and validation RMSE for GraphSAGE (Homogeneous representation) vs. HGT (Heterogeneous representation) over five runs. (a) Case 1, (b) Case 2, (c) Case 3, (d) Case 4. Solid lines denote the mean training RMSE, dashed lines the mean validation RMSE, for both GraphSAGE (blue) and HGT (orange) over 0–1000 epochs. Shaded ribbons show $\pm\,$ one standard deviation across five independent runs.}\label{fig: HGT and SAGE loss curve}
\end{figure}

In this subsection, we compare the performance of the homogeneous (GraphSAGE) with the heterogeneous graph network (HGT) using four test cases introduced in Section \ref{sec4}. The homogeneous graph representations used for comparison follow the approach illustrated in Fig. \ref{fig: Heterogeneous_conceptual_graph}, where structural geometry information, boundary conditions, and external loadings are concatenated into a single vector. For the heterogeneous graph representation, we selected type (d) from Fig. \ref{fig: Heterogeneous_graph_alternatives} due to its superior performance compared to the other configurations (Table \ref{tab: ablation}). The comparison is presented in Table \ref{tab: homo and hetero}. All models were trained five times using the same random seed, with the mean and standard deviation of RMSE (in MPa) reported for each case. The corresponding training and validation curves for each test case are shown in Fig. \ref{fig: HGT and SAGE loss curve}. 

\begin{table}[h]
	\centering
	\caption{Comparison of homogeneous and heterogeneous graph representations for RMSE (MPa) of the stress prediction.}\label{tab: homo and hetero}%
	\footnotesize
	\begin{tabular}{@{}lcc@{}}
		\toprule
		Test case & Homogeneous representation & Heterogeneous representation \\
		\midrule
		Case 1: Isolated stiffened panel \cite{cai2024efficient}   &  6.197$\pm$0.186  & \textbf{6.084}$\pm$0.336 \\
		Case 2: Box beam - four point bending   & 3.260$\pm$0.076  & \textbf{1.756}$\pm$0.079 \\
		Case 3: Box beam - uniform pressure    & 6.553$\pm$0.407   & \textbf{6.230}$\pm$0.085 \\
		Case 4: Box beam - non-uniform pressure   & 13.423$\pm$0.595 & \textbf{5.230}$\pm$0.363 \\
		\toprule
	\end{tabular}
\end{table}

Overall, HGT with the proposed heterogeneous graph representation yields both lower error and consistency than the homogeneous baseline. In all test cases, HGT converges more rapidly and stabilizes in a lower RMSE regime. These differences were particularly significant for cases 2 and 4, with RMSE reductions of 46.1\% and 61.0\%, respectively. The reduced impact in case 1 is likely due to its oversimplified boundary conditions, where introducing heterogeneity into the graph may not significantly benefit the learning process. Similarly, in case 3, the structural behavior for stiffeners on the top of box beam are dominated by uniform pressure rather than edge boundary conditions, leading to a less noticeable improvement. However, in cases with pronounced non-uniformity of boundary DOFs and loads, the heterogeneous graph representation proved more effective.

\subsection{Performance evaluation of the proposed heterogeneous representation on stiffened panels}\label{sec5_3}

In this section, we examine the performance of the heterogeneous graph transformer (HGT) on the test cases introduced in Section \ref{sec4} and compare it with the GraphSAGE with homogeneous representation in \cite{cai2024efficient}. The heterogeneous graph type (f) (see Section \ref{sec3} and \ref{sec5_1}) is used here because of its high accuracy and computational efficiency. For all figures in this section, ‘HGT’ denotes the predicted field from the employed HGT model, ‘GraphSAGE’ denotes the predicted field from the GraphSAGE model with the homogeneous graph representation, ‘FEA’ denotes the corresponding ABAQUS results, and ‘Error’ is the absolute pointwise difference between ABAQUS and each neural network. For each test case, we performed 2000 unique ABAQUS FEM simulations (directly on isolated panels for Case 1 and on parametrically generated box beams for Cases 2–4) and then randomly partitioned the resulting instances into 80\% training, 10\% validation, and 10\% test sets. On our test workstation (Intel i7-12700 CPU, NVIDIA RTX 3090 GPU), generating the 2000 case FEA database took about 10 hours of CPU time and training HGT networks took 6 hours of GPU time, while the model predicts the response of a new panel under 0.1 seconds. By contrast, a single ABAQUS run takes 10–60 seconds on the same CPU. Following the practice in several PINN works \cite{es2024separable,rezaei2022mixed}, four distinct networks were trained separately to predict each of the three translational displacement components and the von Mises stress. The total displacement field presented in this section is a combination of the three predicted translational displacement fields, U1, U2, and U3, align with the x-, y-, and z-axes, respectively (see Fig. \ref{fig: Stiffened_panel_and_box_beam}). To show the best performance of the models implemented in this section, we tuned the hyperparameters of all neural networks using the quasi-random search algorithm. A brief discussion on the influence of hyperparameters on the predictions can be found in \ref{secA1}. 

Table \ref{tab: pred metrics} summarizes the RMSE for stress, total displacement, and each displacement component across cases 2–4.  Compared to GraphSAGE, HGT reduces stress RMSE by 49.0\%, 8.2\%, and 54.4\%, and total displacement RMSE by 60.4\%, 59.2\%, and 41.5\%, for cases 2-4, respectively.  Component‐wise, all U1 to U3 errors decrease by about 55–65\%, except U3 in Case 4. Because U2 magnitudes are roughly an order of magnitude larger than U1 and U3, total‐displacement RMSE closely matches RMSE for U2. These consistent improvements emphasize the advantage of the proposed heterogeneous graph representation. Additionally, for each test case, we chose two representative test examples, aiming to show a broad spectrum of conditions, including elastic versus plastic response, different number of stiffeners, and distinct positions along the box beam. The statistical ranking (percentiles within the dataset) for all test examples are shown in Table \ref{tab: percentile}. The following subsections provide a detailed, case-by-case breakdown of these results.

\begin{table}[htp]
	\centering
	\caption{HGT and GraphSAGE prediction metrics for stress and displacements}
	\label{tab: pred metrics}
	\begin{tabular}{llccccc}
		\toprule
		Test case   & Model       & Stress (MPa)  & Total disp. (mm)&      U1 (mm)   &      U2 (mm)   &      U3 (mm)   \\
		\midrule
		\multirow{2}{*}{Case 2} 
		& HGT         & 0.9710  & 0.07730    & 0.002686  & 0.07821   & 0.01147   \\
		& GraphSAGE   & 1.904   & 0.1907     & 0.005945  & 0.1949    & 0.01773   \\
		\midrule
		\multirow{2}{*}{Case 3} 
		& HGT         & 2.765   & 0.06968    & 0.01081   & 0.07212   & 0.01190   \\
		& GraphSAGE   & 3.012   & 0.1709     & 0.01911   & 0.1817    & 0.02651   \\
		\midrule
		\multirow{2}{*}{Case 4} 
		& HGT         & 3.712   & 0.1504     & 0.01000   & 0.1523    & 0.01378   \\
		& GraphSAGE   & 8.136   & 0.2571      & 0.02821   & 0.2642     & 0.01291   \\
		\bottomrule
	\end{tabular}
\end{table}

\begin{table}[htp]
	\centering
	\caption{Percentile ranks of selected test examples based on RMSE for displacement and von Mises stress predictions for both models.}
	\label{tab: percentile}
	\begin{tabular}{@{}l c cc cc@{}}
		\toprule
		\multirow{2}{*}{Test case} & \multirow{2}{*}{Example} 
		& \multicolumn{2}{c}{HGT} 
		& \multicolumn{2}{c}{GraphSAGE} \\
		\cmidrule(lr){3-4} \cmidrule(lr){5-6}
		&  & Disp (\%) & Stress (\%) & Disp (\%) & Stress (\%) \\
		\midrule
		Case 2: 4-point bending 
		& 1 & 20  & 75 & 40 & 55 \\
		& 2 & 90  & 85 & 60 & 95 \\[3pt]
		Case 3: uniform pressure 
		& 3 & 45  & 75 & 50 & 80 \\
		& 4 & 5   & 1  & 20 & 10 \\[3pt]
		Case 4: non-uniform pressure 
		& 5 & 10  & 25 & 20 & 5 \\
		& 6 & 35  & 50 & 50 & 50 \\
		\bottomrule
	\end{tabular}
\end{table}

\subsubsection{Box beam subjected to 4-point bending}\label{sec5_3_1}

\begin{figure}[h]
	\centering
	\scriptsize
	\setlength\tabcolsep{1mm}
	\resizebox{\textwidth}{!}{%
		\begin{tabular}{cccc}
			\toprule
			Test example   & Model & Displacement field & von Mises field \\
			\midrule
			\multirow{20}{*}{\rotatebox{90}{1 (Top panel 1)}}  
			& HGT  & \makecell[c]{\includegraphics[width=0.4\textwidth]{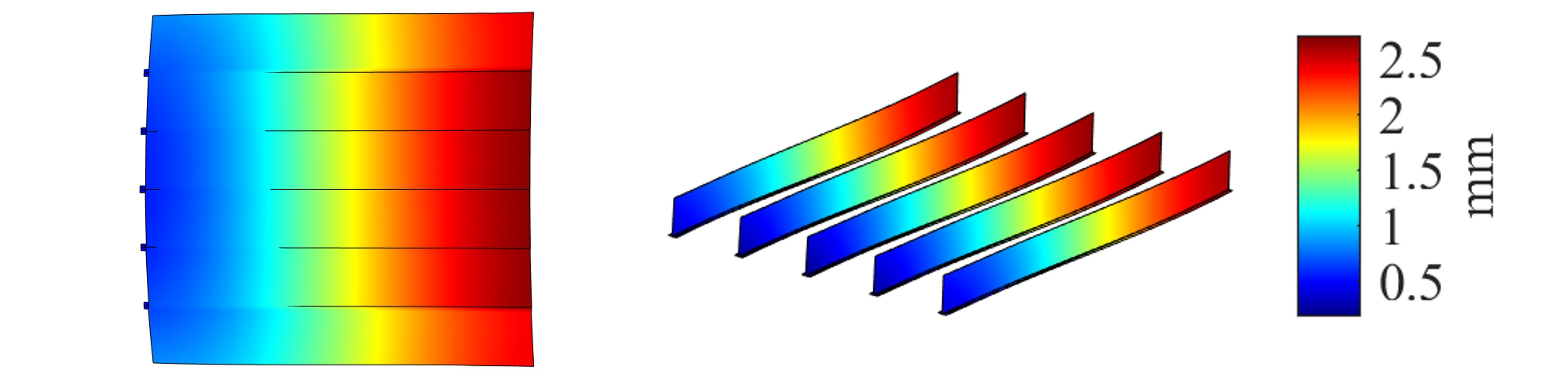}} 
			& \makecell[c]{\includegraphics[width=0.4\textwidth]{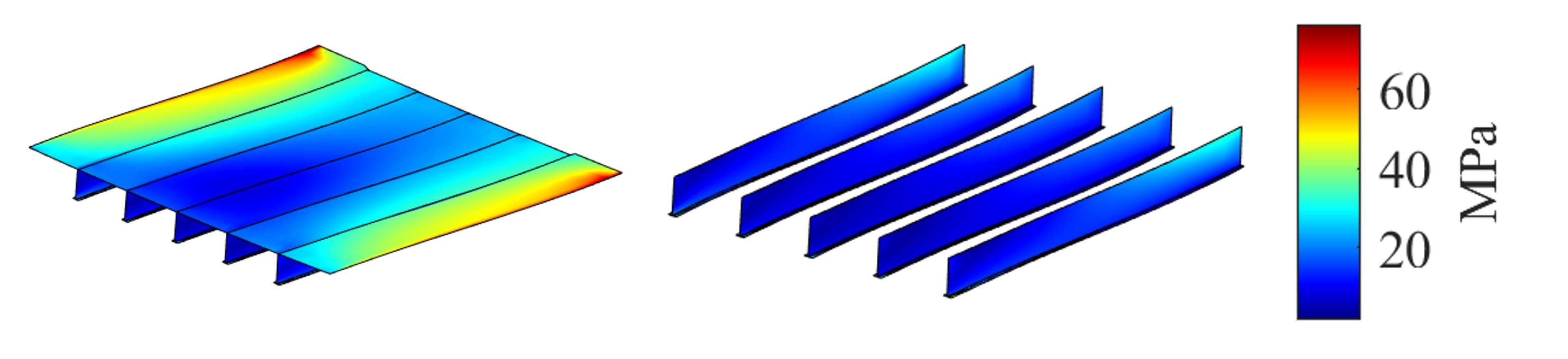}} \\
			& \makecell[c]{HGT Error\\(Disp.\,0.0522 mm; \\Stress\,1.25 MPa)}  
			& \makecell[c]{\includegraphics[width=0.4\textwidth]{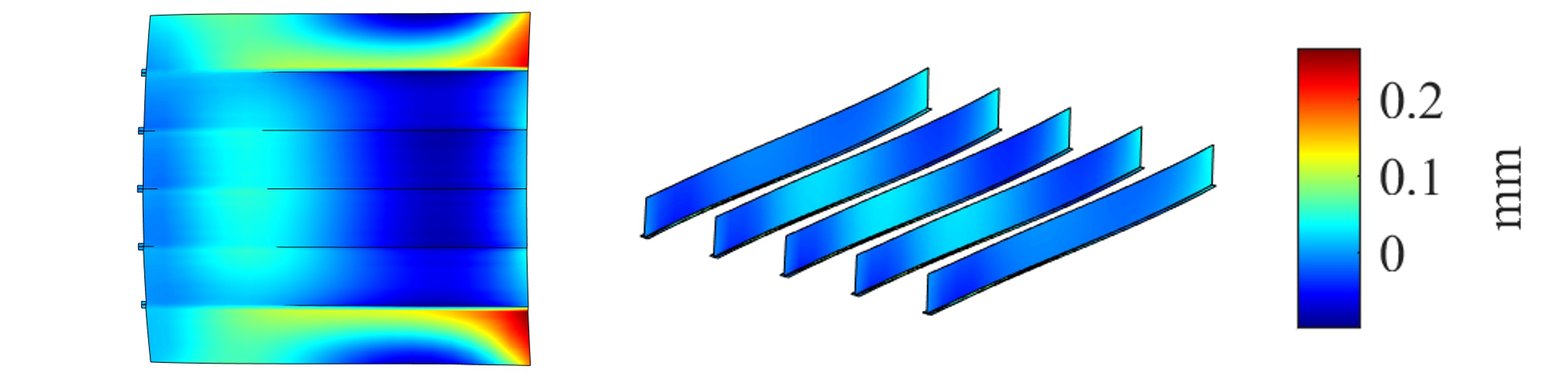}}
			& \makecell[c]{\includegraphics[width=0.4\textwidth]{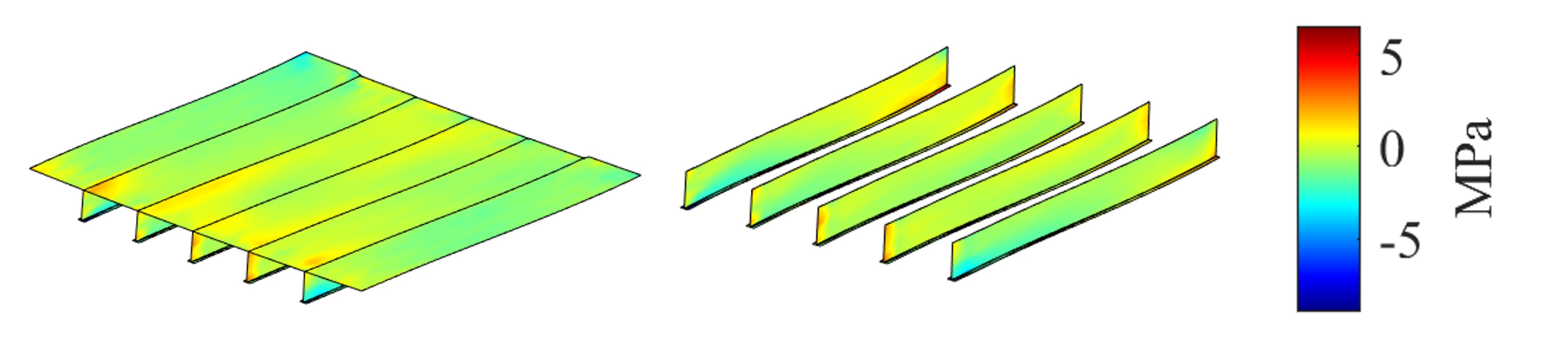}} \\
			& GraphSAGE  
			& \makecell[c]{\includegraphics[width=0.4\textwidth]{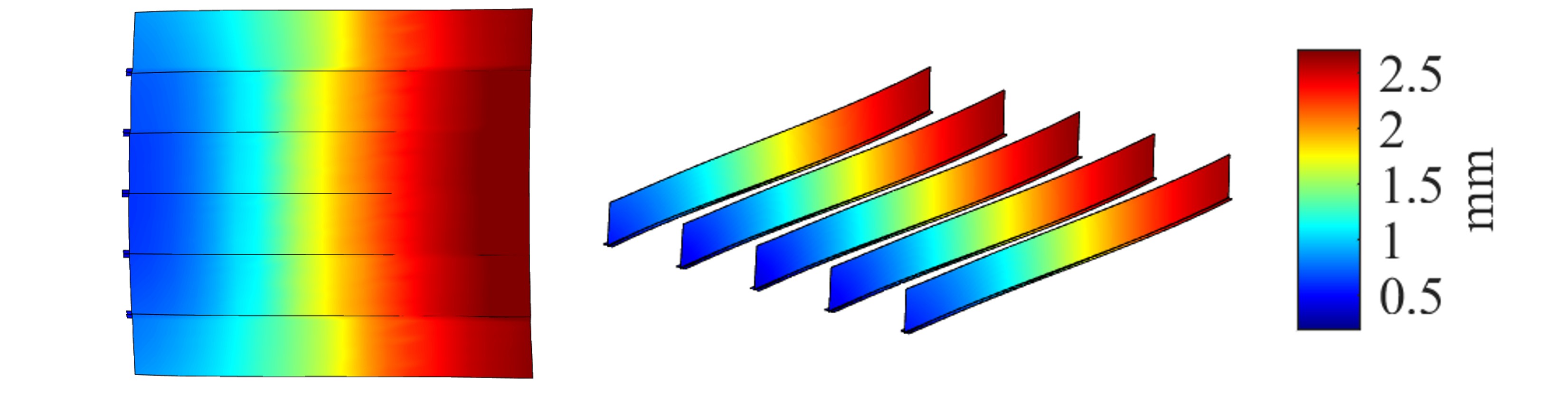}}
			& \makecell[c]{\includegraphics[width=0.4\textwidth]{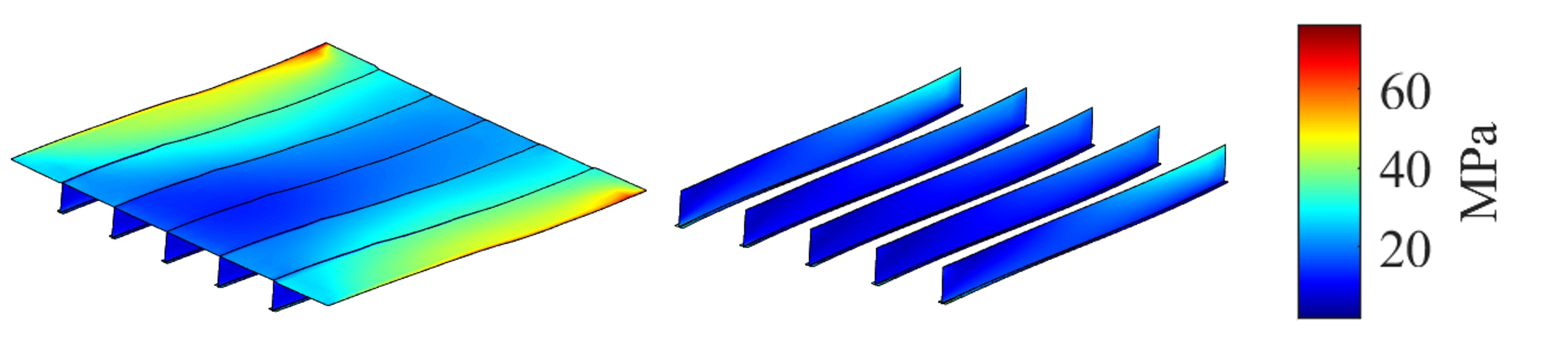}} \\
			& \makecell[c]{GraphSAGE Error\\(Disp.\,0.170 mm;\\Stress\,2.09 MPa)}  
			& \makecell[c]{\includegraphics[width=0.4\textwidth]{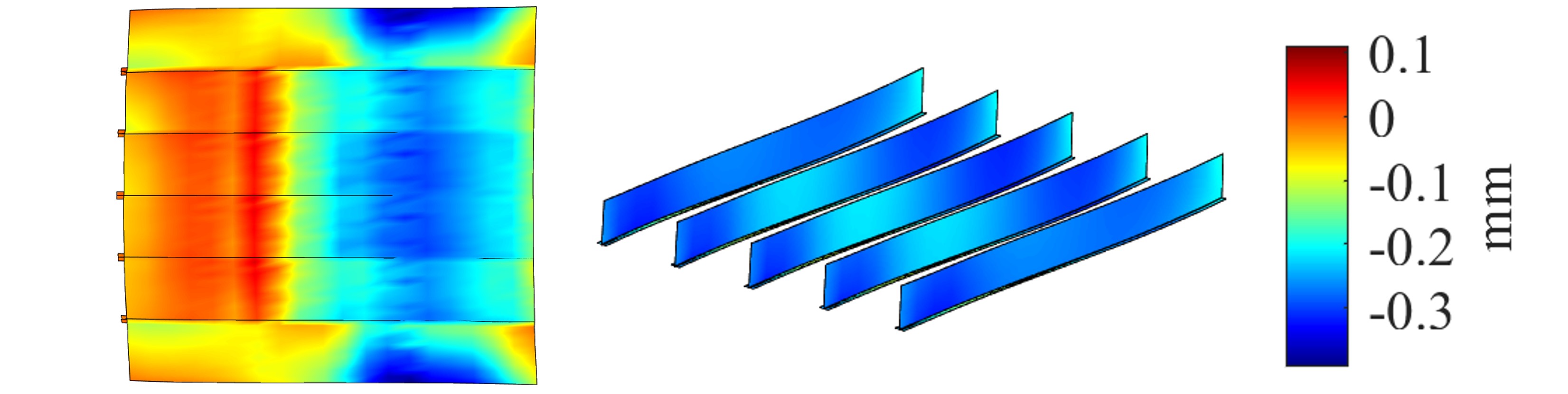}}
			& \makecell[c]{\includegraphics[width=0.4\textwidth]{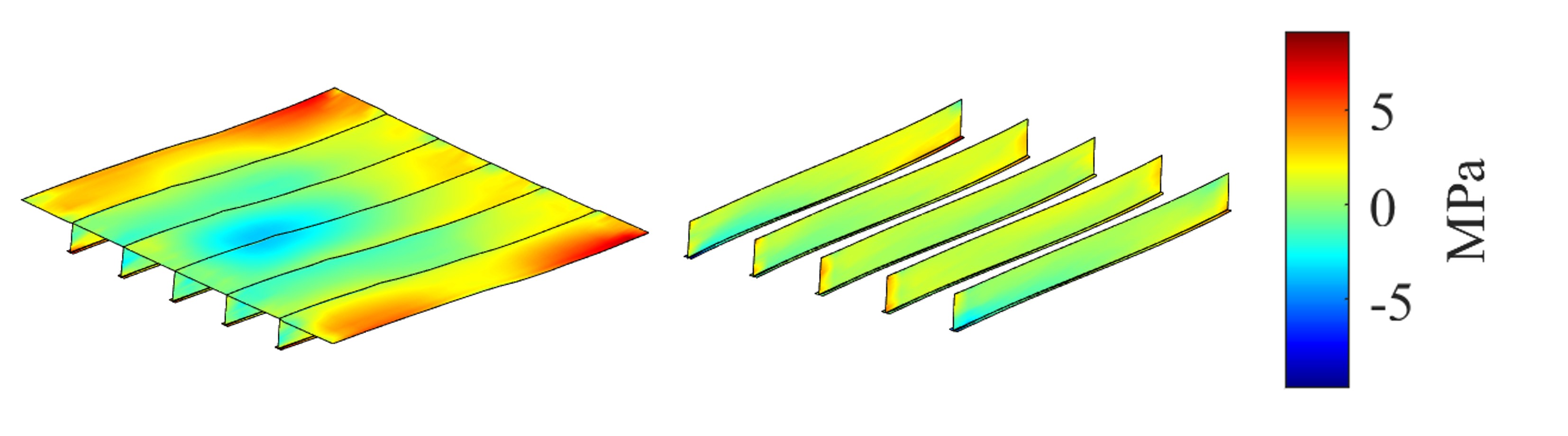}} \\
			& FEA  
			& \makecell[c]{\includegraphics[width=0.4\textwidth]{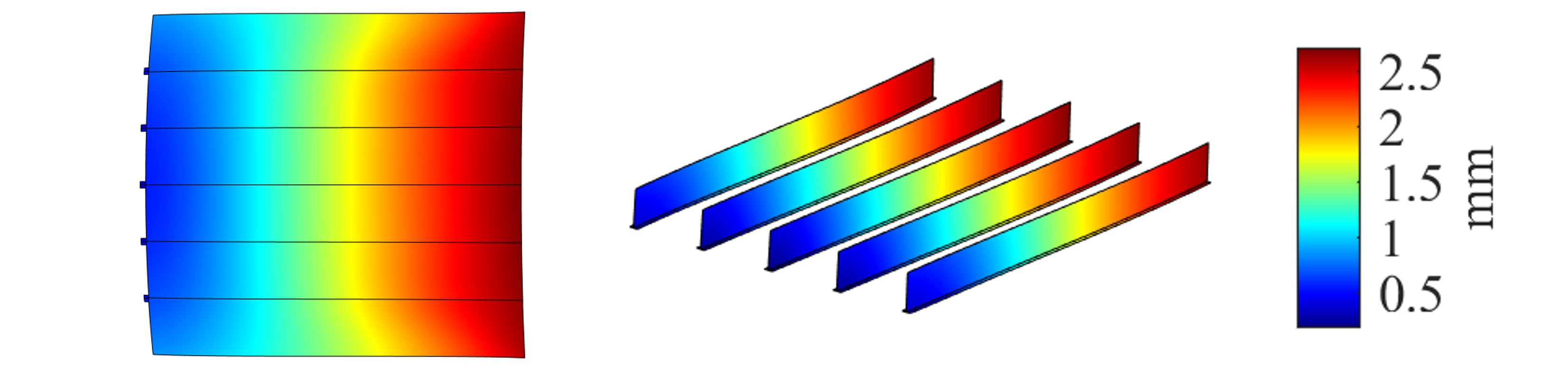}}
			& \makecell[c]{\includegraphics[width=0.4\textwidth]{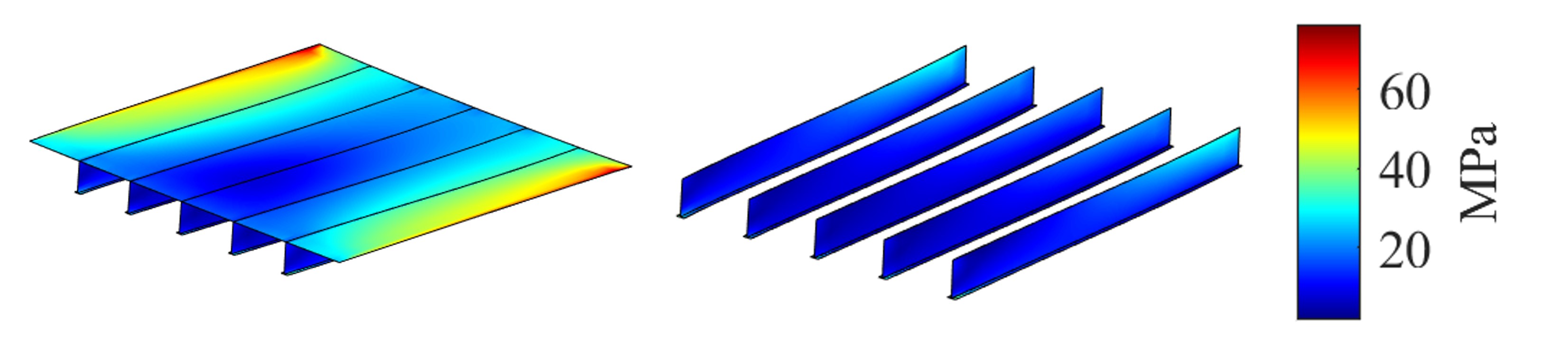}} \\
			\midrule
			\multirow{20}{*}{\rotatebox{90}{2 (Side panel 1)}}  
			& HGT  
			& \makecell[c]{\includegraphics[width=0.4\textwidth]{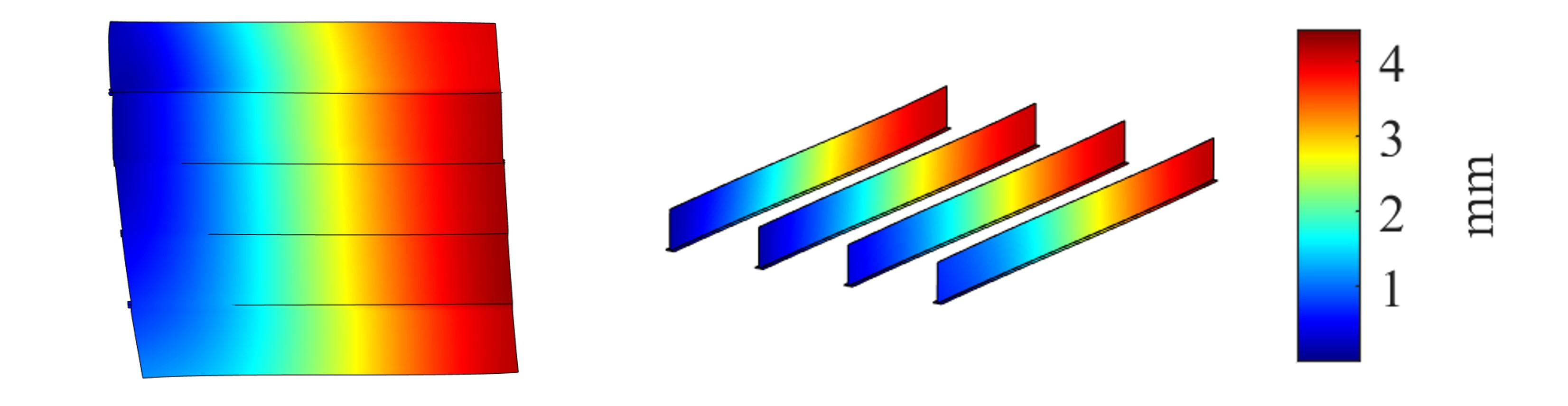}}
			& \makecell[c]{\includegraphics[width=0.4\textwidth]{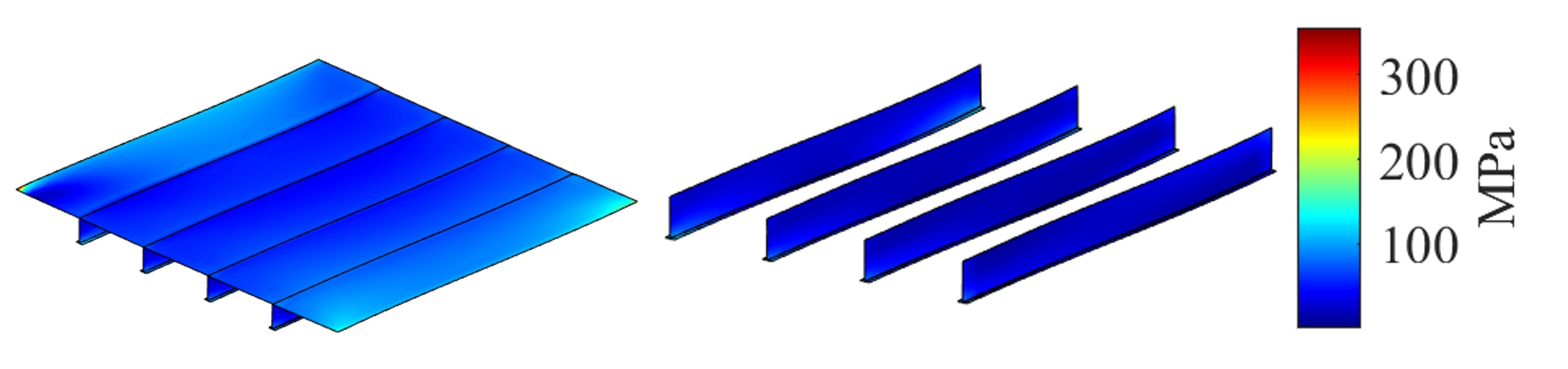}} \\
			& \makecell[c]{HGT Error\\(Disp.\,0.128 mm;\\Stress\,1.92 MPa)}  
			& \makecell[c]{\includegraphics[width=0.4\textwidth]{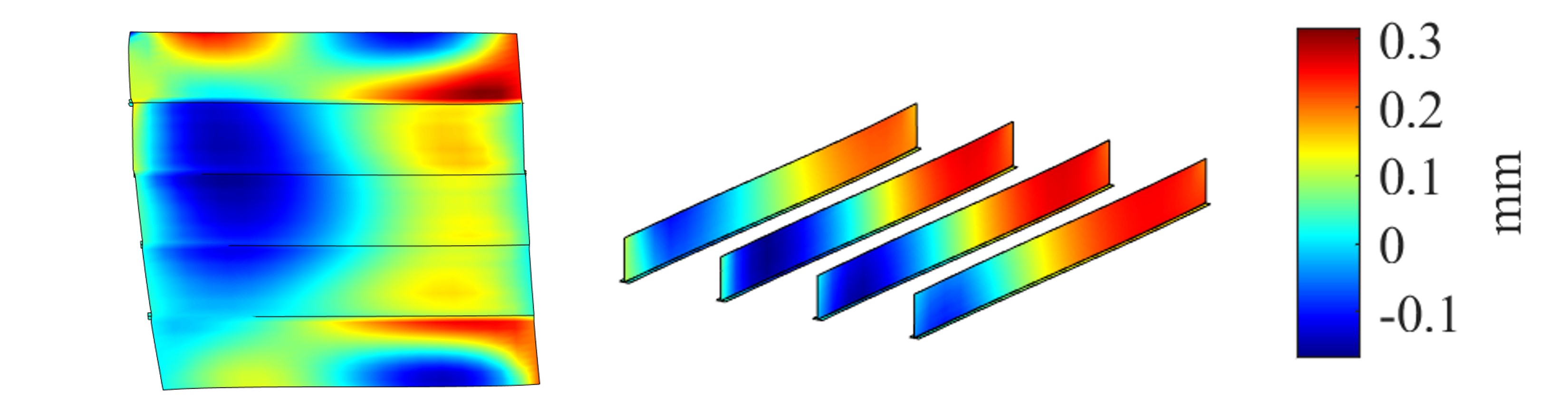}}
			& \makecell[c]{\includegraphics[width=0.4\textwidth]{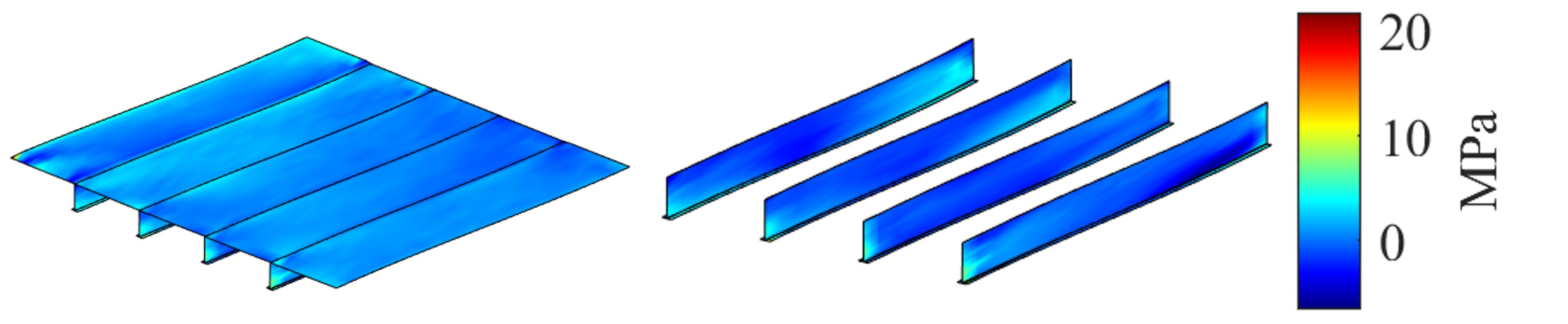}} \\
			& GraphSAGE  
			& \makecell[c]{\includegraphics[width=0.4\textwidth]{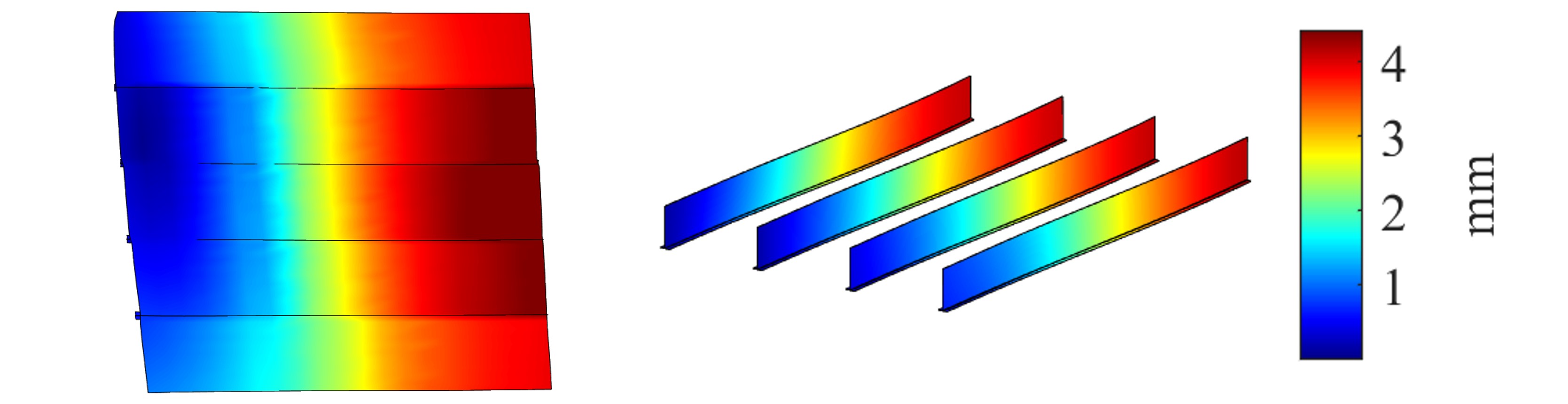}}
			& \makecell[c]{\includegraphics[width=0.4\textwidth]{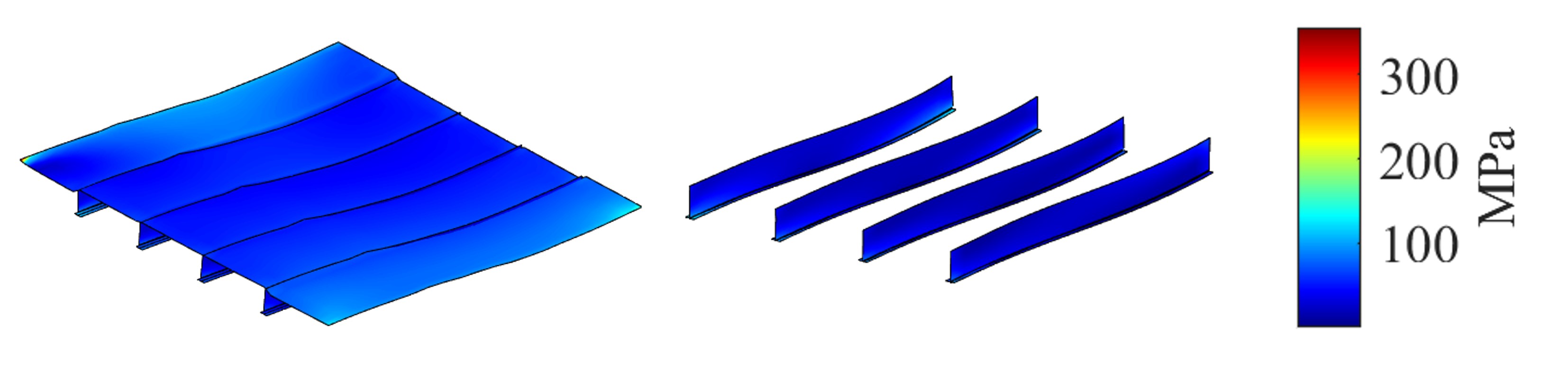}} \\
			& \makecell[c]{GraphSAGE Error\\(Disp.\,0.213 mm;\\Stress\,8.21 MPa)}  
			& \makecell[c]{\includegraphics[width=0.4\textwidth]{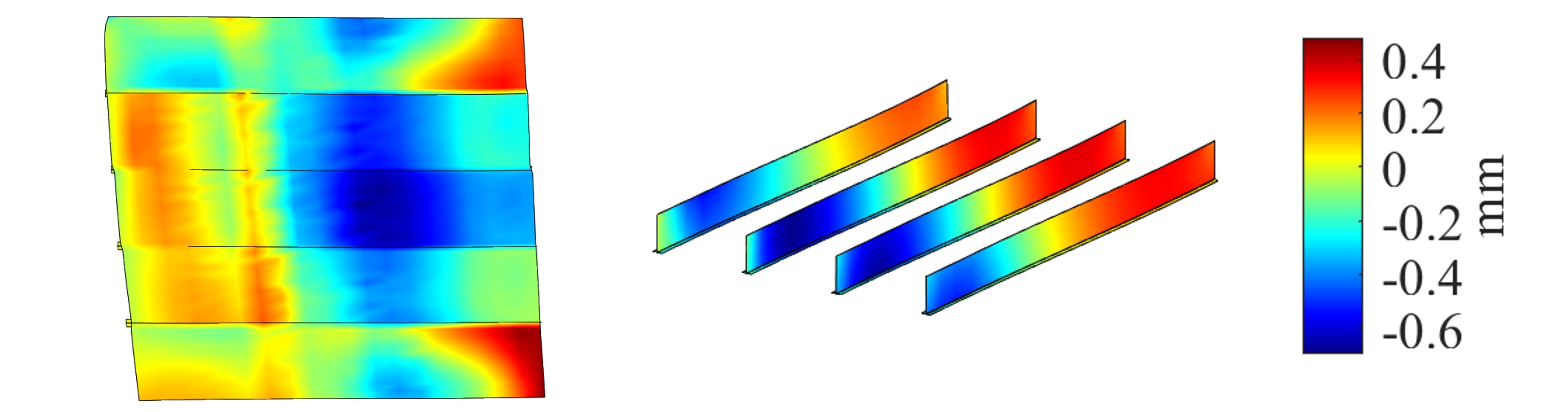}}
			& \makecell[c]{\includegraphics[width=0.4\textwidth]{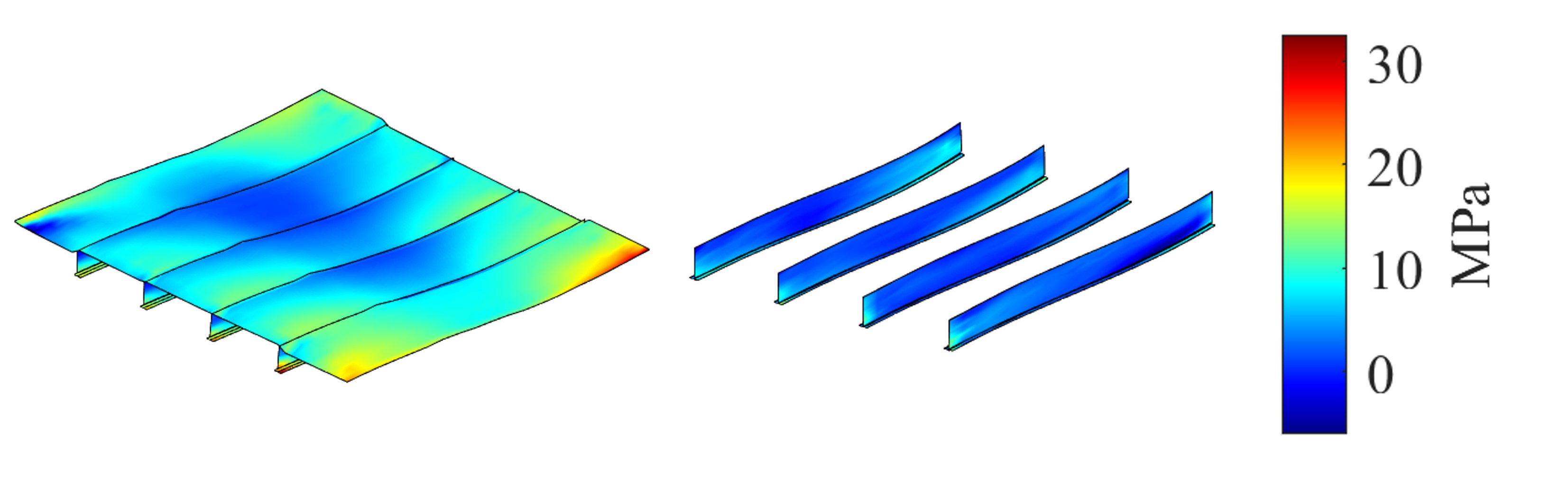}} \\
			& FEA  
			& \makecell[c]{\includegraphics[width=0.4\textwidth]{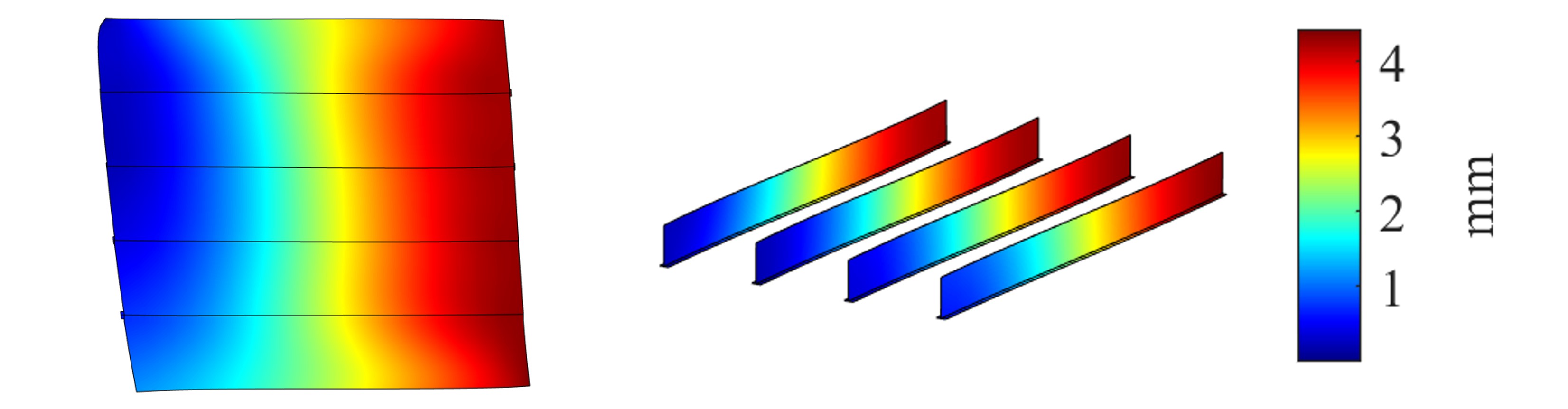}}
			& \makecell[c]{\includegraphics[width=0.4\textwidth]{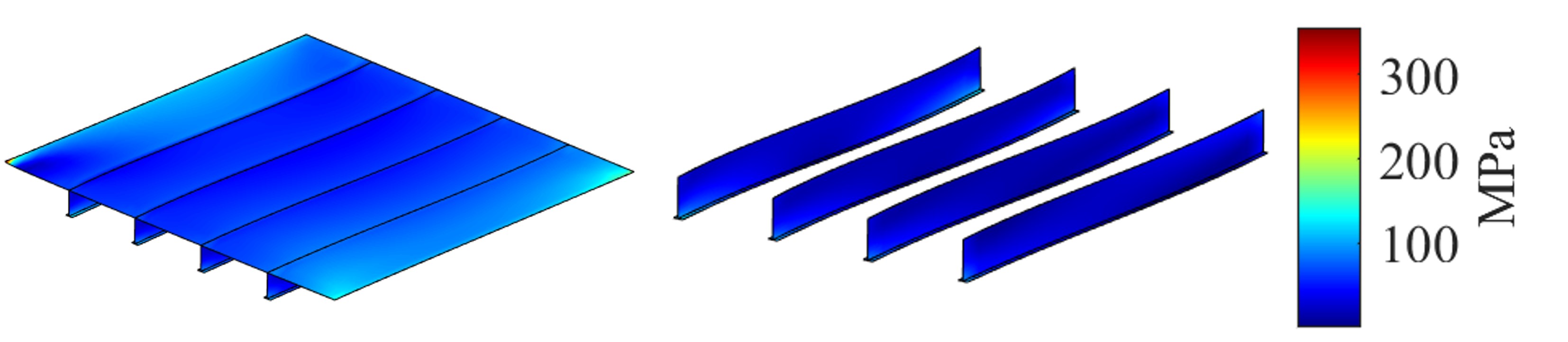}} \\
			\bottomrule
		\end{tabular}%
	}
	\caption{Comparison of HGT and GraphSAGE predictions versus FEA results for test case 2: box beam subjected to 4-point bending.}
	\label{fig: Case1 contour}
\end{figure}

\begin{figure}[!htbp]
	\centering
	\begin{minipage}[b]{0.4\textwidth}
		\includegraphics[width=\textwidth]{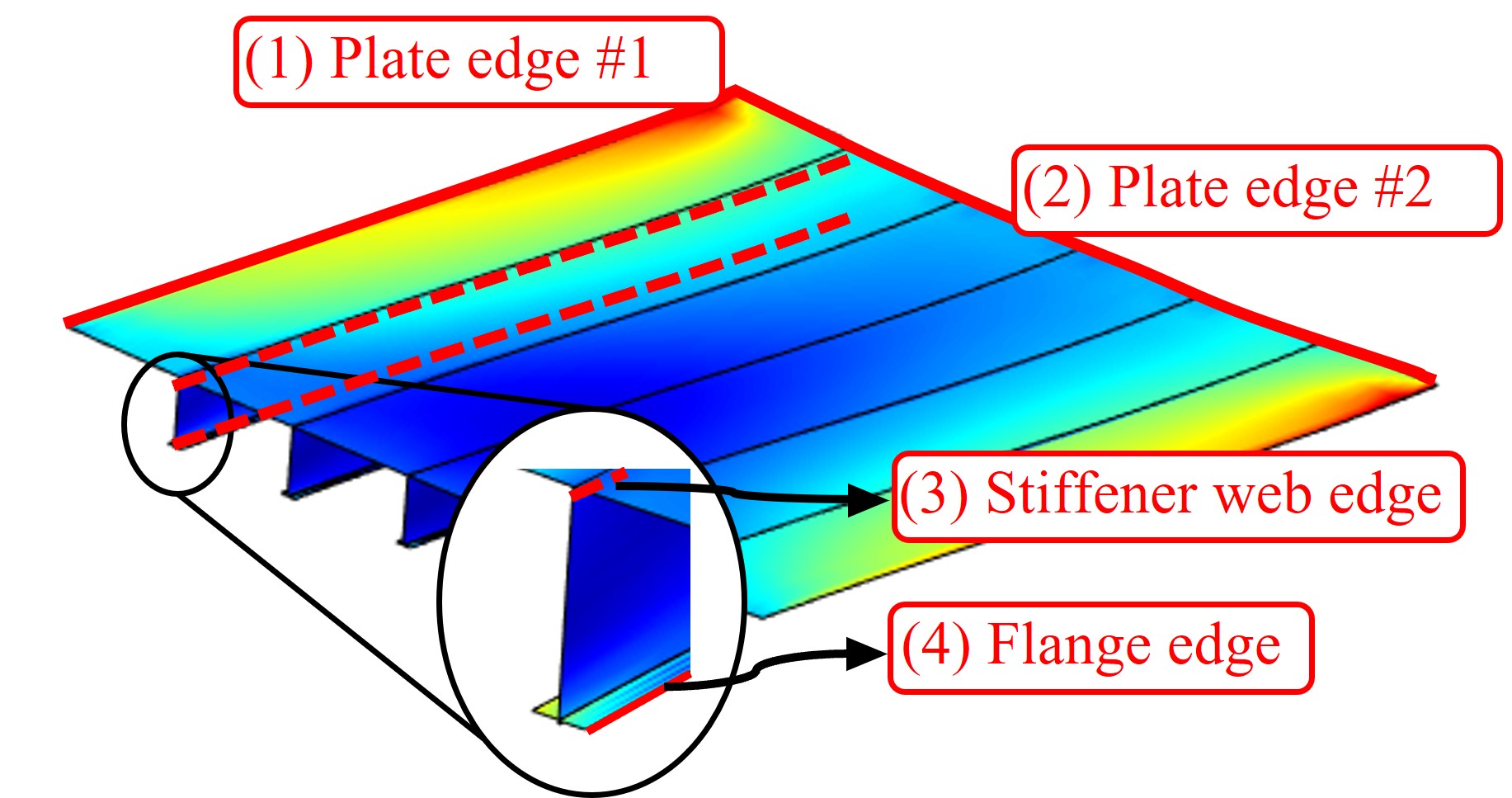}
		\\
		\subfloat[\normalsize Test example 1 (Top panel 1)]{\includegraphics[width=\textwidth]{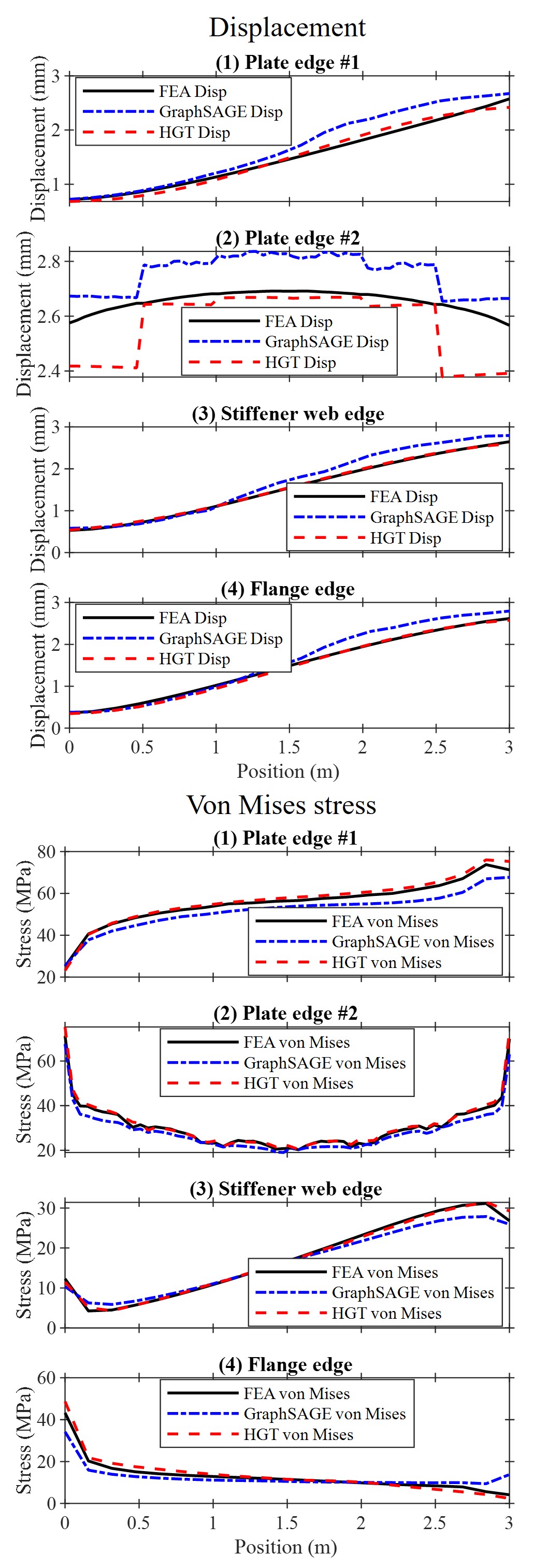}\label{fig: Case 2 plot (a)}}
	\end{minipage}
	\hfill
	\begin{minipage}[b]{0.4\textwidth}
		\includegraphics[width=\textwidth]{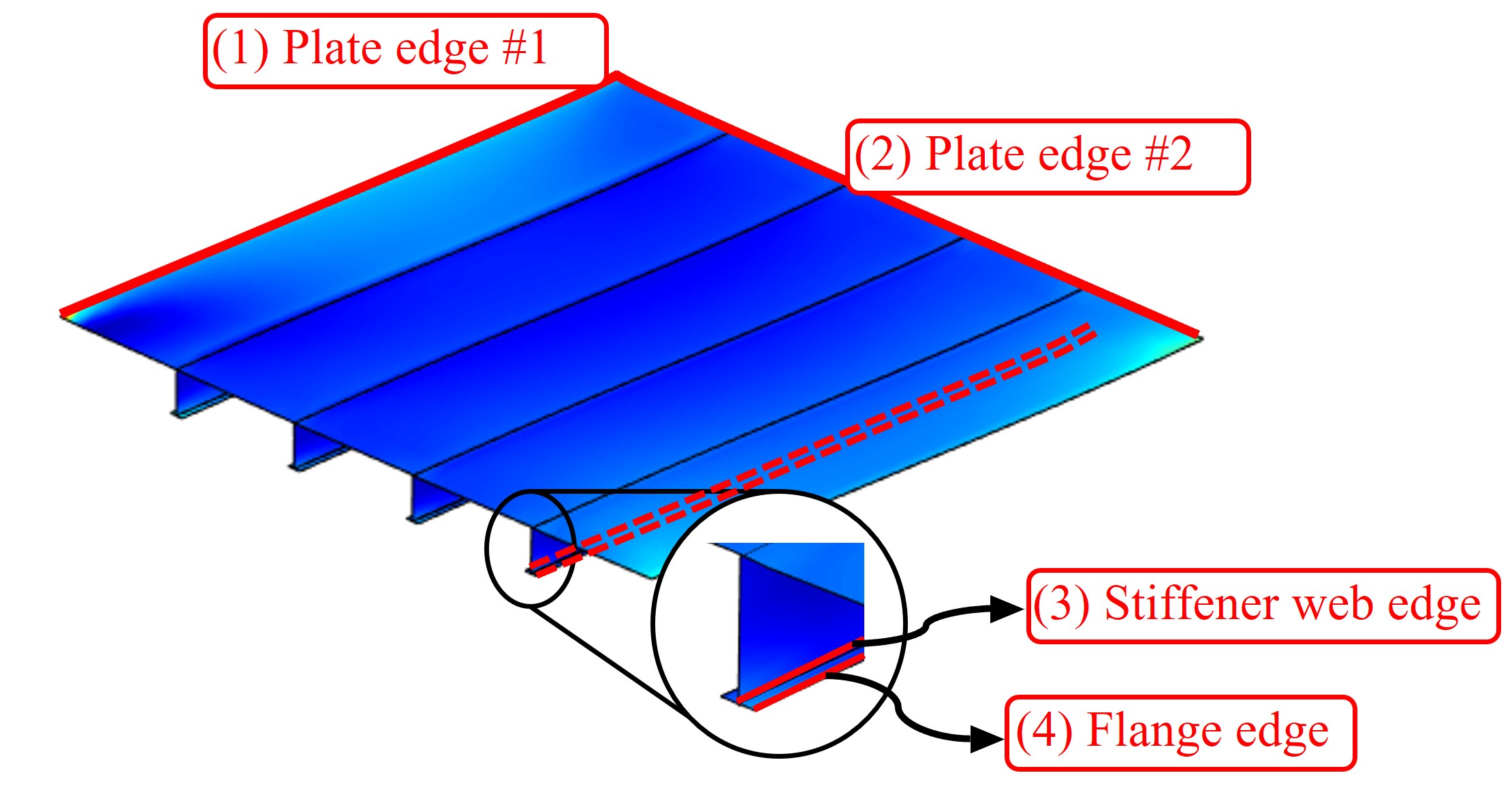}
		\\
		\subfloat[\normalsize Test example 2 (Side panel 1)]{\includegraphics[width=\textwidth]{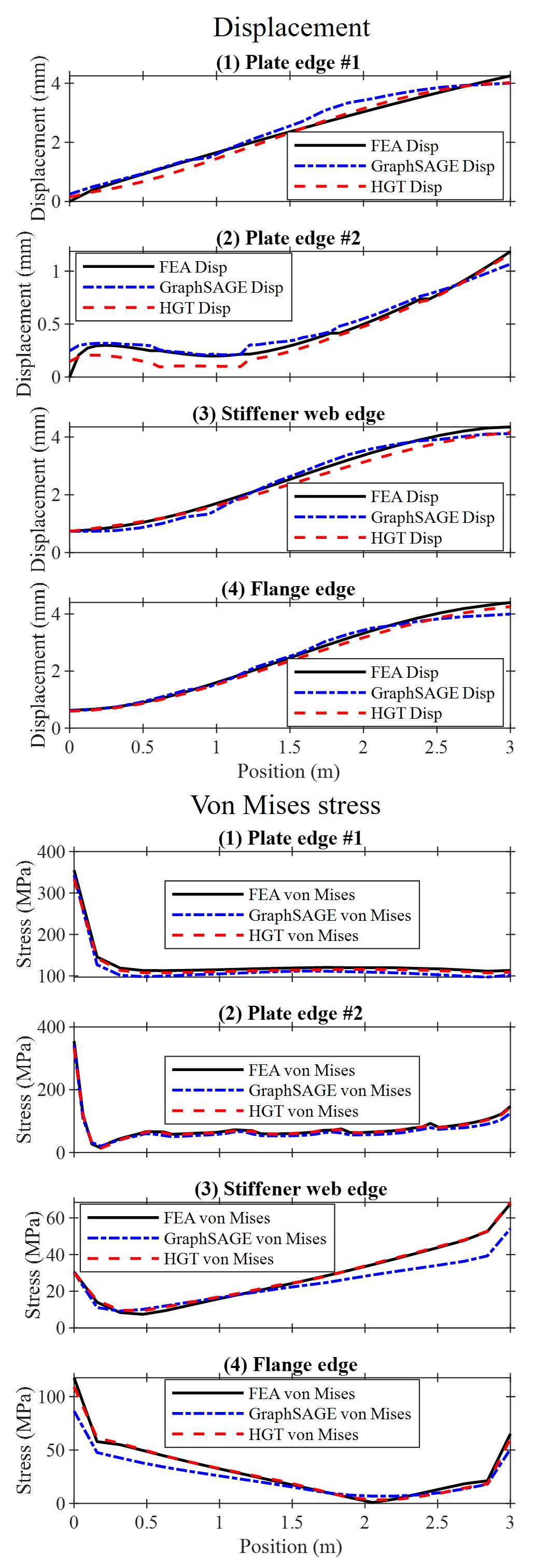}\label{fig: Case 2 plot (b)}}
	\end{minipage}
	\caption{Comparison of stress and total displacement distribution along specified paths for test case 2 examples (box beam subjected to 4-point bending).}\label{fig: Case 2 plot}
\end{figure}

This subsection presents the performance of HGT with the proposed heterogeneous graph representation on test case 2: box beam subjected to 4-point bending. Specifically, this section compares von Mises stress and displacement contours predicted by HGT and GraphSAGE against those simulated by FEA, see Fig. \ref{fig: Case1 contour}. The geometric details for test examples in case 2 are given in Table \ref{tab: Case 2 geometry}. For each test example, we also compare the HGT and GraphSAGE outputs with the FEA simulation results along several specified paths, as depicted in Fig. \ref{fig: Case 2 plot}. These paths were selected to highlight regions with the largest displacements and stresses.

\begin{table}[h]
	\centering
	\caption{Structural details for test case 2 examples (box beam subjected to 4-point bending).}\label{tab: Case 2 geometry}%
	\small
	\begin{tabular}{@{}lccc@{}}
		\toprule
		Category & Test example 1 &  Test example 2 & Unit \\
		\midrule
		Plate thickness    & 11.50   & 10.40  & mm  \\
		Stiffener web thickness    & 18.59   &  16.28 & mm  \\
		Stiffener web height    &  237.9  & 209.9  & mm  \\
		Flange thickness  & 6.32  & 11.22 & mm \\
		Flange width   & 58.19 & 58.36 & mm \\
		Number of stiffeners  & 5 & 4 & $-$ \\
		\toprule
	\end{tabular}
\end{table}

It can be observed that HGT provides accurate predictions of both displacement and von Mises stress distributions. As shown in Fig. \ref{fig: Case 2 plot (a)}, the predictions for the first test example closely follow the FEA results, except for the displacement profile along plate edge \#2, which exhibits discontinuities at the intersections of different structural units. This phenomenon is also observed in GraphSAGE results for both test examples. However, the HGT prediction for test example 2 shows a smoother displacement field than GraphSAGE predictions, with an RMSE of 0.128, 40\% lower than results by GraphSAGE. Fig. \ref{fig: Case 2 plot} shows more detailed comparisons, where it is seen that the greatest error comes from the boundaries between individual structural units. This discrepancy likely arises from the graph representation framework, which treats each structural unit as a separate node and therefore does not account for the smoothness and continuity at their interfaces. In contrast, the stress predictions maintain a high degree of consistency for both test examples on both HGT and GraphSAGE models, successfully capturing not only the general distribution patterns but also the peak stress values. Regarding the maximum stress prediction, in test example 1, HGT accurately identifies the maximum von Mises stress at the plate edge with an accuracy of 96.92\%. Test example 2 (Fig. \ref{fig: Case 2 plot (b)}), which exhibits plastic deformation behavior, shows that HGT correctly locates the maximum stress at the plate corner, achieving 93.92\% accuracy. GraphSAGE also predicted these values correctly: 94.31\% for test example 1 and 94.18\% for test example 2.

\subsubsection{Box beam subjected to uniform pressure}\label{sec5_3_2}

\begin{figure}[h]
	\centering
	\scriptsize
	\setlength\tabcolsep{1mm}
	\resizebox{\textwidth}{!}{%
		\begin{tabular}{cccc}
			\toprule
			Test example   & Model & Displacement field & von Mises field \\
			\midrule
			\multirow{20}{*}{\rotatebox{90}{3 (Top panel 1)}}  
			& HGT  
			& \makecell[c]{\includegraphics[width=0.4\textwidth]{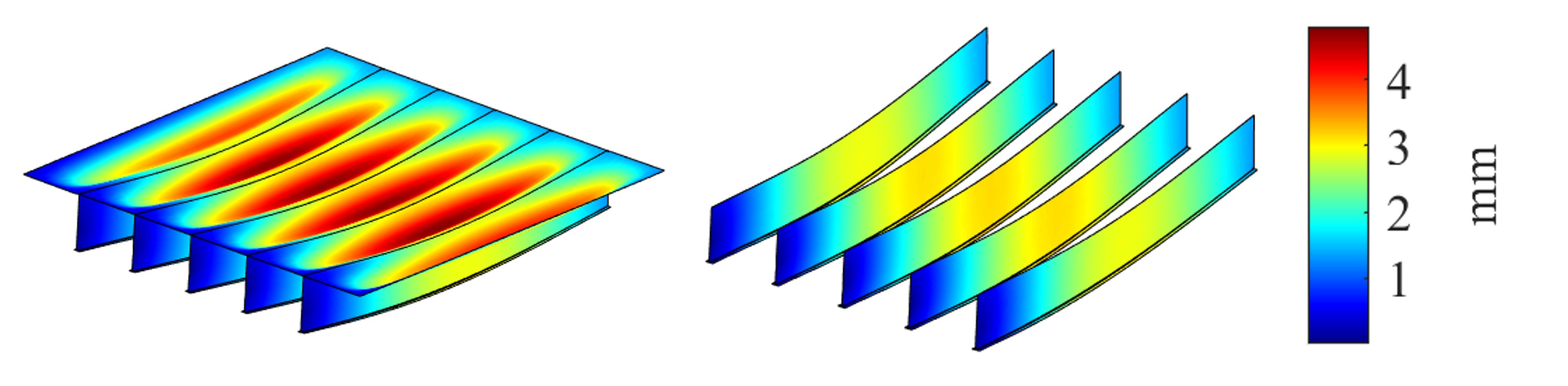}}
			& \makecell[c]{\includegraphics[width=0.4\textwidth]{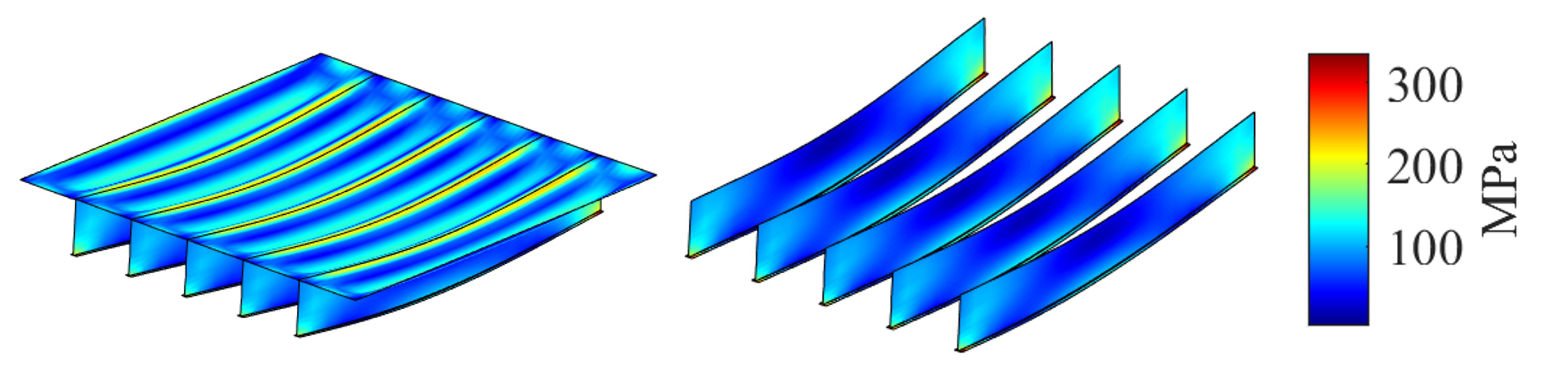}} \\
			& \makecell[c]{HGT Error\\(Disp.\,0.0672 mm;\\ Stress\,5.79 MPa)}
			& \makecell[c]{\includegraphics[width=0.4\textwidth]{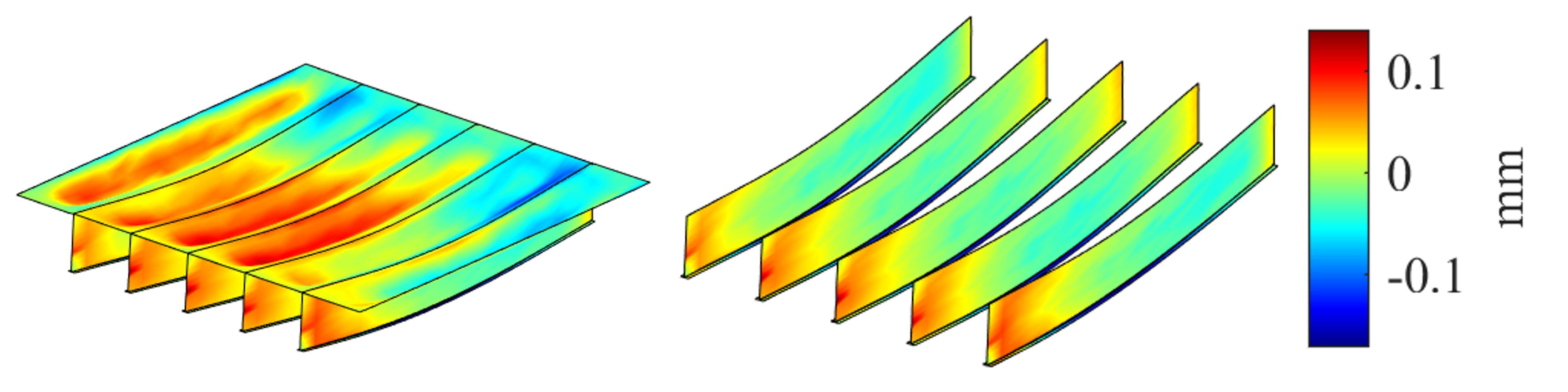}}
			& \makecell[c]{\includegraphics[width=0.4\textwidth]{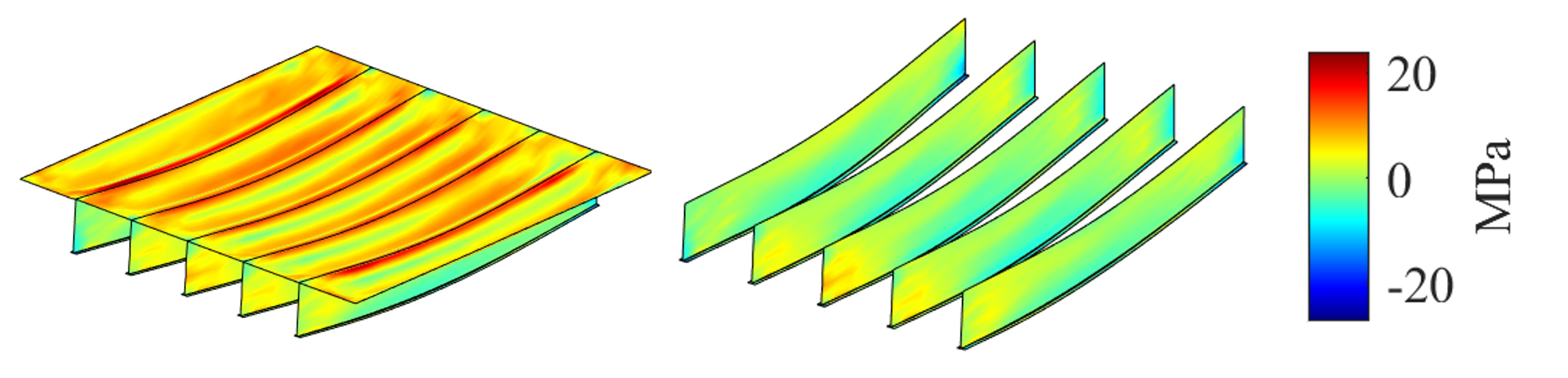}} \\
			& GraphSAGE  
			& \makecell[c]{\includegraphics[width=0.4\textwidth]{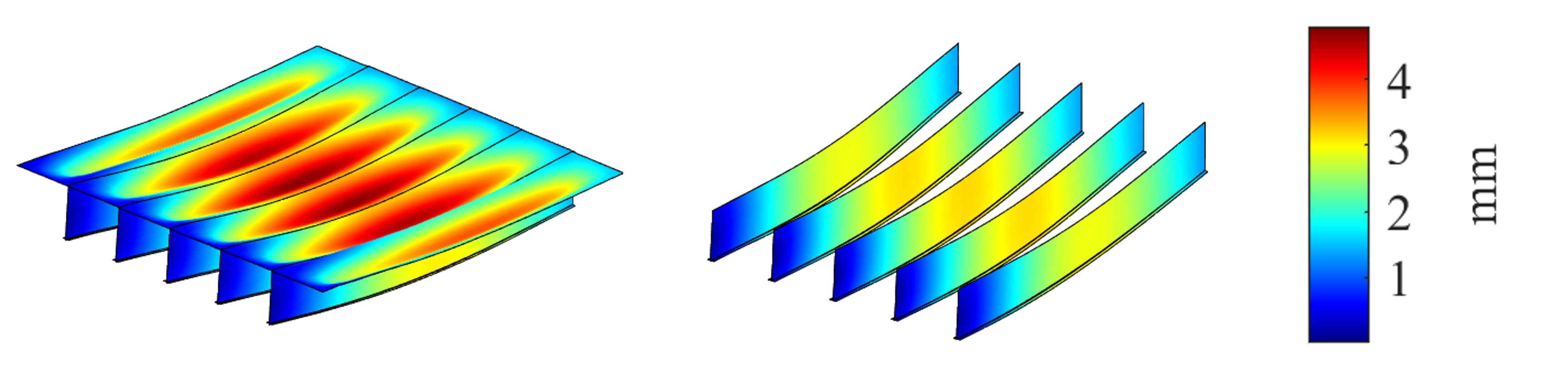}}
			& \makecell[c]{\includegraphics[width=0.4\textwidth]{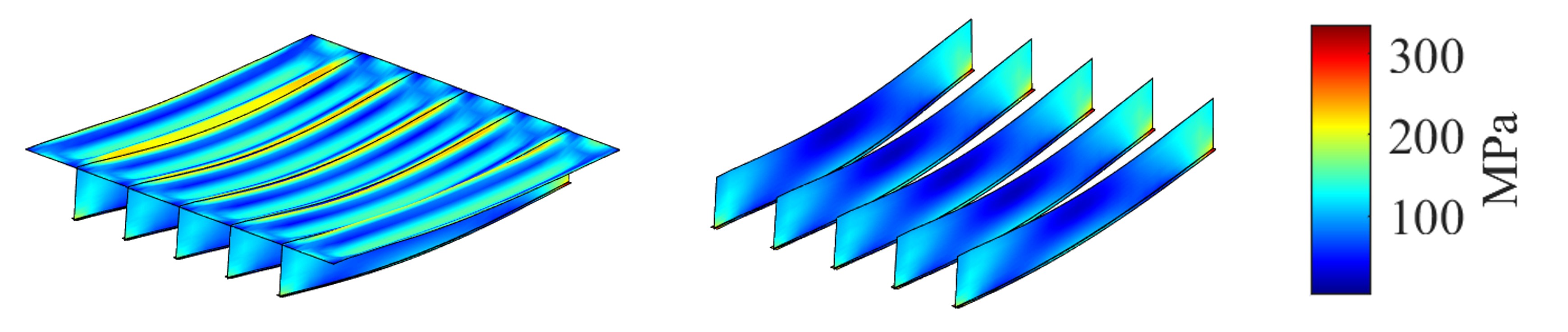}} \\
			& \makecell[c]{GraphSAGE Error\\(Disp.\,0.182 mm;\\Stress\,7.15 MPa)}
			& \makecell[c]{\includegraphics[width=0.4\textwidth]{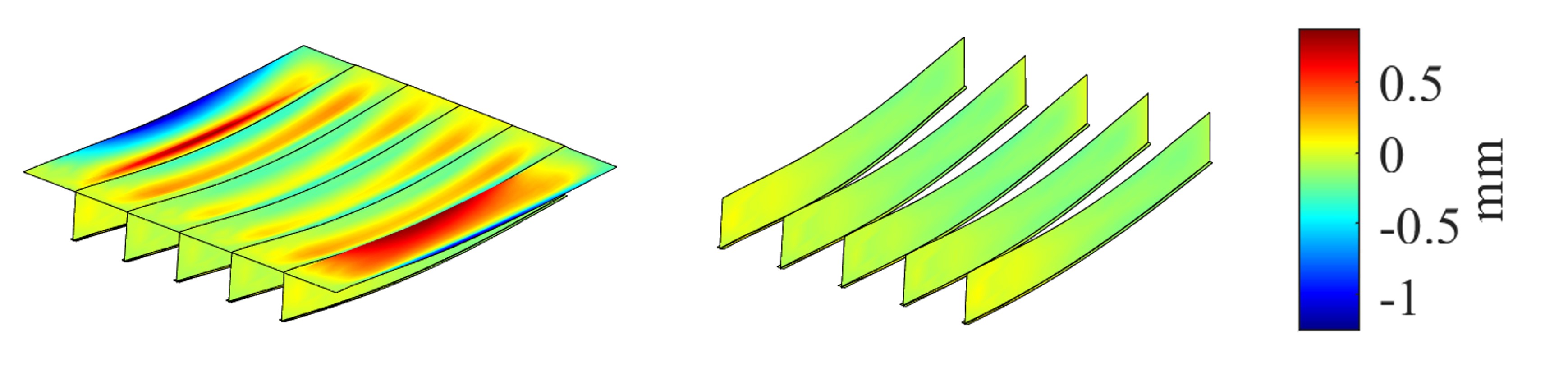}}
			& \makecell[c]{\includegraphics[width=0.4\textwidth]{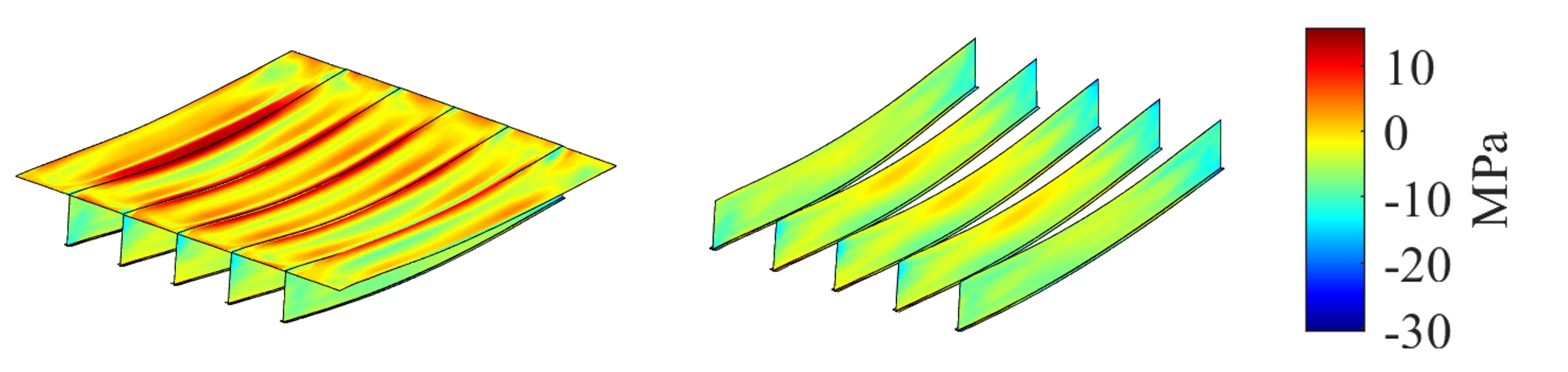}} \\
			& FEA  
			& \makecell[c]{\includegraphics[width=0.4\textwidth]{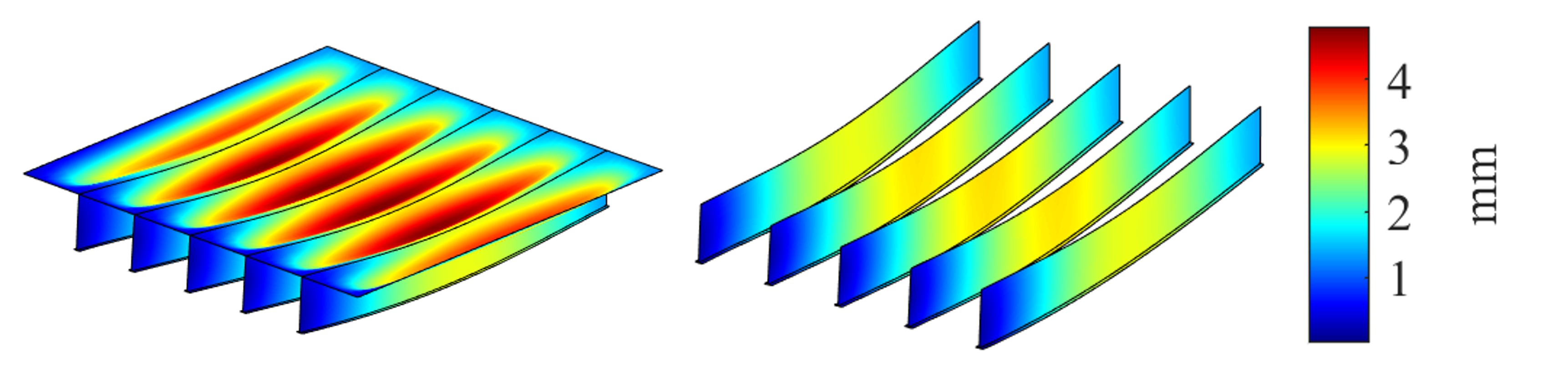}}
			& \makecell[c]{\includegraphics[width=0.4\textwidth]{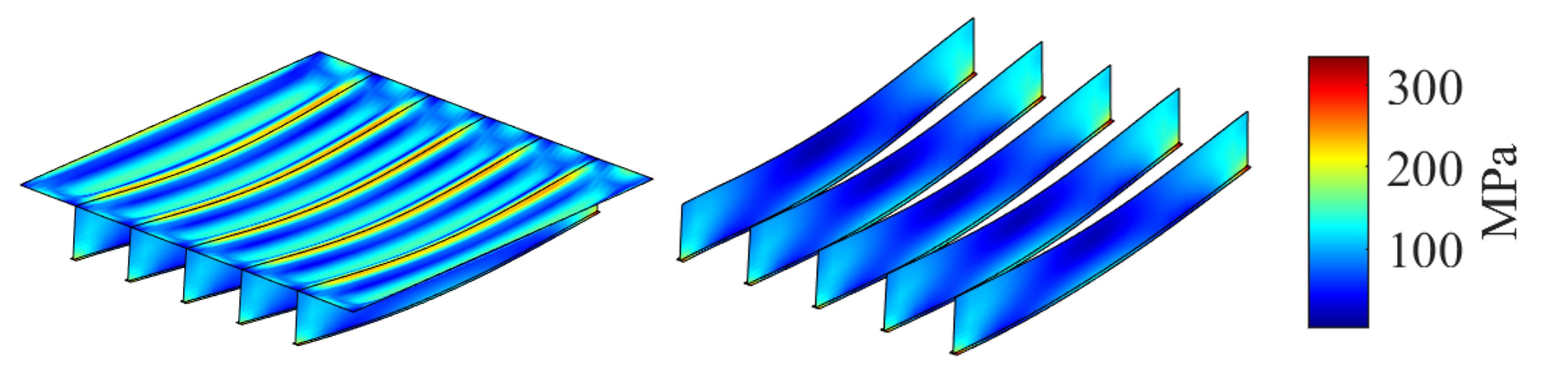}} \\
			\midrule
			\multirow{20}{*}{\rotatebox{90}{4 (Bottom panel 1)}}  
			& HGT  
			& \makecell[c]{\includegraphics[width=0.4\textwidth]{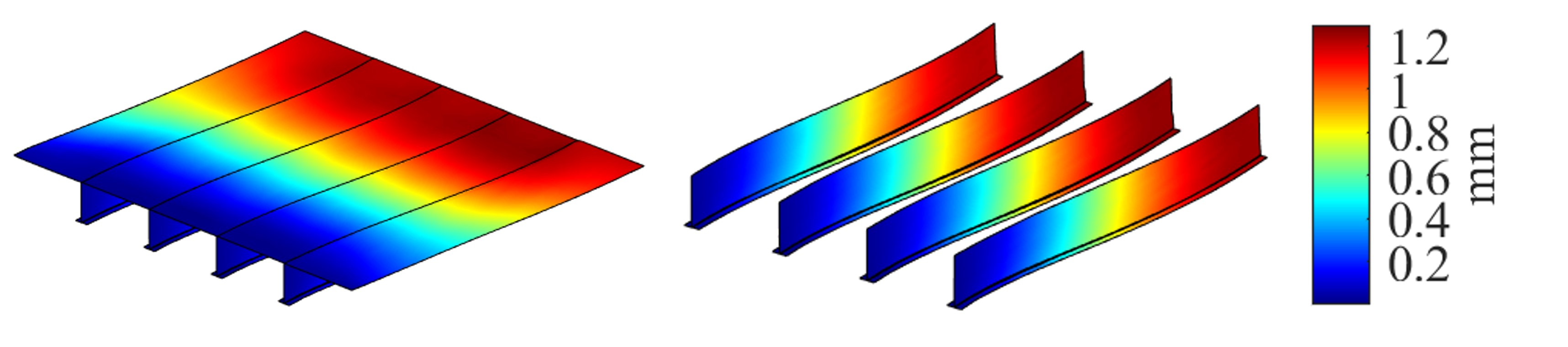}}
			& \makecell[c]{\includegraphics[width=0.4\textwidth]{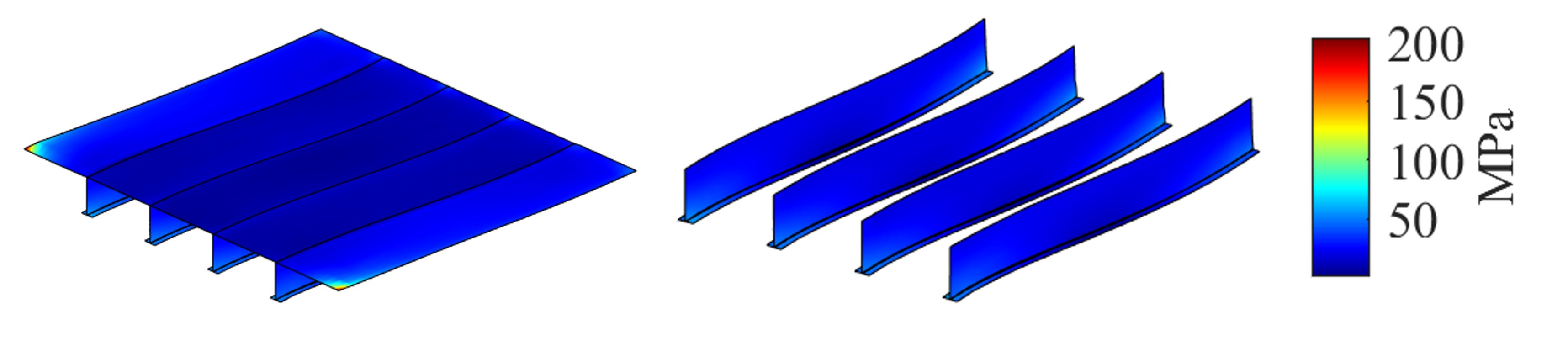}} \\
			& \makecell[c]{HGT Error\\(Disp.\,0.0277 mm;\\Stress\,1.24 MPa)}
			& \makecell[c]{\includegraphics[width=0.4\textwidth]{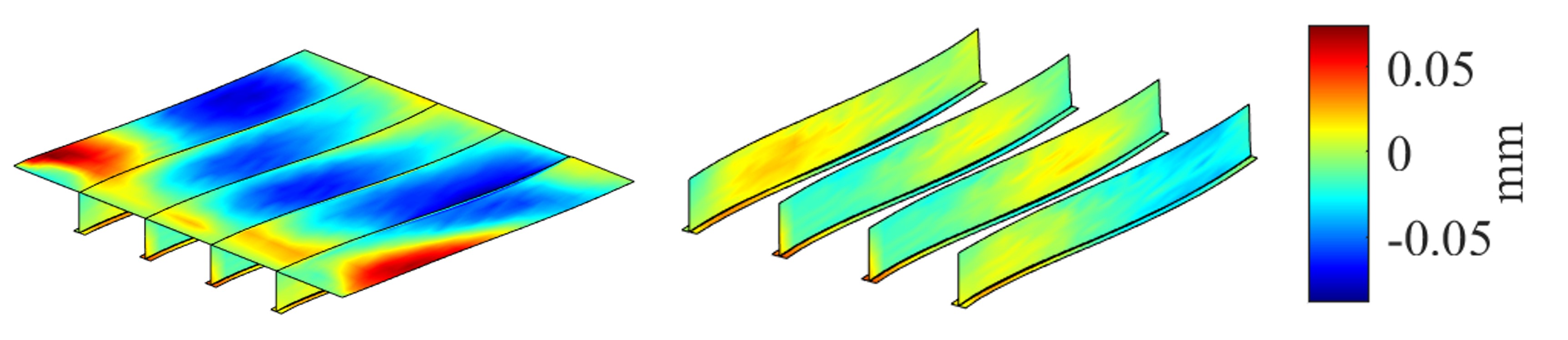}}
			& \makecell[c]{\includegraphics[width=0.4\textwidth]{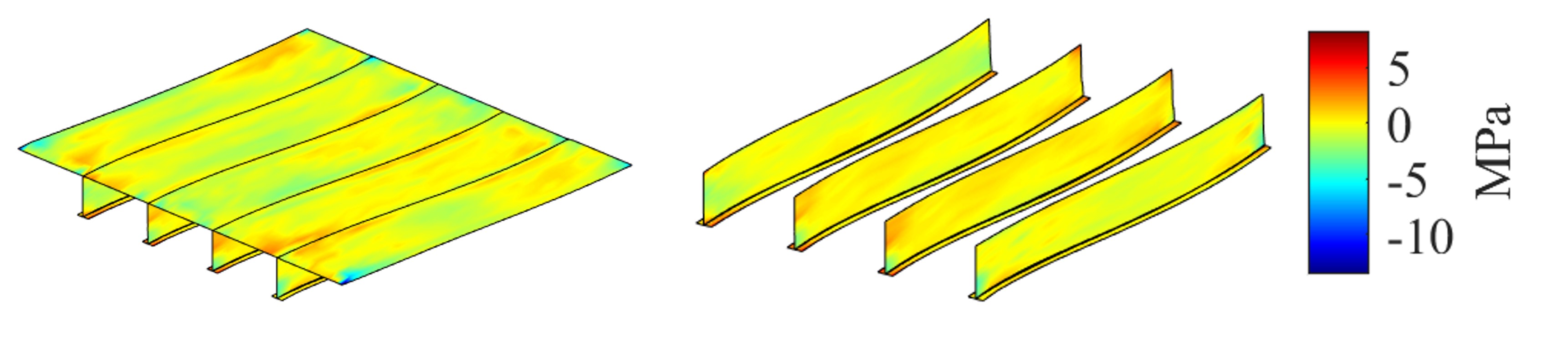}} \\
			& GraphSAGE  
			& \makecell[c]{\includegraphics[width=0.4\textwidth]{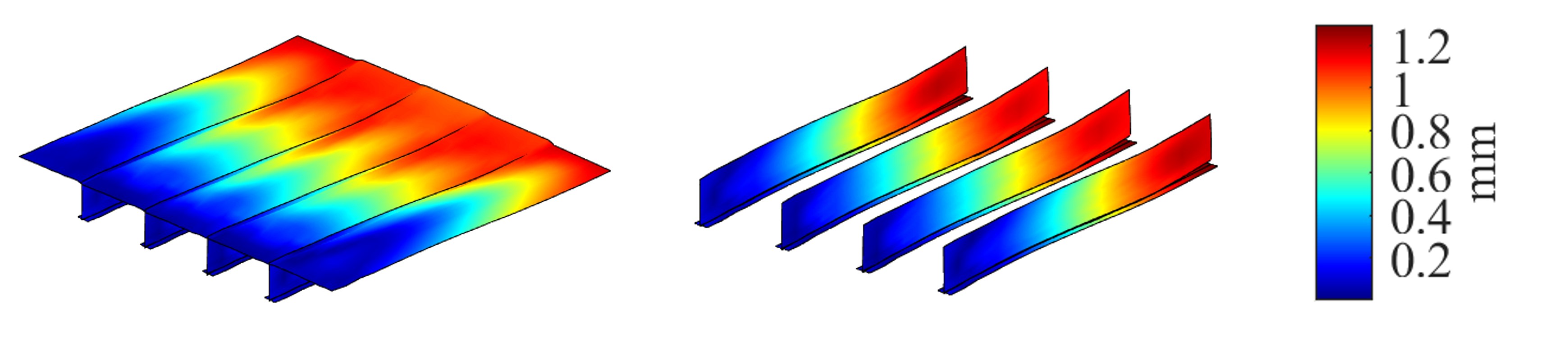}}
			& \makecell[c]{\includegraphics[width=0.4\textwidth]{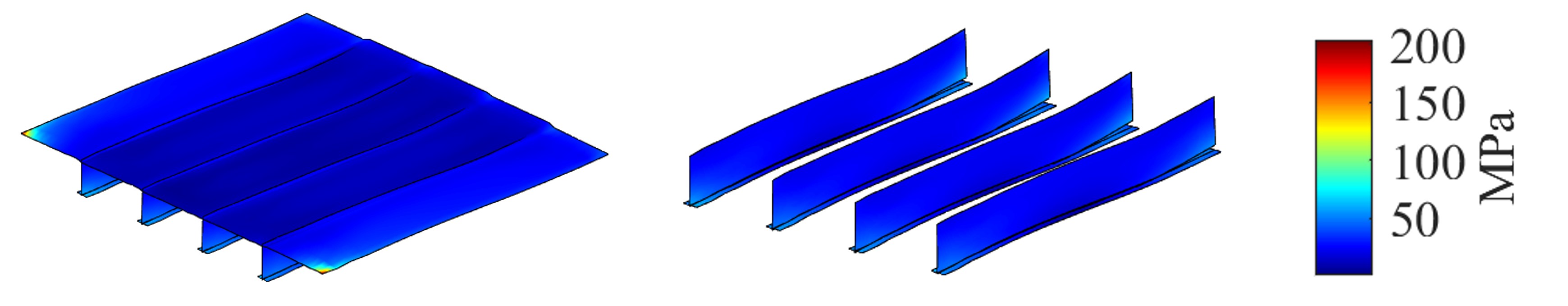}} \\
			& \makecell[c]{GraphSAGE Error\\(Disp.\,0.125 mm;\\Stress\,1.98 MPa)}
			& \makecell[c]{\includegraphics[width=0.4\textwidth]{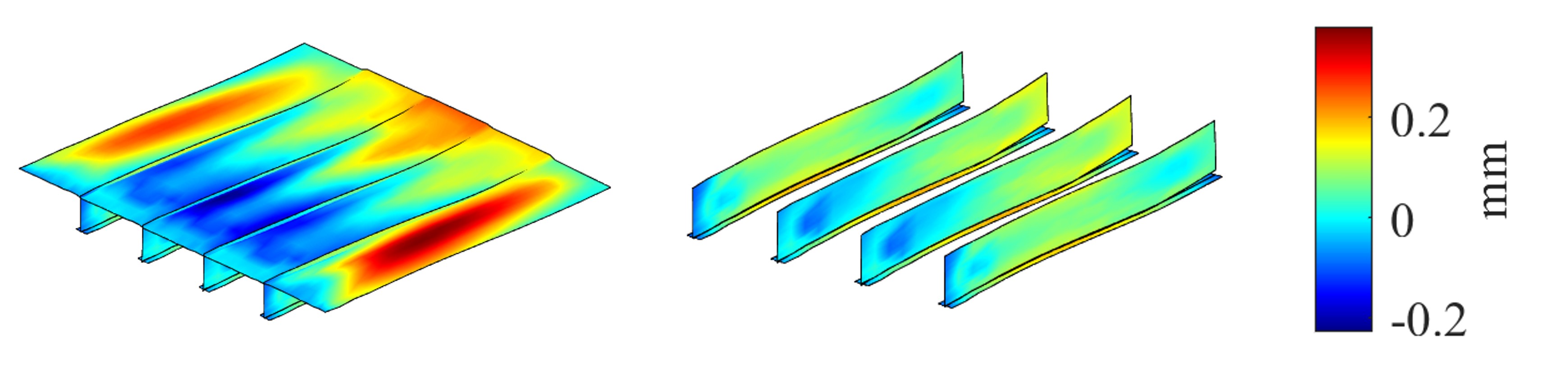}}
			& \makecell[c]{\includegraphics[width=0.4\textwidth]{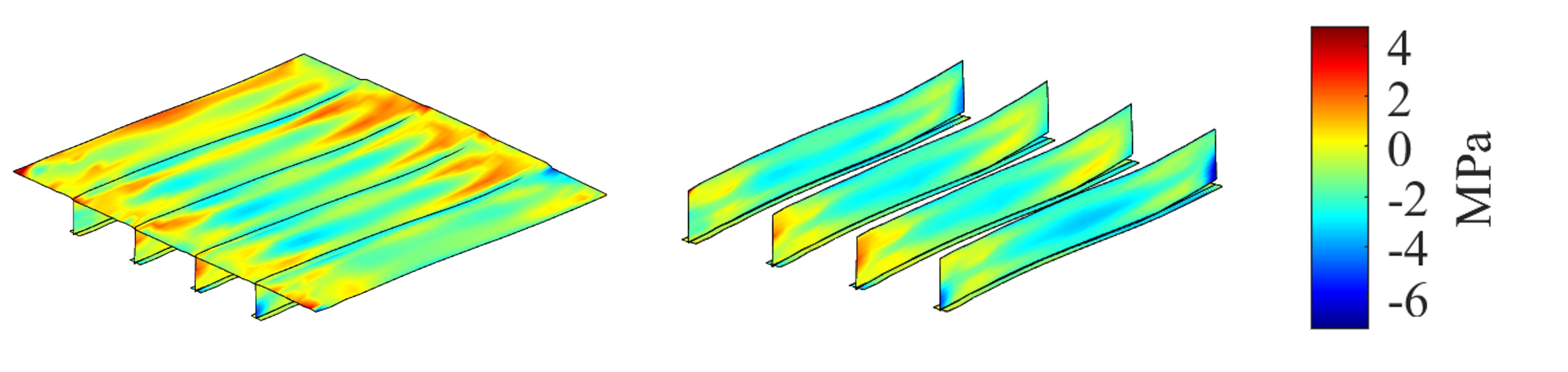}} \\
			& FEA  
			& \makecell[c]{\includegraphics[width=0.4\textwidth]{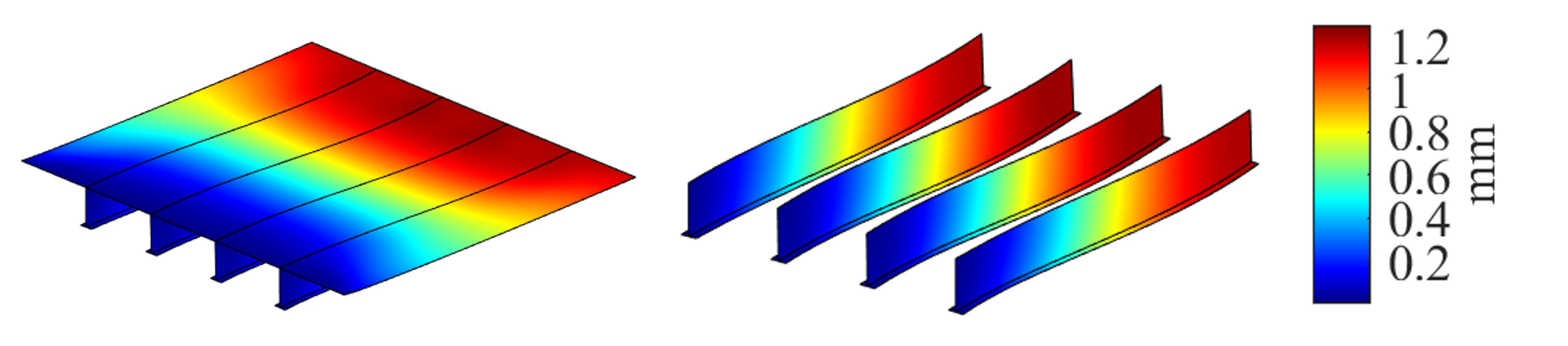}}
			& \makecell[c]{\includegraphics[width=0.4\textwidth]{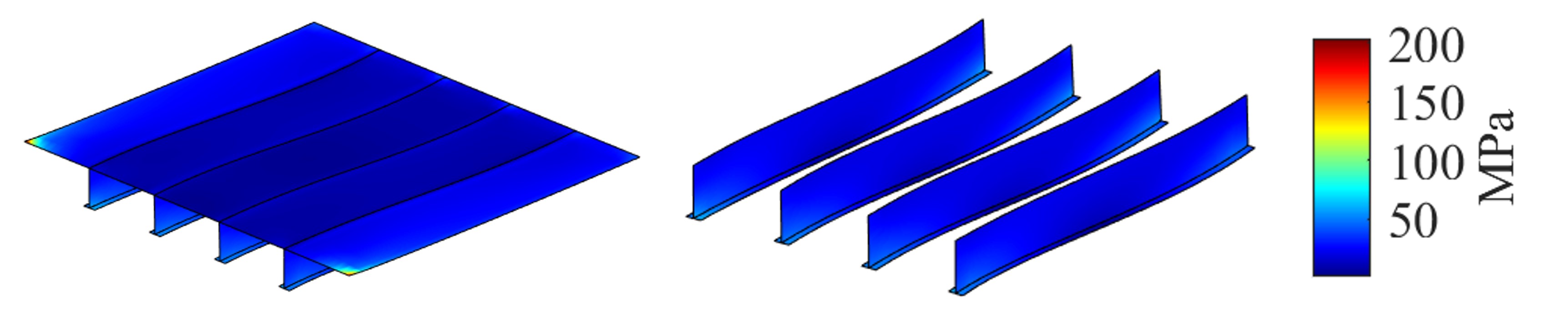}} \\
			\bottomrule
		\end{tabular}%
	}
	\caption{Comparison of HGT and GraphSAGE predictions versus FEA results for test case 3 (box beam subjected to uniform pressure).}
	\label{fig: Case 3 contour}
\end{figure}

\begin{figure}[!htbp]
	\centering
	\begin{minipage}[b]{0.42\textwidth}
		\includegraphics[width=\textwidth]{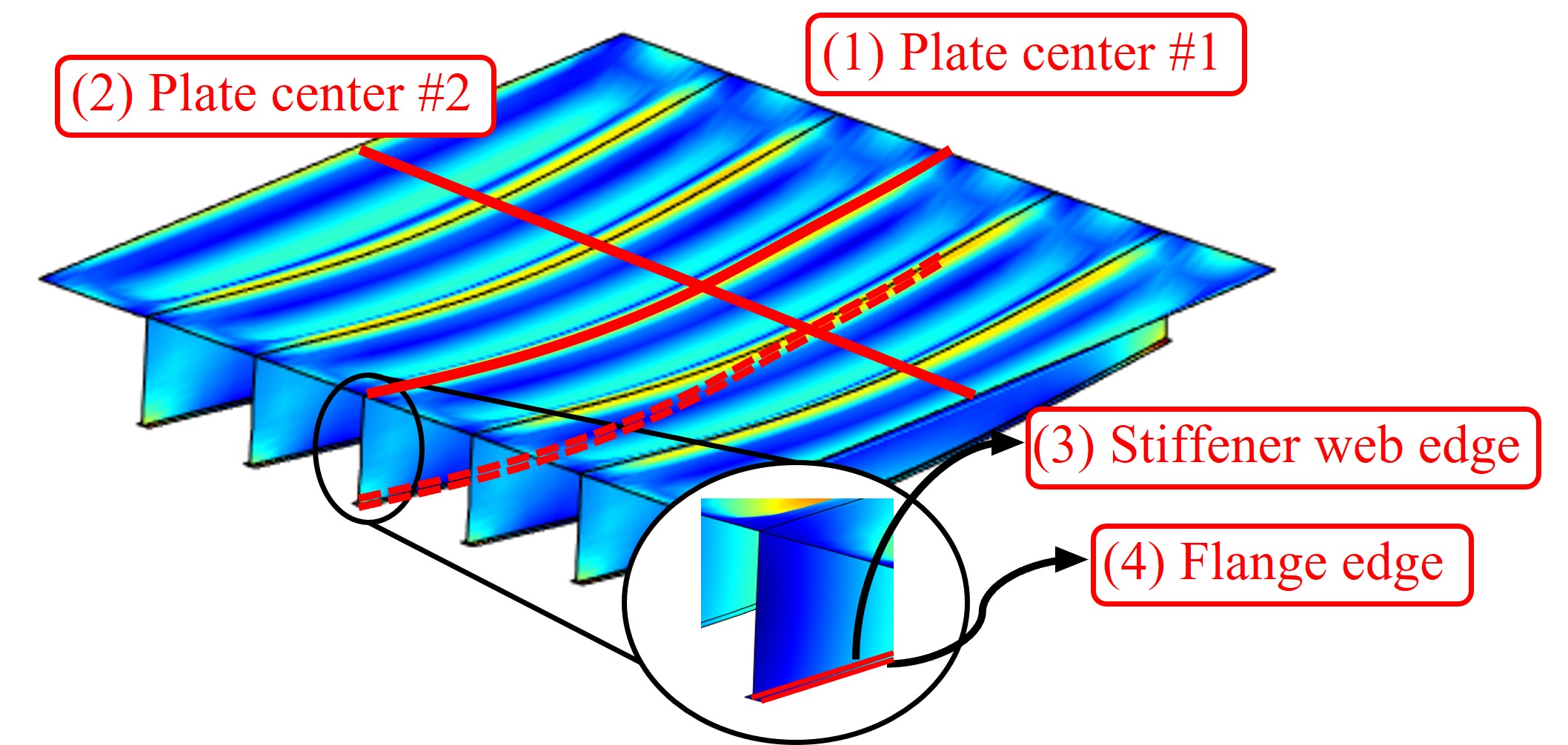}
		\\
		\subfloat[\normalsize Test example 3 (Top panel 1)]{\includegraphics[width=\textwidth]{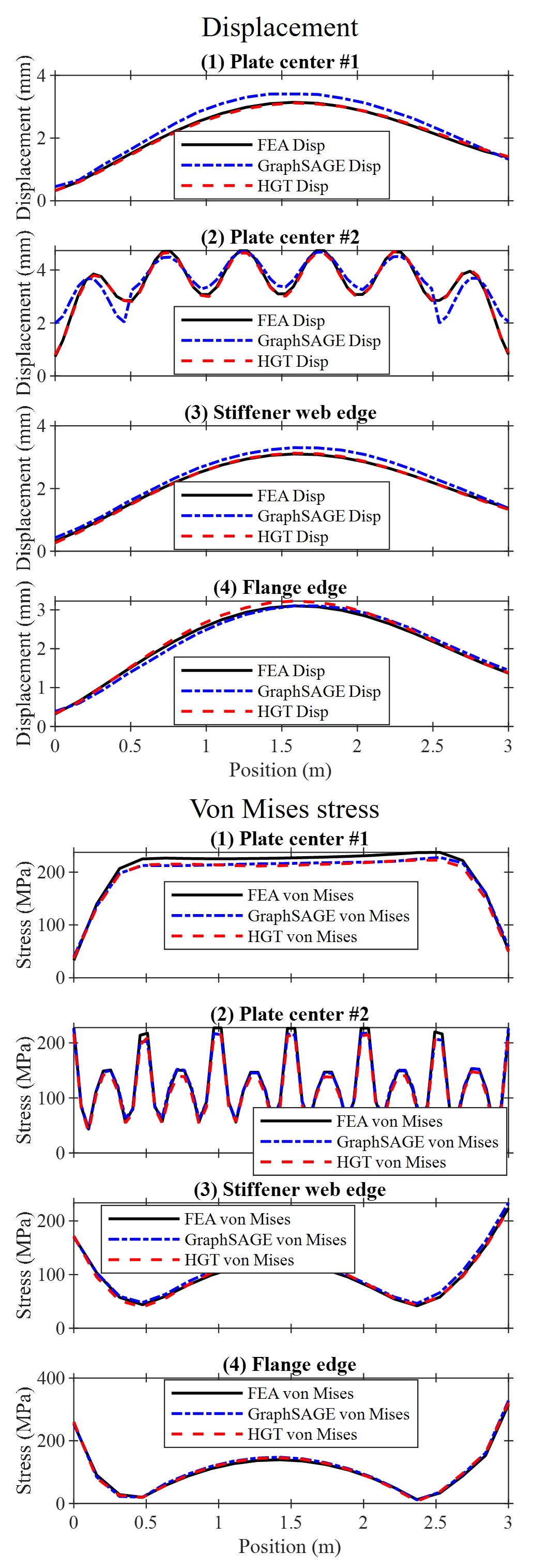}\label{fig: Case 3 plot (a)}}
	\end{minipage}
	\hfill
	\begin{minipage}[b]{0.42\textwidth}
		\includegraphics[width=\textwidth]{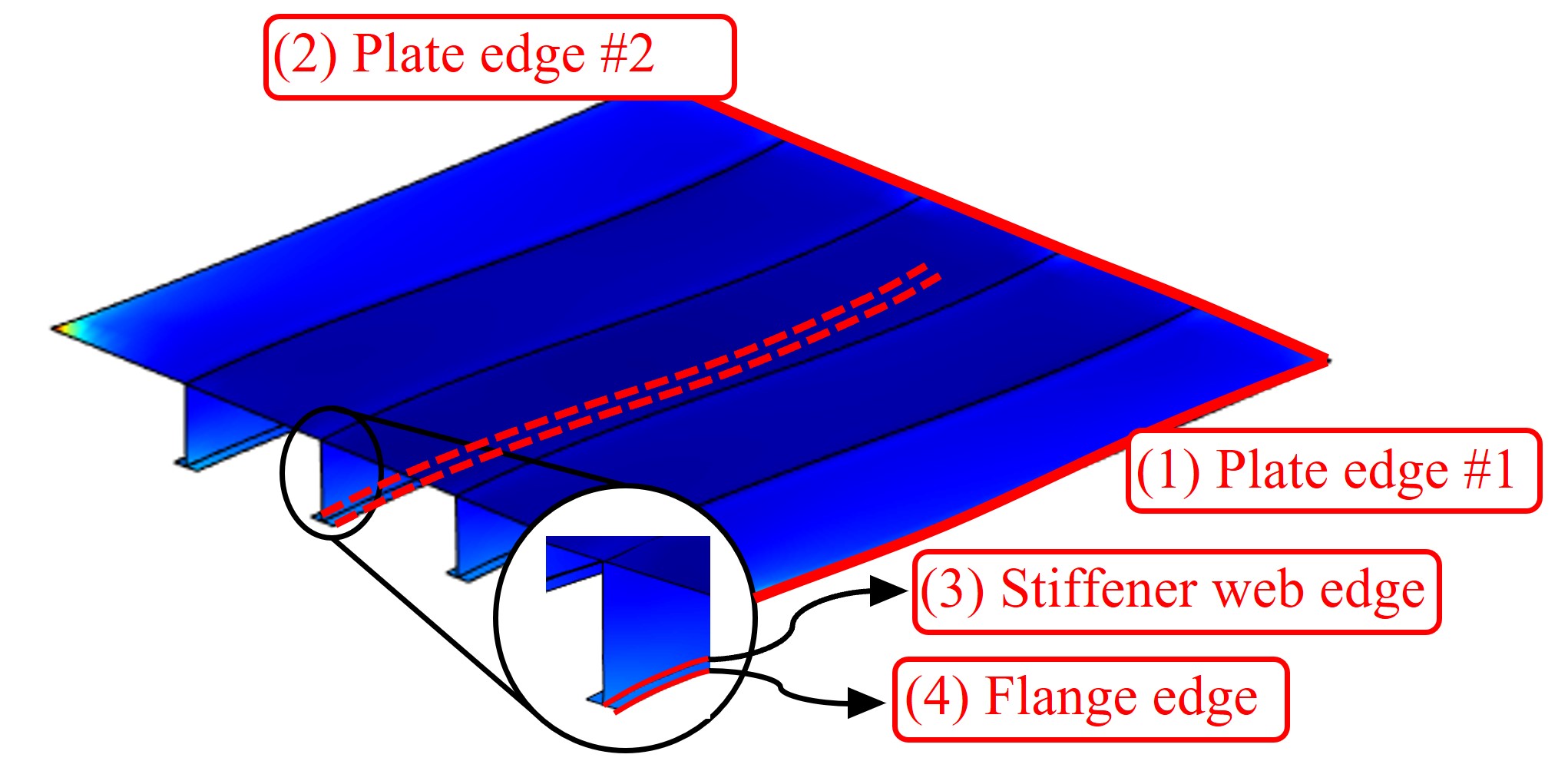}
		\\
		\subfloat[\normalsize Test example 4 (Bottom panel 1)]{\includegraphics[width=\textwidth]{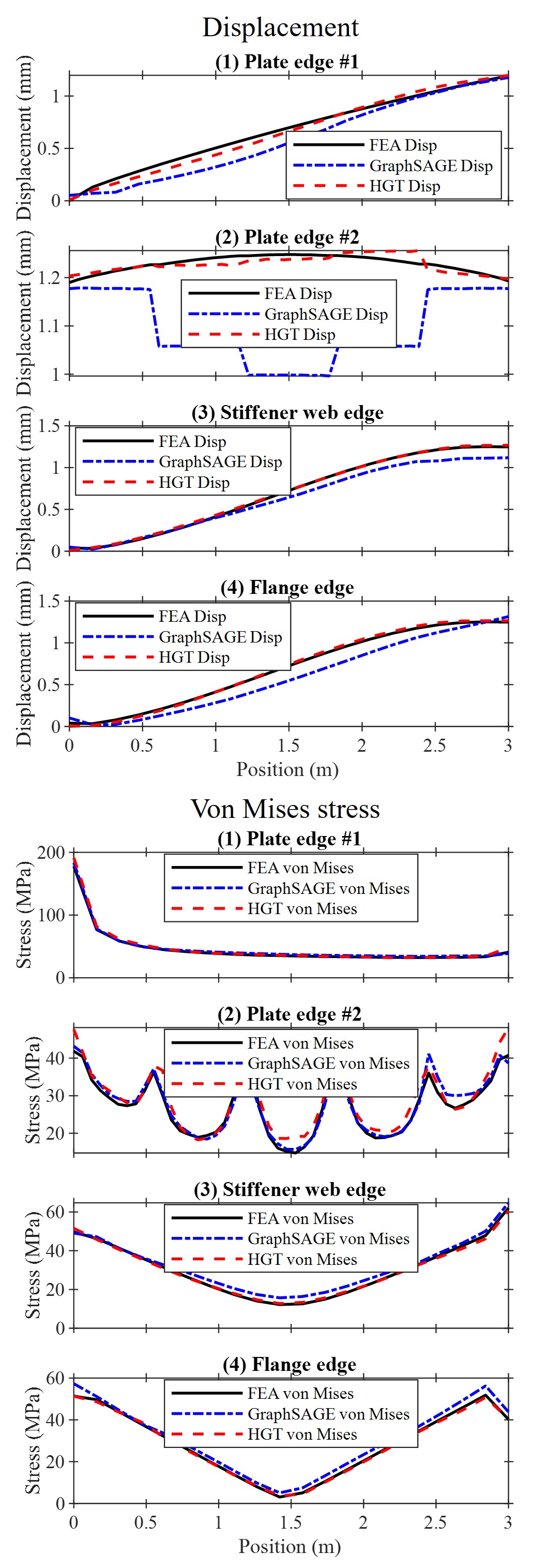}\label{fig: Case 3 plot (b)}}
	\end{minipage}
	\caption{Comparison of total displacement and stress distribution along specified paths for test case 3 examples (box beam subjected to uniform pressure).}\label{fig: Case 3 plot}
\end{figure}

This subsection presents the performance of HGT on test case 3: box beam subjected to uniform pressure. In addition to the non-uniformly distributed edge boundary conditions, some panels may also be subjected to uniform pressures if the stiffened panel is selected from the top of the box beam. Fig. \ref{fig: Case 3 contour} presents a comparison of the displacement field and von Mises stress field contours for two representative examples. Fig. \ref{fig: Case 3 plot} provides detailed comparisons for these examples, along a few specified paths. The geometric details for both test examples are provided in Table \ref{tab: Case 3 geometry}.

\begin{table}[h]
	\centering
	\caption{Structural details for two test case 3 examples (box beam subjected to uniform pressure).}\label{tab: Case 3 geometry}%
	\small
	\begin{tabular}{@{}lccc@{}}
		\toprule
		Category & Test example 3 &  Test example 4 & Unit \\
		\midrule
		Plate thickness    & 13.27   & 17.84  & mm  \\
		Stiffener web thickness    & 11.20   &  11.58 & mm  \\
		Stiffener web height    &  387.5  & 320.1  & mm  \\
		Flange thickness  & 7.75  & 9.96 & mm \\
		Flange width   & 52.61 & 98.64 & mm \\
		Number of stiffeners  & 5 & 4 & $-$ \\
		\toprule
	\end{tabular}
\end{table}

Despite the increased complexity of this test case compared to case 2, both HGT and GraphSAGE still exhibit good prediction performance. In test example 3, the panel is located on the top of the box beam and is subjected to uniform pressure. The maximum displacement and von Mises stress occur in the plate center and flange edge. HGT achieves accuracies of 98.37\% (displacement) and 98.13\% (stress), while GraphSAGE attains similar accuracies of 99.15\% and 93.98\%, respectively. The von Mises prediction performance from both models differ by less than 20\%, and they both accurately captured the trend of the stress distribution across the panel, see Fig. \ref{fig: Case 3 plot}, where the red (HGT) and blue (GraphSAGE) curves align well with FEA simulations. The structure in test example 4 is the bottom panel of the box beam, thus not subjected directly to uniform pressure. The maximum von Mises stress (stress concentration) occurs at the two corners of the panel due to the boundary conditions applied to the box beam (simply supported). Nevertheless, both HGT and GraphSAGE accurately capture the stresses in that area, achieving accuracies of 92.93\% and 98.58\%, respectively, as illustrated in Fig. \ref{fig: Case 3 plot (b)}. However, displacement predictions in test example 4 are less accurate. Along plate edge \#2, both models deviate noticeably from FEA, though HGT still tracks the overall trend, whereas GraphSAGE fails to capture it. Across the remaining paths, HGT maintains good performance, while GraphSAGE shows systematic offsets, resulting in an average displacement error of 0.125 mm, around 351\% higher than HGT’s error.

\subsubsection{Box beam subjected to non-uniform pressure}\label{sec5_3_3}

\begin{figure}[h]
	\centering
	\scriptsize
	\setlength\tabcolsep{1mm}
	\resizebox{\textwidth}{!}{%
		\begin{tabular}{cccc}
			\toprule
			Test example & Model & Displacement field & von Mises field \\
			\midrule
			\multirow{20}{*}{\rotatebox{90}{5 (Side panel 2)}}  
			& HGT  
			& \makecell[c]{\includegraphics[width=0.4\textwidth]{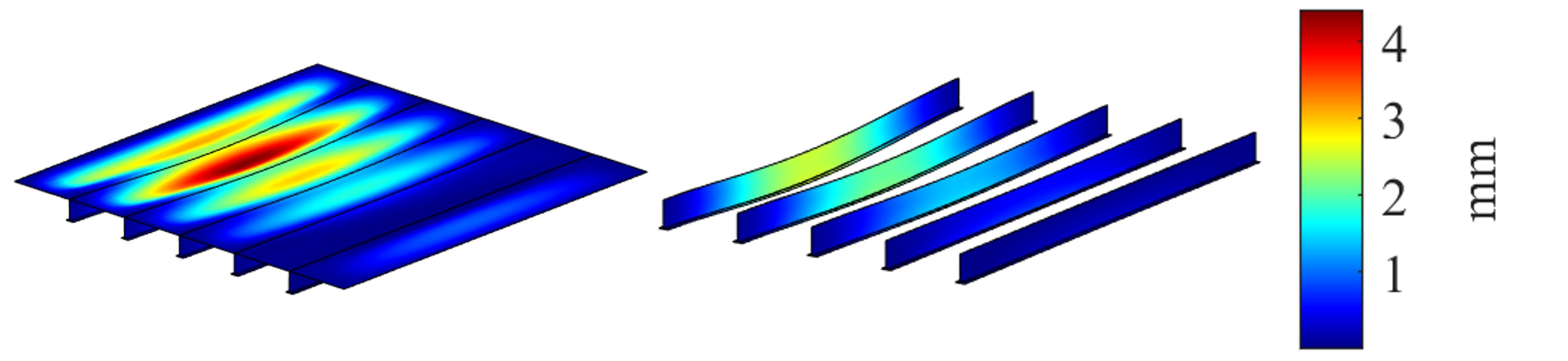}} 
			& \makecell[c]{\includegraphics[width=0.4\textwidth]{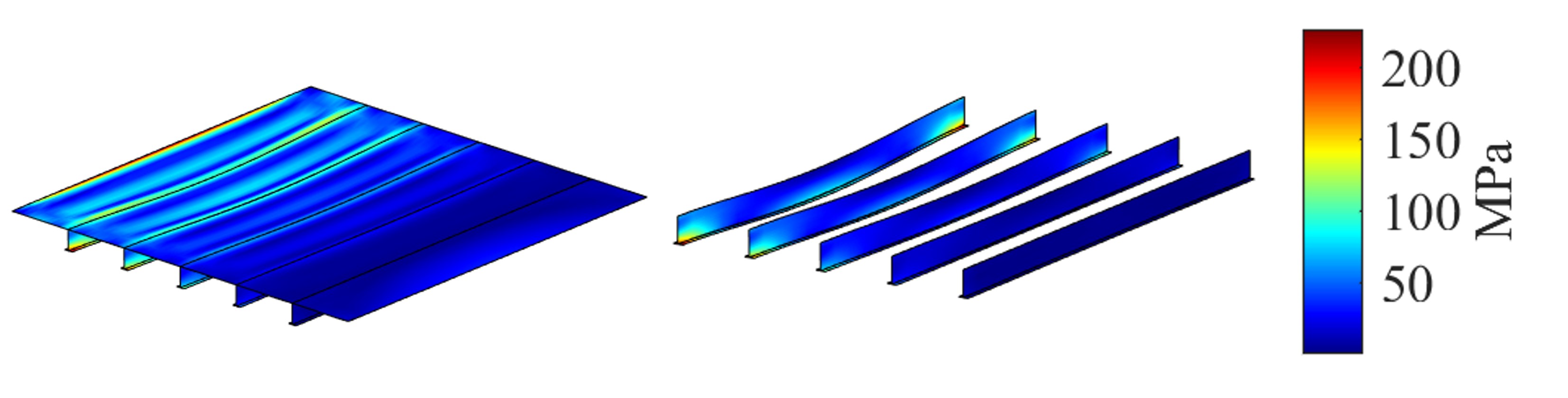}} \\
			& \makecell[c]{HGT Error\\(Disp.\,0.0967 mm;\\ Stress MPa\,2.66)}
			& \makecell[c]{\includegraphics[width=0.4\textwidth]{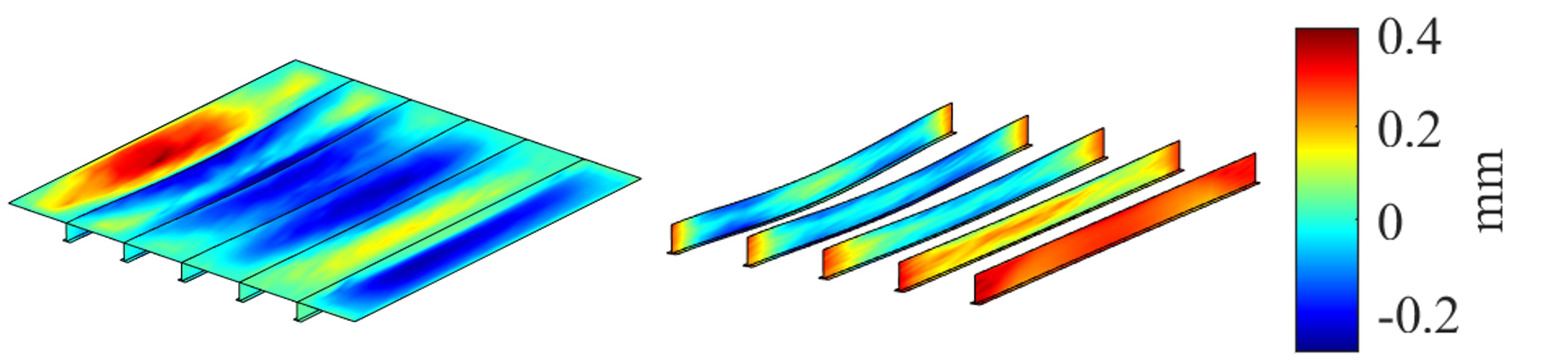}}
			& \makecell[c]{\includegraphics[width=0.4\textwidth]{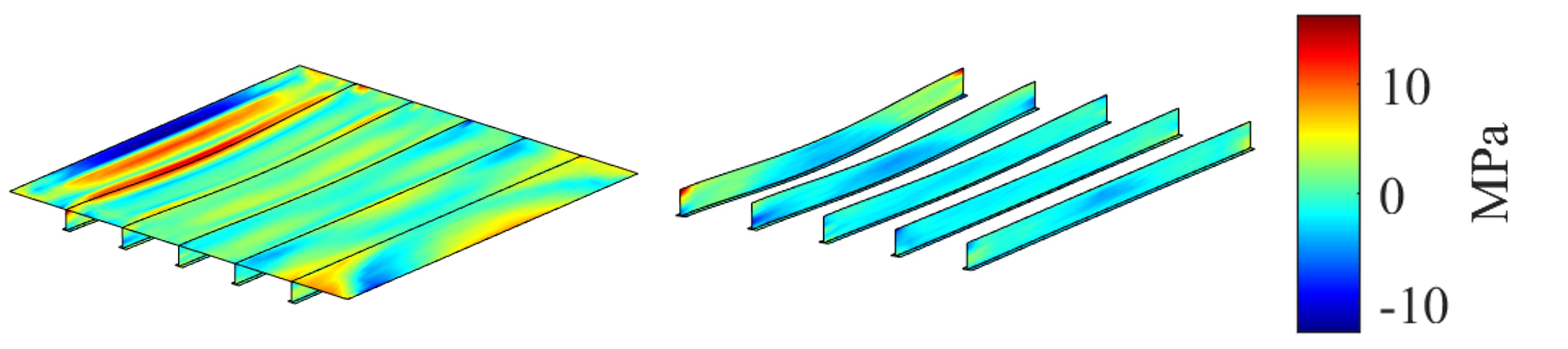}} \\
			& GraphSAGE  
			& \makecell[c]{\includegraphics[width=0.4\textwidth]{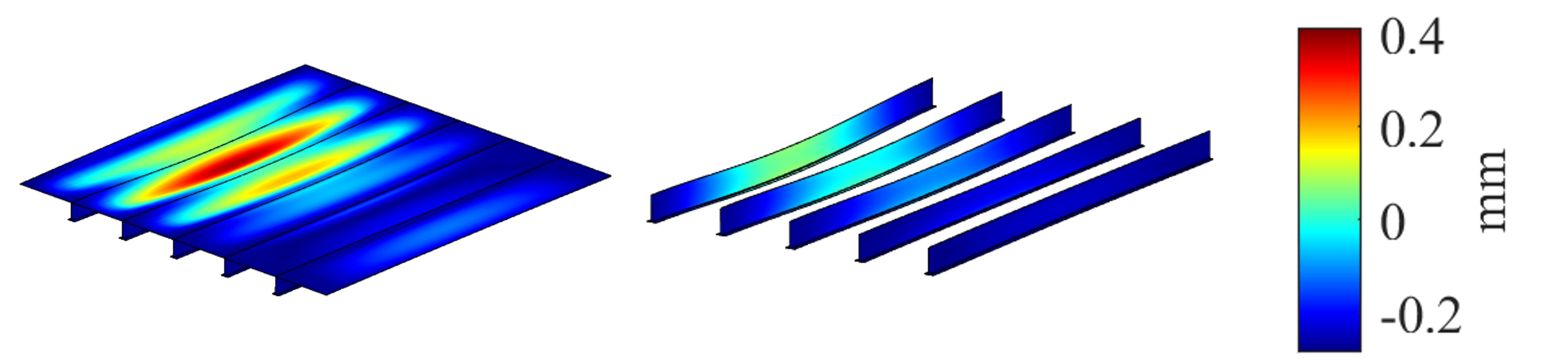}}
			& \makecell[c]{\includegraphics[width=0.4\textwidth]{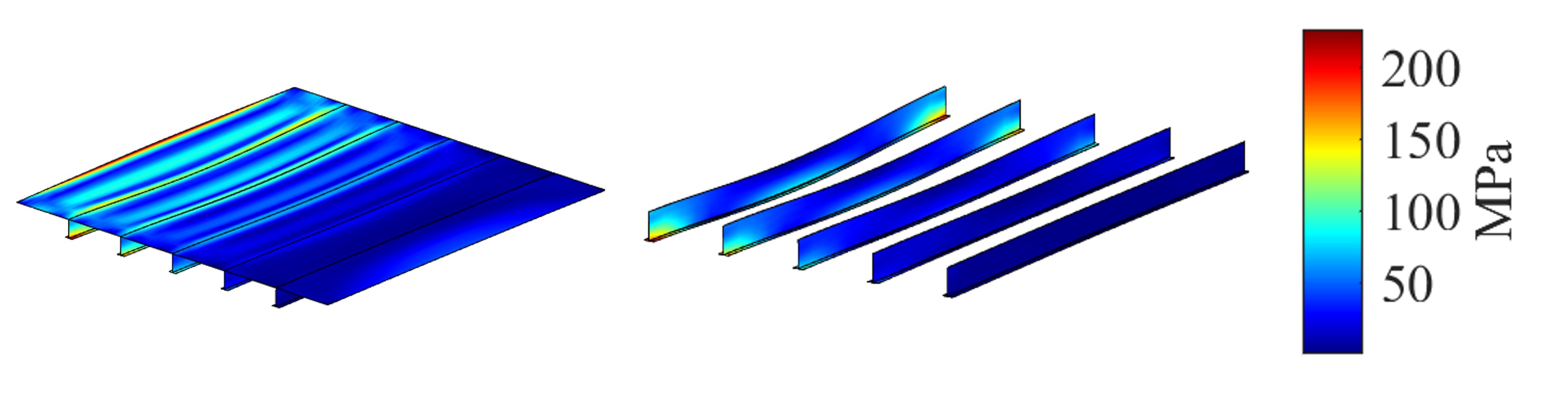}} \\
			& \makecell[c]{GraphSAGE Error\\(Disp.\,0.187 mm;\\ Stress\,4.43 MPa)}
			& \makecell[c]{\includegraphics[width=0.4\textwidth]{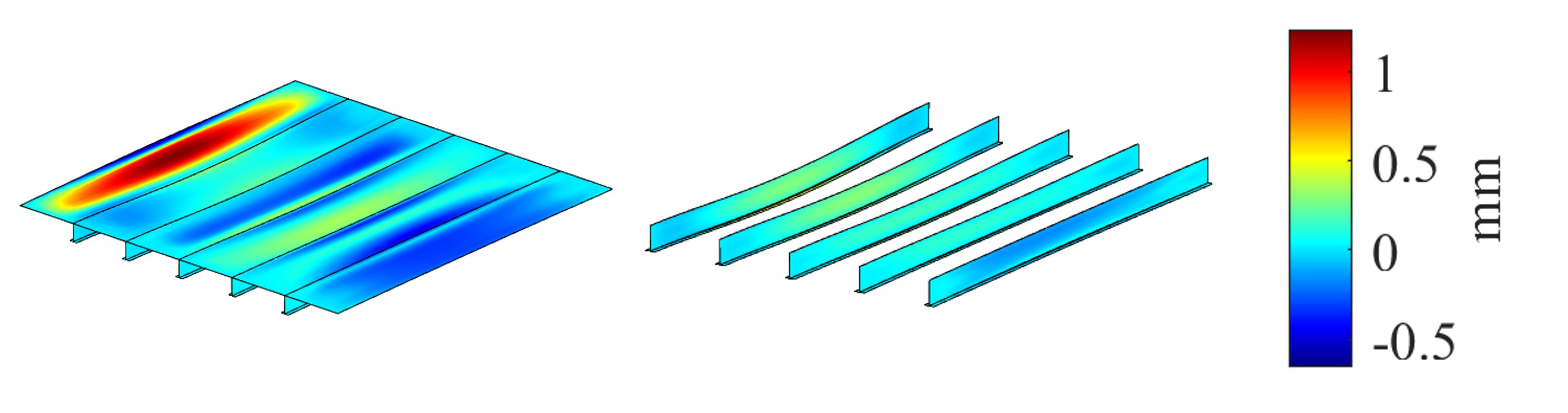}}
			& \makecell[c]{\includegraphics[width=0.4\textwidth]{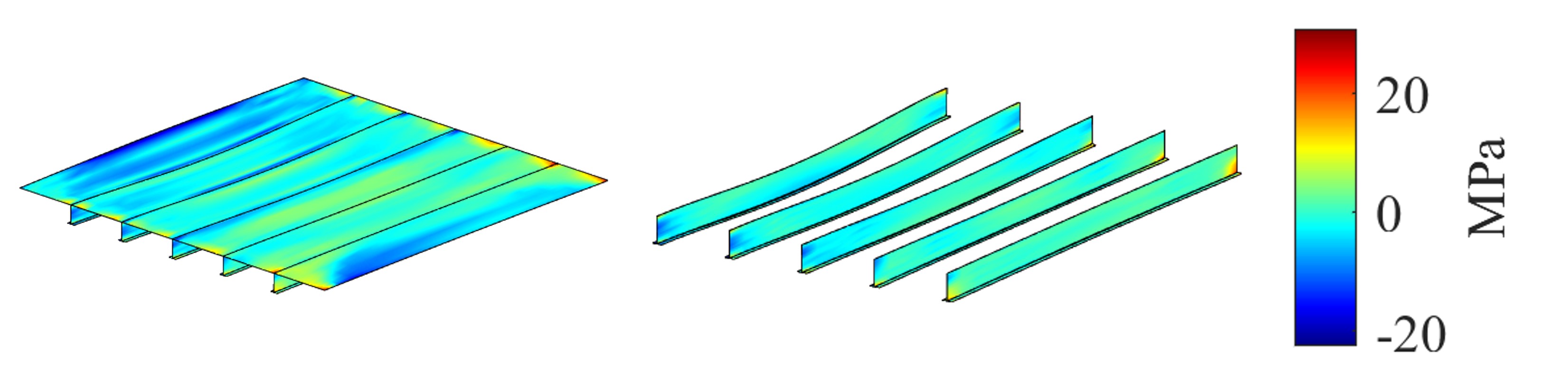}} \\
			& FEA  
			& \makecell[c]{\includegraphics[width=0.4\textwidth]{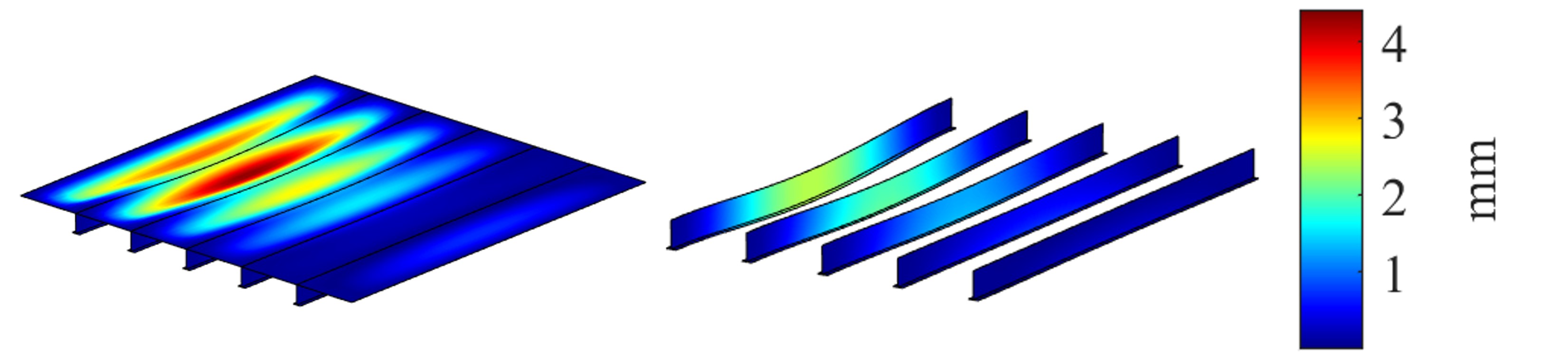}}
			& \makecell[c]{\includegraphics[width=0.4\textwidth]{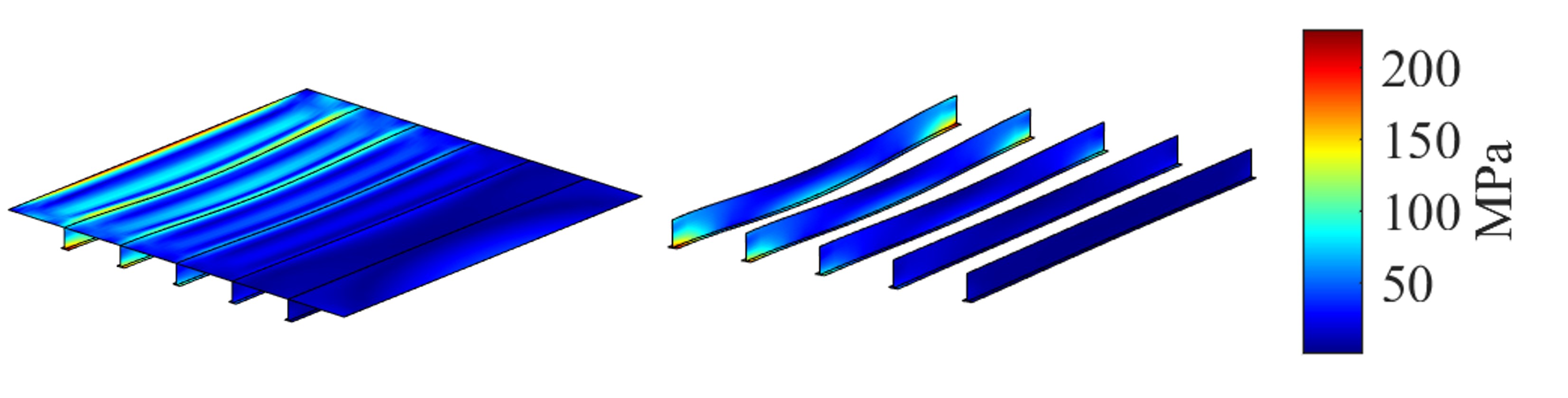}} \\
			\midrule
			\multirow{20}{*}{\rotatebox{90}{6 (Top panel 2)}}  
			& HGT  
			& \makecell[c]{\includegraphics[width=0.4\textwidth]{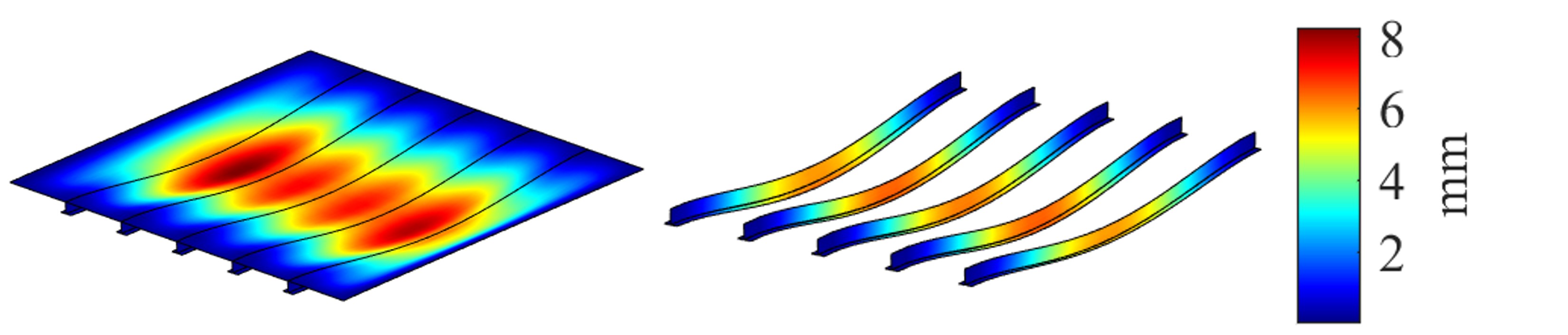}}
			& \makecell[c]{\includegraphics[width=0.4\textwidth]{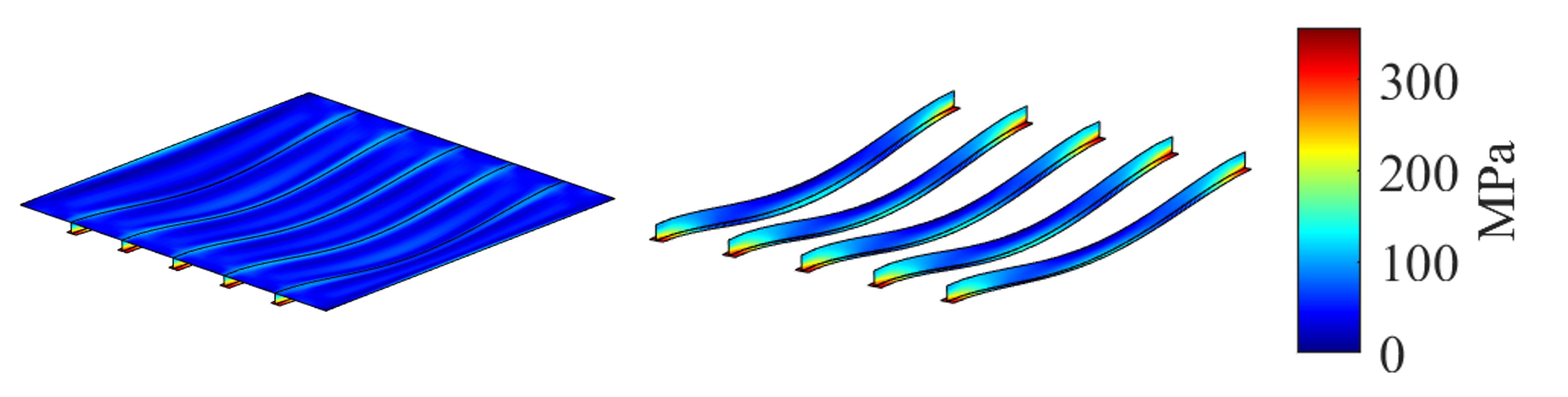}} \\
			& \makecell[c]{HGT Error\\(Disp.\,0.131  ;\\ Stress\,3.70 MPa)}
			& \makecell[c]{\includegraphics[width=0.4\textwidth]{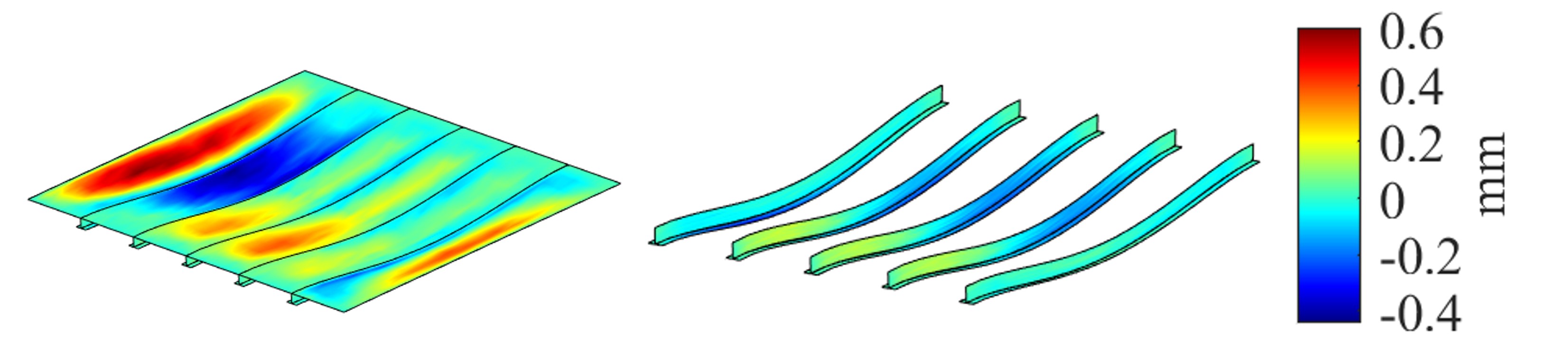}}
			& \makecell[c]{\includegraphics[width=0.4\textwidth]{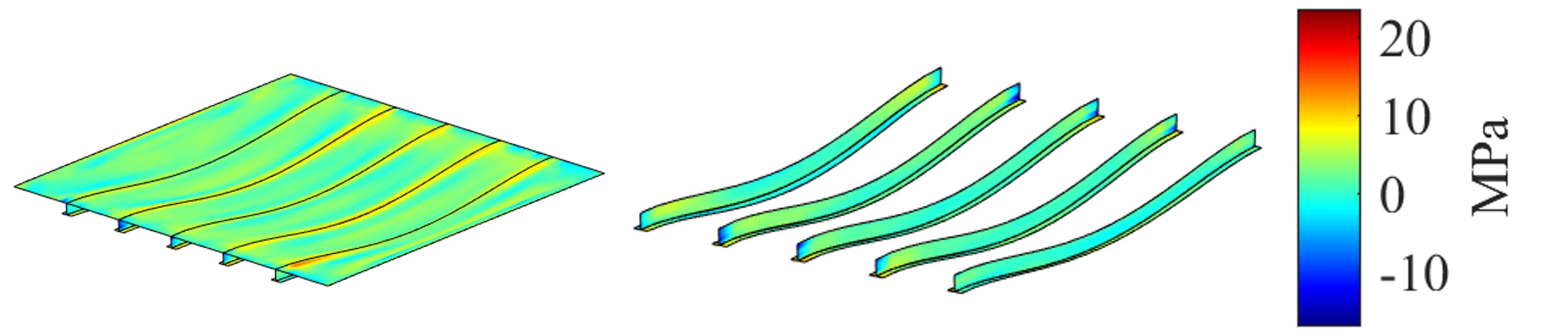}} \\
			& GraphSAGE  
			& \makecell[c]{\includegraphics[width=0.4\textwidth]{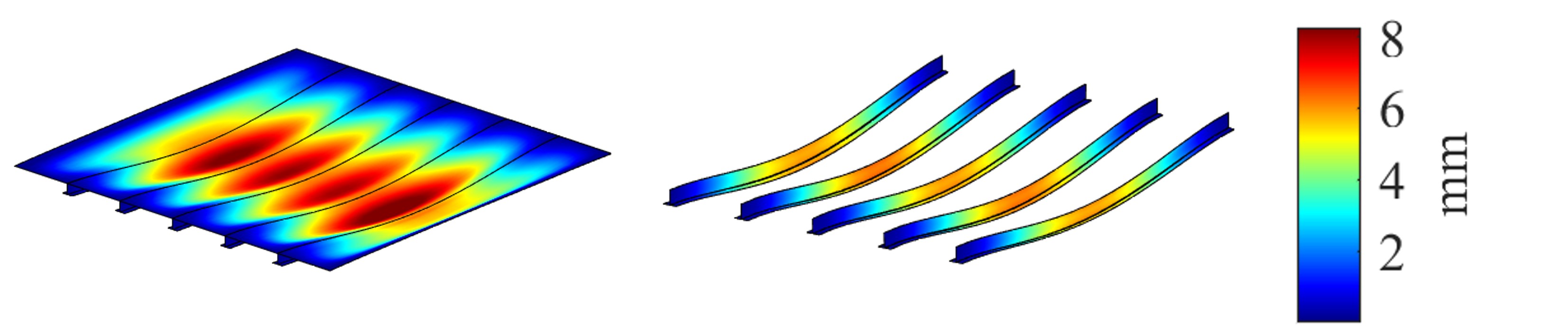}}
			& \makecell[c]{\includegraphics[width=0.4\textwidth]{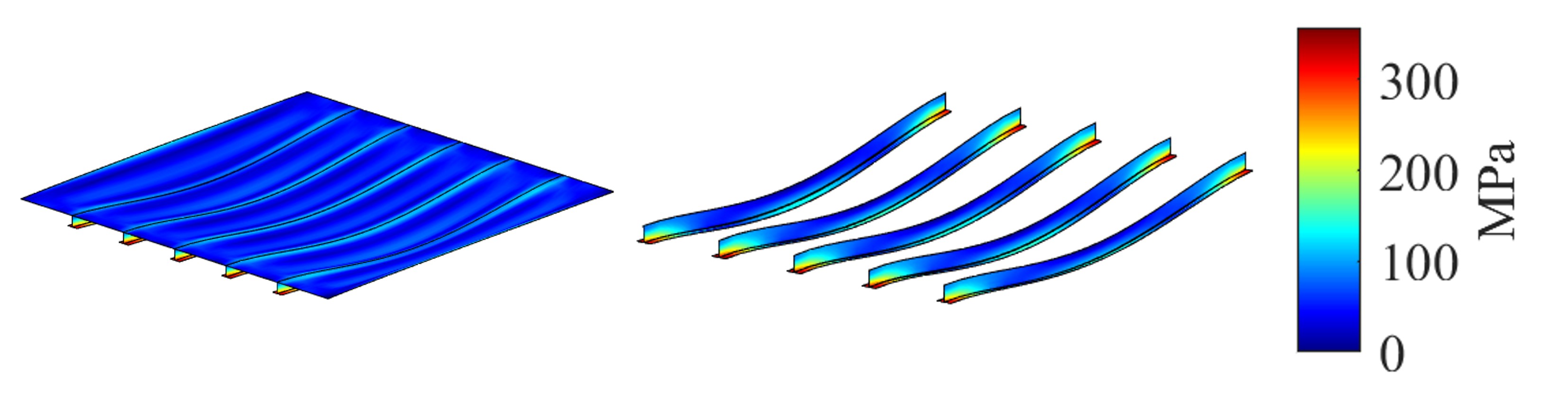}} \\
			& \makecell[c]{GraphSAGE Error\\(Disp.\,0.267 mm;\\ Stress\,8.28 MPa)}
			& \makecell[c]{\includegraphics[width=0.4\textwidth]{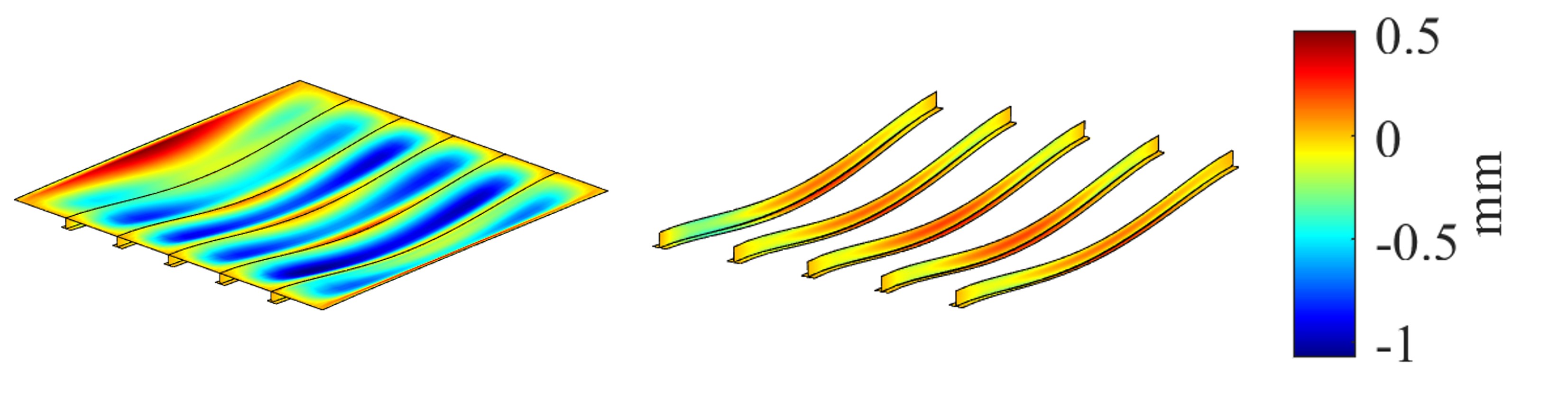}}
			& \makecell[c]{\includegraphics[width=0.4\textwidth]{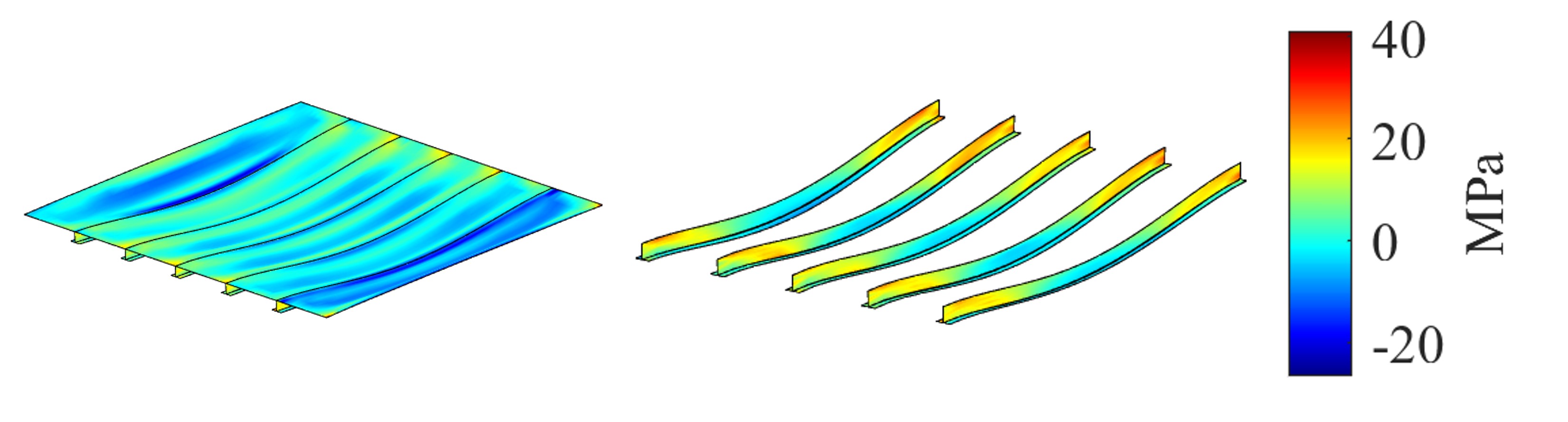}} \\
			& FEA  
			& \makecell[c]{\includegraphics[width=0.4\textwidth]{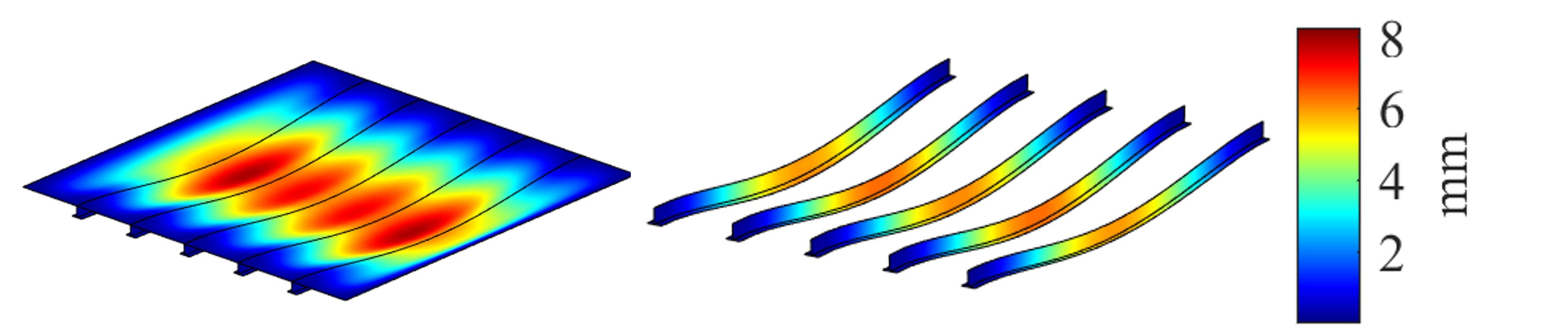}}
			& \makecell[c]{\includegraphics[width=0.4\textwidth]{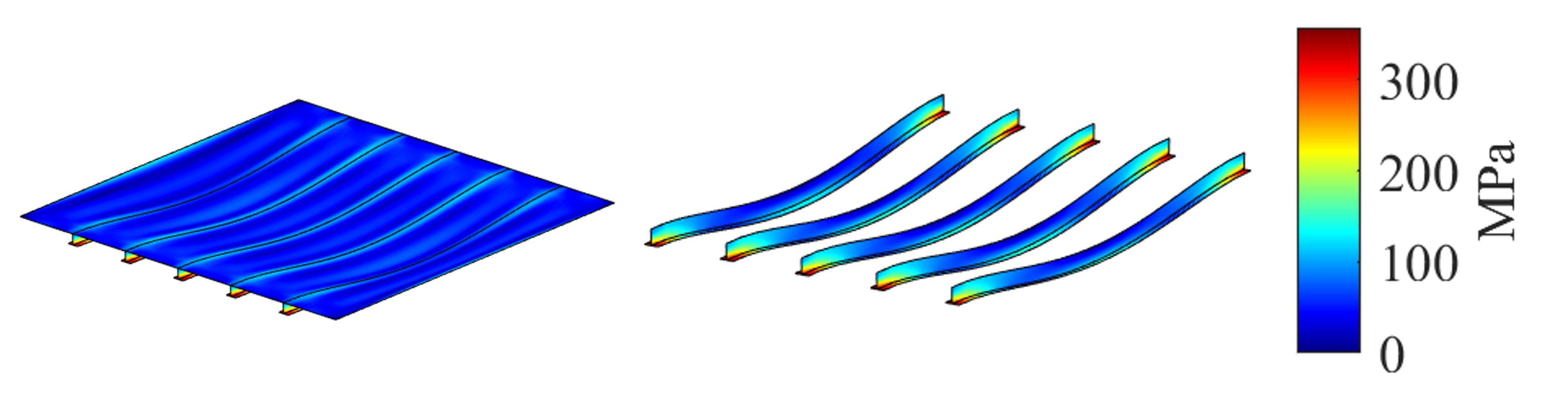}} \\
			\bottomrule
		\end{tabular}%
	}
	\caption{Comparison of HGT and GraphSAGE predictions versus FEA results for test case 4 (box beam subjected to non-uniform pressure).}
	\label{fig: Case 4 contour}
\end{figure}

\begin{figure}[!htbp]
	\centering
	\begin{minipage}[b]{0.42\textwidth}
		\includegraphics[width=\textwidth]{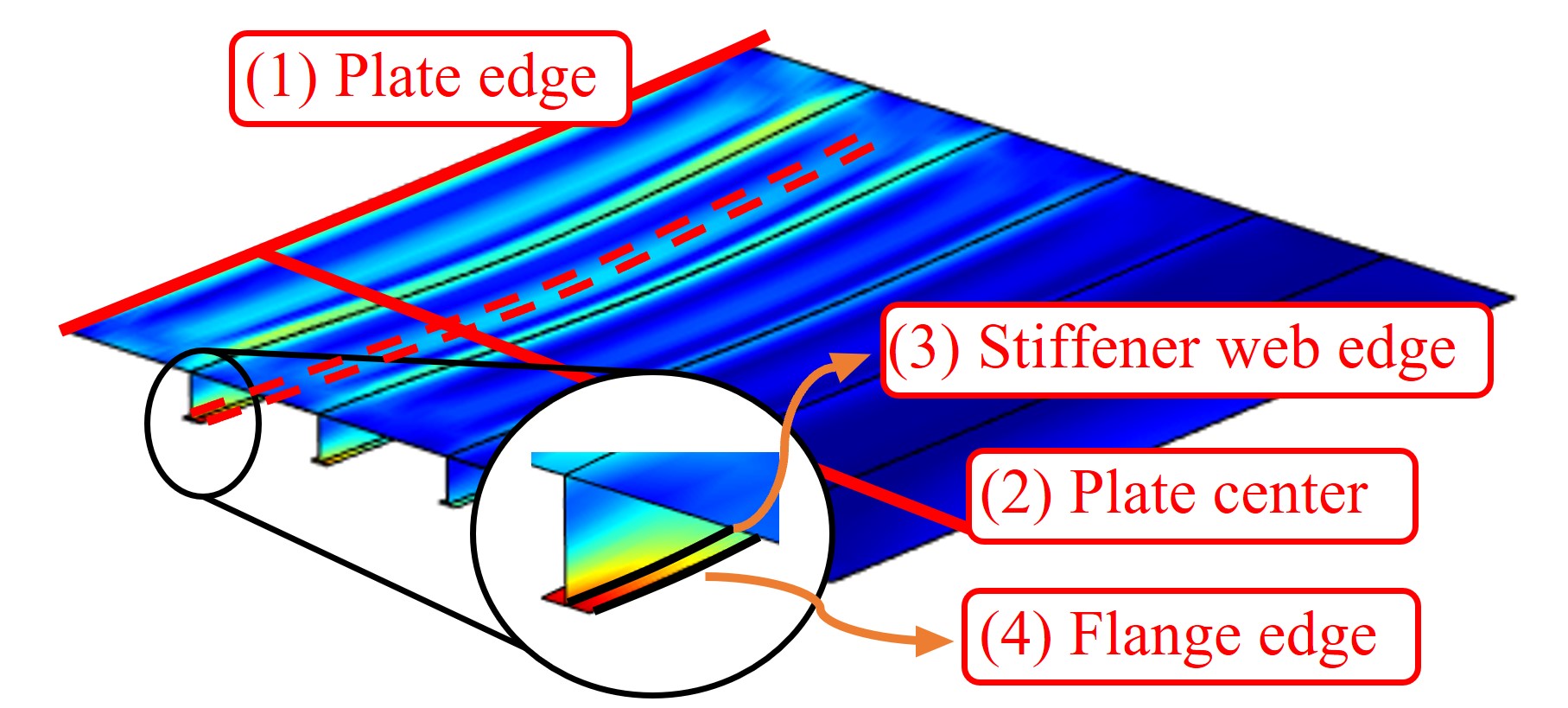}
		\\
		\subfloat[\normalsize Test example 5 (Side panel 2)]{\includegraphics[width=\textwidth]{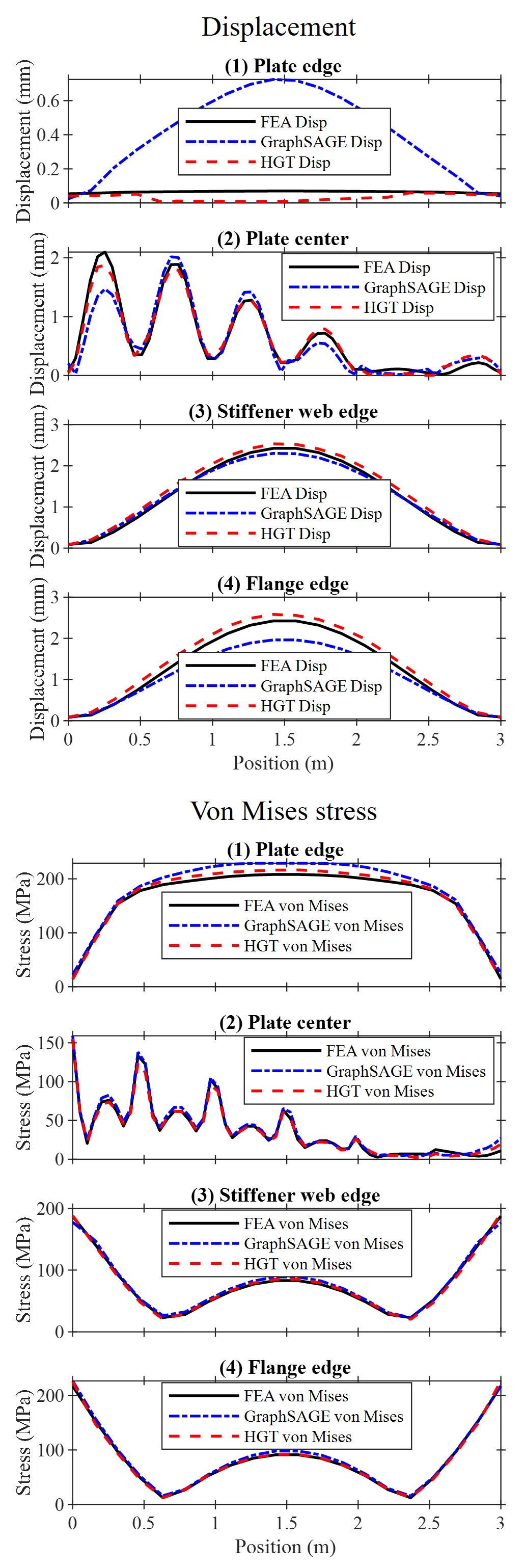}\label{fig: Case 4 plot (a)}}
	\end{minipage}
	\hfill
	\begin{minipage}[b]{0.42\textwidth}
		\includegraphics[width=\textwidth]{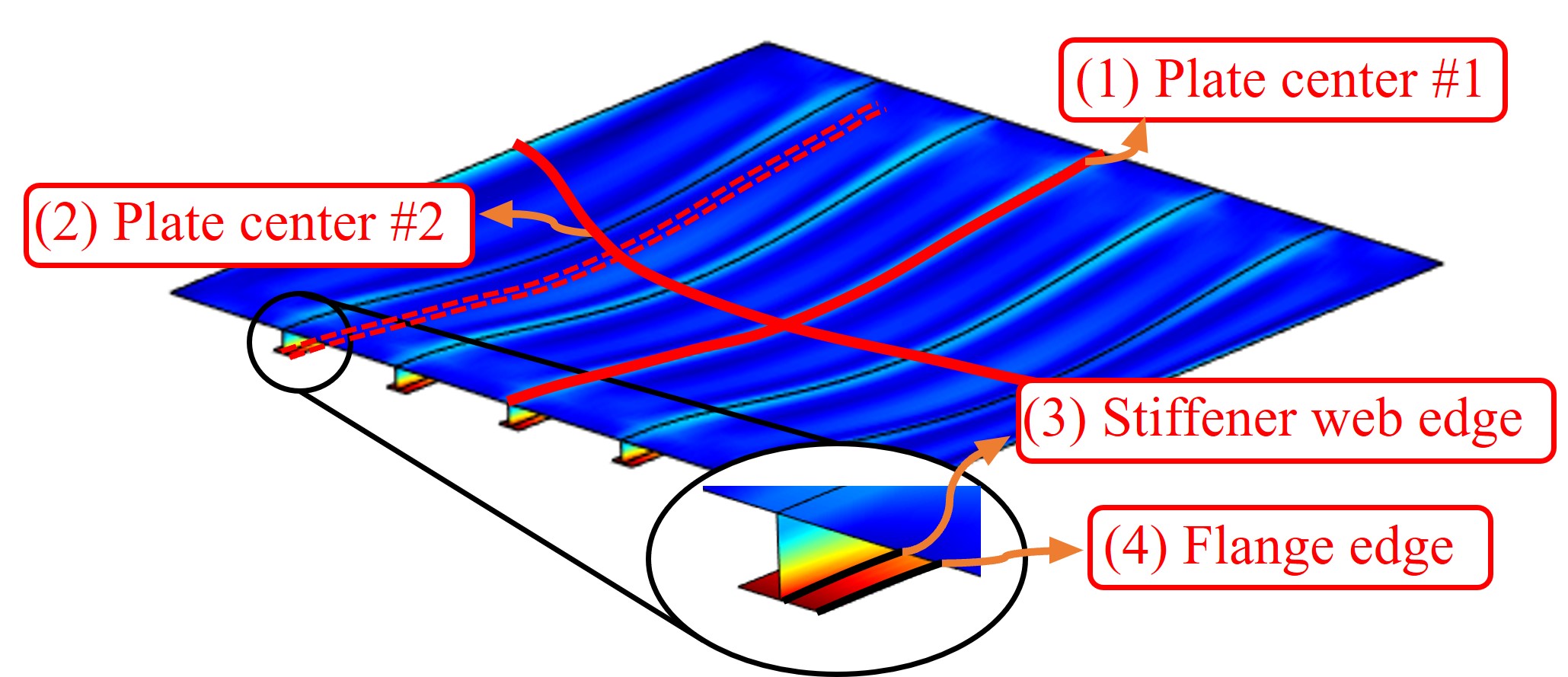}
		\\
		\subfloat[\normalsize Test example 6 (Top panel 2)]{\includegraphics[width=\textwidth]{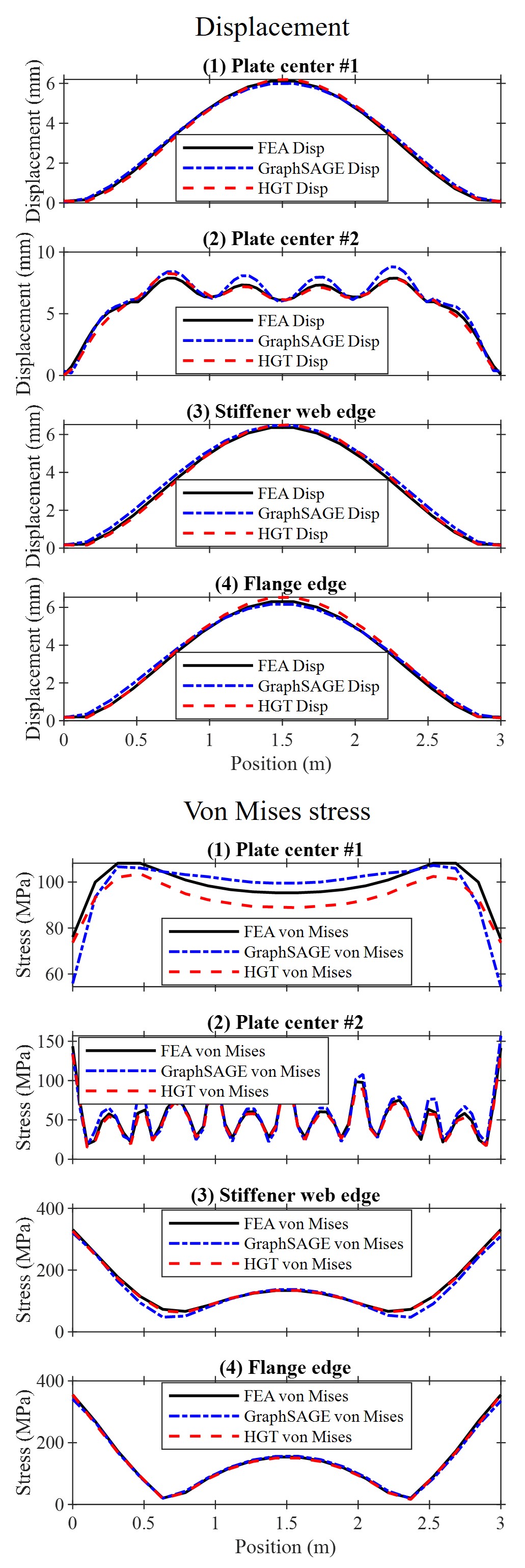}\label{fig: Case 4 plot (b)}}
	\end{minipage}
	\caption{Comparison of total displacement and stress distribution along specified paths for test case 4 examples (box beam subjected to non-uniform pressure).}\label{fig: Case 4 plot}
\end{figure}

In comparison to previous box beam cases, test case 4 features non-uniformly distributed pressure. Due to variations in the deck pressure and structural weight (resulting from the geometry of the box beam), the waterline of the box beam varies. Consequently, the area of the side panel subjected to non-uniform pressure becomes an additional variable in this test case. Fig. \ref{fig: Case 4 contour} shows the displacement and von Mises stress field contours for two representative examples. Fig. \ref{fig: Case 4 plot} provides detailed comparisons for these examples, and the geometric details for both test examples are provided in Table \ref{tab: Case 4 geometry}.

\begin{table}[h]
	\centering
	\caption{Structural details for test case 4 examples (box beam subjected to non-uniform pressure).}\label{tab: Case 4 geometry}%
	\small
	\begin{tabular}{@{}lccc@{}}
		\toprule
		Category & Test example 5 &  Test example 6 & Unit \\
		\midrule
		Plate thickness    & 6.05   & 7.63  & mm  \\
		Stiffener web thickness    & 7.79   &  6.91 & mm  \\
		Stiffener web height    &  134.6  & 79.1  & mm  \\
		Flange thickness  & 2.92  & 3.89 & mm \\
		Flange width   & 64.44 & 88.86 & mm \\
		Number of stiffeners  & 5 & 5 & $-$ \\
		\toprule
	\end{tabular}
\end{table}

Under non-uniform boundary conditions and loading, HGT clearly outperforms GraphSAGE especially for the displacement predictions. In Fig. \ref{fig: Case 4 plot (a)}, HGT slightly underpredicts the plate-edge displacement, whereas GraphSAGE overestimates it by almost an order of magnitude. Along the other sample paths, both methods capture the overall trend, but HGT more closely matches the displacements with an overall RMSE of 0.0967 mm versus 0.187 mm for GraphSAGE. In test example 6, GraphSAGE exhibits better performance on displacement predictions that align more closely with the FEA curves along all four critical paths, yet its RMSE remains nearly twice as large as HGT’s. This performance gap reflects HGT’s ability to leverage non-uniform boundary DOFs and loading patterns. Similar discrepancies can also be observed from the stress predictions for both models. HGT could better captures the contour trends, achieving an RMSE of 3.70 MPa, about 55\% lower than GraphSAGE. Nonetheless, both models exhibit similar accuracy for maximum stress prediction. In examples 5 and 6, HGT obtains maximum stress prediction accuracies of 96.22\% and 98.84\%, while GraphSAGE achieves 94.63\% and 99.27\%, respectively. These high percentages arise because RMSE emphasizes absolute errors, which translate into small relative errors for large stress predictions, and vice versa for smaller values. Thus, both models perform well at predicting maximum values.

\section{Conclusion}\label{sec13}

In this study, we proposed a heterogeneous graph representation for modeling stiffened panels. Panels were subjected to non-uniform generalized loads and boundary conditions. The proposed approach separates structural information into distinct types of nodes: geometry, boundary, and loading nodes, allowing for a precise representation of complex interactions within the structure. This separation of information not only addresses the limitations of previous homogeneous graph representations but also enhances the prediction accuracy of the Heterogeneous Graph Transformer (HGT) model, especially in cases with non-uniformly distributed boundary conditions and loads. We introduced and evaluated several heterogeneous graph representations with different levels of heterogeneity. The case studies involved stiffened panels subjected to patch loading and box beams composed of stiffened panels under various loading conditions. Through comparison analysis with homogeneous graph-based GraphSAGE model, ablation studies, and comprehensive evaluation, we draw the following conclusions:
\begin{itemize}
  \item Defining the structure into distinct node types provides effective representation of stiffened panels for heterogeneous graph neural networks.
  \item The proposed heterogeneous graph representation shows better versatility than the homogeneous graph representations in terms of modeling stiffened panels subjected to non-uniform boundary conditions and loadings.
  \item Increasing graph heterogeneity allows for more accurate neural network predictions, although this may require greater computational resources.
  \item HGT integrated with the proposed embedding efficiently handles various variables across multiple test cases, including material behaviors, geometric variations, and non-uniform generalized loadings and displacements.
  \item The predicted structural behaviors exhibit high consistency with Finite Element Analysis (FEA) results.
\end{itemize}

The findings of this study suggest that incorporating graph heterogeneity into the modeling of stiffened panels with complicated interactions is crucial for capturing the complex relationships between geometric, boundary, and loading information. This approach holds promise for broader applications in structural analysis and optimization, particularly in fields where accurate and efficient modeling of complex structures is essential. Future work could explore the extension of this method to other types of structural systems, as well as the integration of more advanced machine learning techniques to further enhance predictive capabilities.

%%\backmatter

\section*{Acknowledgments}

The authors acknowledge the financial support from the Natural Sciences and Engineering Research Council of Canada (NSERC) [grant number IRCPJ 550069-19]. This research was supported in part through computational resources and services provided by Advanced Research Computing at The University of British Columbia.

\section*{Declarations}

The authors declare that they have no competing financial interests or personal relationships that could have appeared to influence the work reported in this paper.

\section*{Authorship contribution statement}

\textbf{Yuecheng Cai}: Conceptualization, Methodology, Modelling, Validation, Writing - original draft, Writing - review, Writing - editing. \\
\textbf{Jasmin Jelovica}: Conceptualization, Supervision, Funding acquisition, Writing - review, Writing - editing.

%%===========================================================================================%%
\bibliographystyle{elsarticle-num.bst}
\bibliography{Bib}% common bib file
%% if required, the content of .bbl file can be included here once bbl is generated
%%\input sn-article.bbl
\appendix
	
\renewcommand{\thefigure}{\arabic{figure}} % Continue figure 
\renewcommand{\thesection}{Appendix \Alph{section}}

\section{Influence of hyperparameters on the HGT performance with the proposed heterogeneous graph representation}\label{secA1}

\begin{figure}[htp]
	\centering
	\includegraphics[width=0.7\textwidth]{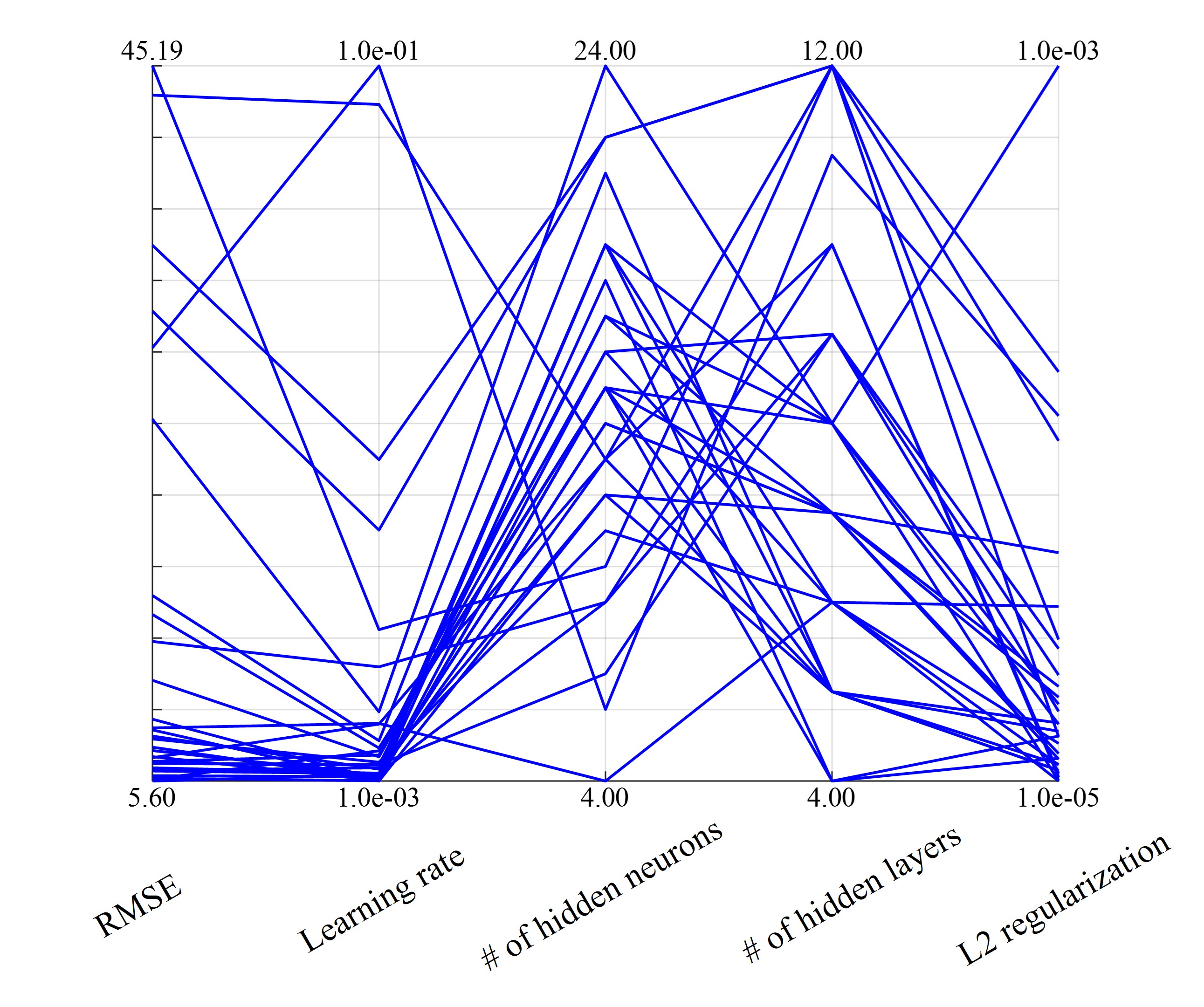}
	\caption{Influence of hyperparameters on HGT with the proposed heterogeneous graph representation for test case 4 (box beam subjected to non-uniform pressure).}\label{fig:hyperparameter}
\end{figure}

We evaluated the influence of hyperparameters on the performance of the Heterogeneous Graph Transformer (HGT) using the proposed type (e) heterogeneous graph representation. This test is conducted on test case 4 with von Mises stress predictions. The results are visualized in the parallel coordinates plot shown in Fig. \ref{fig:hyperparameter}, where each line represents a unique model configuration. The hyperparameters analyzed include the learning rate, the number of hidden layers, the number of hidden neurons, and the L2 regularization. The range of each hyperparameter is displayed across the respective columns.

A clear trend emerges where models with lower learning rates tend to produce lower RMSE values, suggesting that a more conservative learning rate helps prevent overshooting and improves model stability. Similarly, configurations with lower L2 regularization also demonstrate improved performance, indicating that the model is not prone to overfitting in these cases. In contrast, the number of hidden layers and hidden neurons exhibits more variability, with no clear correlation between these hyperparameters and the RMSE, implying that neural network architecture may not be a major influencing factor.

\section{Stress prediction for test case 4: Box beam subjected to non-uniform pressure with smaller training data set}\label{secA2}

\begin{figure}[htp]
	\centering
	\includegraphics[width=1\textwidth]{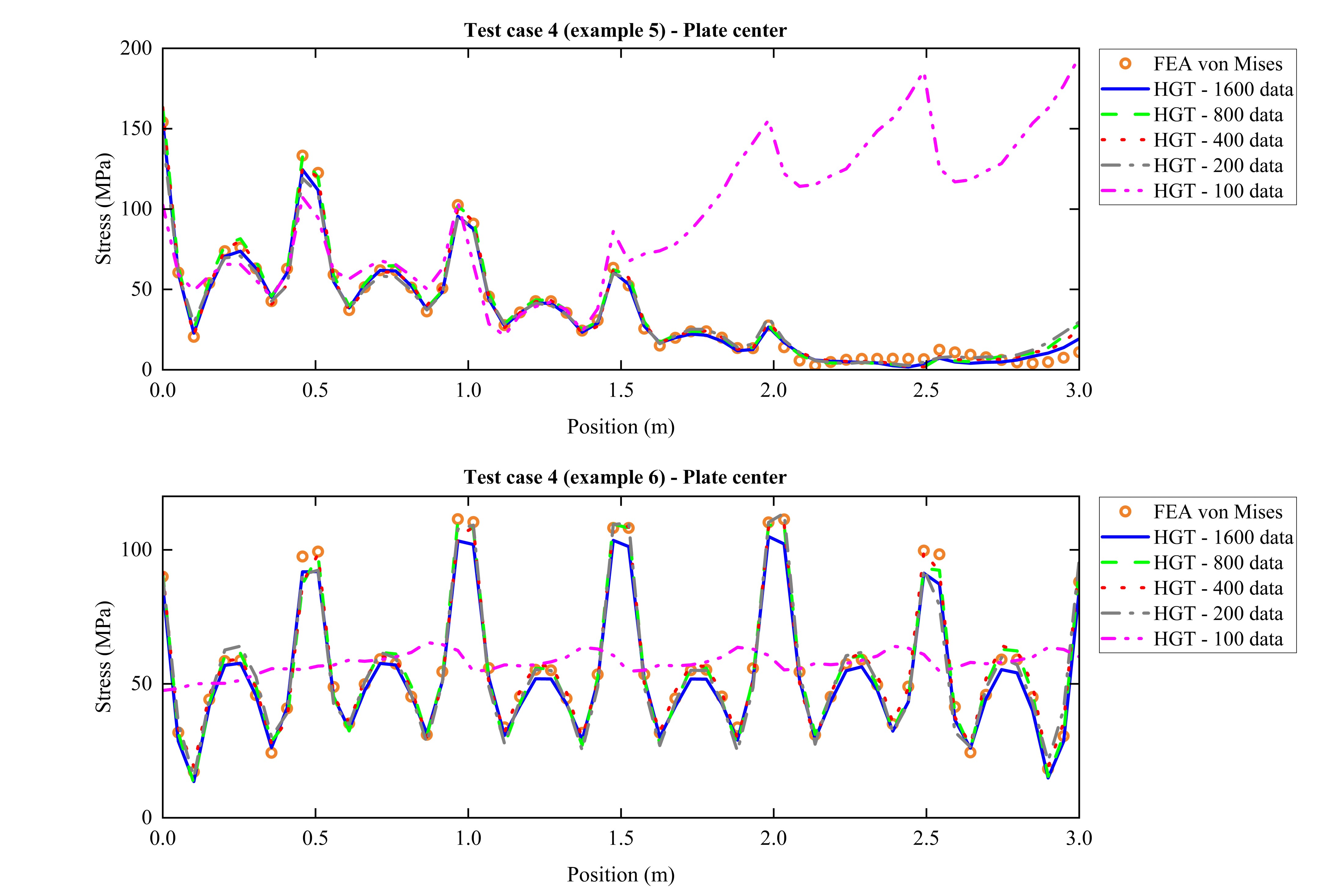}
	\caption{Effect of training dataset size on the performance of HGT for von Mises stress prediction.}\label{fig:paper2_training_data}
\end{figure}

\begin{figure}[htp]
	\centering
	\includegraphics[width=0.5\textwidth]{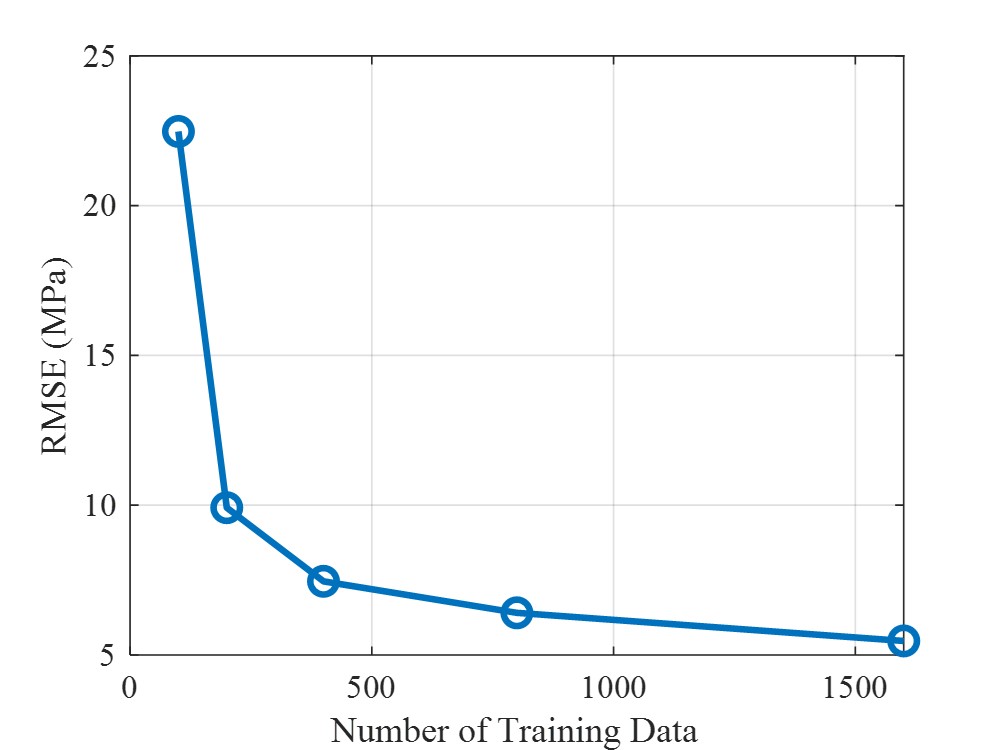}
	\caption{Effect of training dataset size on RMSE performance.}\label{fig:paper2_training_data_curve}
\end{figure}

The effect of the training dataset size on the performance of HGT is investigated in this section. We present von Mises stress predictions along the plate center path for test case 4, selected as it represents the most complex dataset within this work. The plots in Fig. \ref{fig:paper2_training_data} display the predictions by HGT model trained on datasets comprising 1600, 800, 400, 200, and 100 data points. Fig. \ref{fig:paper2_training_data_curve} shows the corresponding RMSE values for each training data set size.

From the results, it is evident that the model trained on 1600 data points closely aligns with the finite element analysis (FEA) von Mises stress distribution, exhibiting minimal deviations. This indicates that, with sufficient data, the HGT model is capable of accurately capturing complex stress patterns. Both figures show that the prediction accuracy gradually deteriorates with data reduction, but interestingly, remains relatively stable until the training dataset is reduced to 100 data points, at which point signs of overfitting become apparent. The other test example in the figure exhibits the same, highlighting that the proposed heterogeneous graph representation maintains strong generalization performance across the datasets used in this study.

\end{document}